\documentclass[12pt,a4paper]{book}
\usepackage{graphicx}
\usepackage{longtable}
\usepackage{amssymb}
\makeindex
\begin{document}

\def\LambdaTensor{{\tt LambdaTensor}}
\def\cV{{\mathcal V}}
\def\cA{{\mathcal A}}
\def\cB{{\mathcal B}}
\def\cC{{\mathcal C}}
\def\cD{{\mathcal D}}
\def\cL{{\mathcal L}}
\def\cM{{\mathcal M}}
\def\cN{{\mathcal N}}
\def\cP{{\mathcal P}}
\def\cQ{{\mathcal Q}}
\def\cJ{{\mathcal J}}
\def\dA{{\dot A}}
\def\dB{{\dot B}}
\def\dC{{\dot C}}
\def\beq{\begin{equation}}
\def\eeq{\end{equation}}
\def\bea{\begin{eqnarray}}
\def\eea{\end{eqnarray}}
\def\Eee{E_{8(+8)}}
\def\tr{{\rm tr\,}}
\def\IJ{{\underline{[IJ]}}}
\def\KL{{\underline{[KL]}}}
\def\MN{{\underline{[MN]}}}
\def\phnu{{\phantom+0}}

\def\Arg{{\rm Arg\,}}
\def\atan{{\rm atan\,}}
\def\Arcosh{{\rm Arcosh\,}}

\long\def\ignore#1{}


{
\thispagestyle{empty}
\begin{flushright}
AEI-2003-046\\
{\tt hep-th/0305176}
\end{flushright}

\begin{center}
\vskip1cm
{\bf\Large Mapping the vacuum structure of gauged maximal supergravities: an application of high-performance symbolic algebra}

\bigskip\bigskip\bigskip

{\bf Thomas Fischbacher}
\vskip.1cm
{\em 
Max-Planck-Institut f\"ur Gravitationsphysik\\
Albert-Einstein-Institut\\
M\"uhlenberg 1\\
14476 Golm, Germany
}
\vskip.1cm
{\tt tf@aei.mpg.de}

\vskip2cm
{\bf Abstract}
\end{center}

\medbreak

\noindent
  The analysis of the extremal structure of the scalar potentials of
  gauged maximally extended supergravity models in five, four, and
  three dimensions, and hence the determination of possible vacuum
  states of these models is a computationally challenging task due
  to the occurrence of the exceptional Lie groups $E_6$,
  $E_7$, $E_8$ in the definition of these potentials.
  At present, the most promising approach to gain information about
  nontrivial vacua of these models is to perform a truncation
  of the potential to submanifolds of the $G/H$ coset manifold
  of scalars which are invariant under a subgroup of the gauge
  group and of sufficiently low dimension to make an analytic treatment
  possible.
  
  New tools are presented which allow a systematic and highly
  effective study of these potentials up to a previously unreached
  level of complexity. Explicit forms of new truncations of the
  potentials of four- and three-dimensional models are given, and for
  $N=16$, $D=3$ supergravities, which are much more rich in structure
  than their higher-dimensional cousins, a series of new nontrivial
  vacua is identified and analysed.
}

\newpage

\ignore{
\thispagestyle{empty}
\small
\begin{center}
\textsc{Mapping the vacuum structure of gauged maximal supergravities:\break
an application of high-performance symbolic algebra}

\bigskip
\bigskip

{\tt Dissertation}

\medskip

zur Erlangung des akademischen Grades

\medskip

{\tt doctor rerum naturalium}

\medskip

(Dr. rer. nat.)

\medskip

im Fach Physik

\medskip

eingereicht an der

\bigskip

Mathematisch-Naturwissenschaftlichen Fakult\"at I

\medskip

der Humboldt-Universit\"at zu Berlin

\medskip

von

\medskip

Herrn Dipl. Phys. Univ. Thomas Fischbacher\break
Geboren am 27.06.1975 in D-83278 Traunstein

\bigskip\bigskip

Pr\"asident der Humboldt-Universit\"at zu Berlin\break
Prof. Dr. J\"urgen Mlynek

\medskip

Dekan der Mathematisch-Naturwissenschaftlichen Fakult\"at I\break
Prof. Dr. Michael Linscheid

\end{center}

\bigskip
\bigskip

\begin{tabular}{lcl}
Gutachter/innen:&&1. $\underline{\mbox{\kern6cm}}$\\
&&\\
&&2. $\underline{\mbox{\kern6cm}}$\\
&&\\
&&3. \underline{\mbox{\kern6cm}}\\
&&\\
&&\\
Tag der m\"undlichen Pr\"ufung:&&\phantom{0.} $\underline{\mbox{\kern6cm}}$
\end{tabular}

\newpage
\clearpage{\thispagestyle{empty}\cleardoublepage}
\newpage

\newpage

{\bf Abstract}

\medbreak

  The analysis of the extremal structure of the scalar potentials of
  gauged maximally extended supergravity models in five, four, and
  three dimensions, and hence the determination of possible vacuum
  states of these models is a computationally challenging task due
  to the occurrence of the exceptional Lie groups $E_6$,
  $E_7$, $E_8$ in the definition of these potentials.
  At present, the most promising approach to gain information about
  nontrivial vacua of these models is to perform a truncation
  of the potential to submanifolds of the $G/H$ coset manifold
  of scalars which are invariant under a subgroup of the gauge
  group and of sufficiently low dimension to make an analytic treatment
  possible.
  
  New tools are presented which allow a systematic and highly
  effective study of these potentials up to a previously unreached
  level of complexity. Explicit forms of new truncations of the
  potentials of four- and three-dimensional models are given, and for
  $N=16$, $D=3$ supergravities, which are much more rich in structure
  than their higher-dimensional cousins, a series of new nontrivial
  vacua is identified and analysed.

\bigskip

{\bf Zusammenfassung}

\medbreak

  Die Analyse der extremalen Struktur der
  Skalarfeldpotentiale der ge\-eichten maximal erweiterten
  Supergravitation in f\"unf, vier, und drei Dimensionen
  -- und damit das Auffinden m\"oglicher Vakua -- wird durch das
  Auftreten der exceptionellen Lie-Gruppen $E_6$, $E_7$, $E_8$ in der
  Definition dieser Potentiale zu einer rechnerischen
  Herausforderung. Der derzeit erfolgversprechendste Ansatz
  um nichttriviale Vakua aufzusp\"uren ist, eine Ein\-schr\"an\-kung
  des Potentials auf Untermannigfaltigkeiten des symmetrischen
  Raums $G/H$ der Skalare vorzunehmen, welche bez\"uglich
  einer Untergruppe der Eichgruppe invariant und hinreichend
  niedrigdimensional sind, um eine ana\-lyti\-sche Behandlung
  zu erm\"oglichen.
  
  Neue Werkzeuge werden vorgestellt, die eine systematische und
  hoch\-gradig effektive Untersuchung dieser Potentiale bis zu einer
  bisher uner\-reichten Komplexit\"at erm\"oglichen. Neue
  Trunkierungen der Potentiale der vier- und dreidimensionalen
  Supergravitationstheorien werden explizit ange\-geben und f\"ur die
  $N=16$, $D=3$ Theorien, welche an Struktur wesentlich reichhaltiger
  sind als ihre h\"oherdimensionalen Verwandten wird eine Reihe neuer
  nichttrivialer Vakua aufgefunden und analysiert.
}

\newpage
\[\]
\newpage

\[
\]
\[
\]

\begin{center}
Dedicated to the memory of\break
Yvonne Schie\ss{}ler\break
1976--2002\break
A great woman, a brilliant dentist, and my closest friend.
\end{center}

\newpage

\tableofcontents
\newpage

\chapter{Introduction}


While three of the four fundamental interactions -- namely the strong,
weak, and electromagnetic force -- can be cast into the framework of a
unified quantum theory, one immediate problem that has to be overcome
to also include the last interaction, gravity, is that the machinery
of non-abelian gauge theory, which serves so well for the other three
forces, cannot be applied directly. From the particle physicist's
point of view, one important point is that the quantum of the field
mediating gravity, the graviton, is not a spin-1 (vector) particle, as
for the other interactions, but a spin-2 (traceless symmetric tensor)
particle. Since an unified `theory of everything' ultimately should
not only quantitatively describe the observed particles and forces,
but also give a good explanation why just this spectrum of matter and
interaction particles is found in nature, theoretical approaches
that unify particles of different spin deserve special interest.

While a continuous symmetry whose generators form a Lie algebra cannot
connect particles of different spin in an interacting theory (due to
the Coleman-Mandula theorem \cite{Coleman:ad}), this can be achieved
by employing a symmetry that contains {\em anticommuting} spinorial
generators \cite{Golfand, Haag:1974qh, Volkov:jx, Wess:tw}; such a
symmetry, which connects bosons with fermions, has been dubbed {\em
  supersymmetry}. Soon after the advent of supersymmetry, it has been
realized that a supersymmetric version of Einstein's theory of gravity
exists \cite{Freedman:1976xh, Deser:1976eh}\footnote{see e.g.
\cite{VanNieuwenhuizen:ae} for a comprehensive introduction}. This
can just as well be regarded as a theory with local (i.e.  gauged)
supersymmetry -- superficially, since the anticommutator of two
supersymmetry transformations is a translation, local supersymmetry
gives rise to spacetime diffeomorphisms. Vice versa, since
supersymmetry is not an internal symmetry, it must be made a local
symmetry when going to curved spacetime.

As a single supersymmetry transformation maps bosons to fermions and
vice versa by changing the spin by $1/2$, and has vanishing square due
to the anticommuting property, particles are grouped into
supermultiplets containing a fermionic superpartner for every boson
and vice versa. Now, it is possible to introduce more than one
independent supersymmetry transformation \cite{Ferrara:1976fu}, and
hence make these multiplets spawn a larger range of spins, up to a
maximal number of $N=8$ independent supersymmetry transformations
\cite{Cremmer:1980zb} combining the helicity $+2$ graviton with the
helicity $-2$ graviton in one unique supermultiplet that entirely
fixes the particle content of the theory.

Two observations deserve special attention here: first, the total
number of spinorial components of eight four-dimensional real
(Majorana, hence four-component) spinors is just the same as that of a
fundamental eleven-dimensional spinor, indicating the possibility to
construct a theory of an `extremal' eleven-dimensional $N=1$
supergravity \cite{Cremmer:1978km}, which gives rise to various other
supergravity theories upon dimensional reduction, but does not produce
massless particles with spin greater than 2 (which, as common lore
tells, cannot consistently be coupled to gravity \cite{Berends:wu})
upon reduction to four dimensions.  Second, the equations of motion of
maximal extended supergravity in $D=4$ possesses a global $E_{7(+7)}$
symmetry, and in fact, this symmetry conjectured from the study of
dimensional reduction of $D=11$ supergravity to various dimensions
considerably facilitated the construction of maximal $N=8$ $D=4$
supergravity. In general, the supergravity obtained from toroidal
compactification of $D=11$ supergravity on a $d$-torus will, after
dualization of all emerging $p$-forms to lowest possible rank, exhibit
a global $E_{11-d(11-d)}$ symmetry \cite{Cremmer:1997ct}.  For the
particular case of $d=8$, this gives the largest finite-dimensional
exceptional group $E_{8}$.

Soon after the construction of extended supergravity, it was realized
that part of the extra global symmetry can be made
local\cite{Freedman:1976aw}; in particular, in $D=4$, one can
introduce a local $SO(8)$ gauge symmetry under which the 28 vector
fields of the supermultiplet transform in the adjoint representation
\cite{deWit:1981eq,deWit:1982ig}.
The construction is quite remarkable in many aspects: first, it
contains gravity as well as nonabelian gauge symmetry (albeit with a
gauge group too small to accommodate the fundamental interactions of
the standard model), second, it is, up to the size of the nonabelian
coupling constant, a totally rigid construction (just as $D=11$
supergravity itself) that does not feature any additional arbitrary
parameters, like the number of particles per type, or their couplings,
third, re-establishing supersymmetry after gauging introduces a
potential for the scalars with a very rich extremal structure, giving
rise to a large number of different possible vacua for such a theory.

While the quantization of general relativity suffers from the
fundamental problem that in the summation over all histories leading
from a given initial to a given final state, contributions from
intermediate (virtual) particles with arbitrarily high momentum cannot
be brought under control by the adjustment of finitely many
parameters, as is required for a physical theory to have predictive
power; that is, general relativity suffers from nonrenormalizability.
Since intermediate bosons and fermions give similar high-energy
contributions of opposite sign, it was hoped that supergravity may
also provide a renormalizable, or rather, {\em finite} quantum theory
of gravity, due to cancellation of positive and negative divergences.

The state of affairs reported so far roughly mirrors the focus of very
fruitful research on quantum gravity and unified theories during the
first half of the 1980's; in particular the study of siblings of
$D=11$ $N=1$ supergravity obtained by dimensional reduction to all
possible spaces of lower dimension turned out to provide valuable
insights into the general underlying structure of such theories. With
the advent of superstring theory, which contains supergravity as a
low-energy limit and seemingly bears a better prospect to eventually
lead to a viable `Theory of Everything', research activity on
supergravity generally declined, but was revived by recent major
discoveries; of prime importance is of course the observation that the
five possible consistent ten-dimensional superstring theories do not
stand in isolation, as was initially believed, but are interrelated by
a `web of dualities' which also connects them to {\em
  eleven}-dimensional supergravity \cite{Witten:1995ex} and is
regarded as compelling evidence for the existence of a fundamental
eleven-dimensional `mother' theory which has supergravity as its low
energy limit and was preliminarily named {\em M-theory}. At present,
it is hard to tell what will emerge from the up to now fragmentary
knowledge of the structure of M-theory, but it is conjectured that
discrete remnants of the hidden global $E_{11-d(11-d)}$ symmetries
that emerge in toroidal compactifications of $D=11$ supergravity are
already present in the original theory, and hence, obtaining a better
understanding of these exceptional symmetries in supergravity is of
major importance.

Results on vacua of gauged supergravities presented in this thesis,
which mainly deals with $D=3$, are of considerable direct relevance to
the celebrated conjectured AdS/CFT duality originally proposed by
Maldacena \cite{Maldacena:1997re} for which a large body of evidence
has been collected by now. It is claimed that supergravity in
Anti-deSitter (AdS) space has a dual description in terms of a
conformal field theory (CFT) living on the boundary of this AdS
space. Thus, for example, supergravity solutions interpolating between
different vacua are in particular believed to encode a renormalization
group flow for the corresponding CFT
\cite{Freedman:1999gp,Girardello:1999bd,Bianchi:2001de,Berg:2001ty}.

Even by itself, the three-dimensional maximal gauged supergravity
models on which we focus here are quite remarkable. First, the
underlying exceptional symmetry is the maximal one, $\Eee$, second,
and in marked contrast to higher-dimensional supergravity theories,
the vector fields appear in these models via non-abelian Chern-Simons
terms rather than the usual Yang-Mills terms. Since these terms which
correspond to a non-abelian duality between vectors and scalars that
does not have an analogue in higher dimensions cannot be obtained by
any known type of dimensional reduction, it is impossible to get the
$N=16$ $D=3$ gauged supergravity models from $D=11$ supergravity, in
contrast to higher-dimensional supergravity as well as the ungauged
and half-maximal ($N=8$) $D=3$ models. This might hint at a new
supergravity theory beyond the known $D=11$ one, cf.
\cite{Nicolai:2001sv}. Furthermore, these models, which are the `most
symmetric' ones in three dimensions known, exhibit a vast richness in
structure; since the choice of a subgroup of $\Eee$ as gauge group is
far less constrained in $D=3$ than in higher dimensions, as an
arbitrary number of unwanted vectors can be dualized away into
scalars, this theory allows many more gaugings than its
higher-dimensional cousins, some of which, like $G_{2(-14)}\times
F_{4(-20)}$ or $E_{7(+7)}\times SL(2)$ even possess exceptional group
factors\footnote{and in fact, all exceptional groups occur here}, none
of which are possible isometry groups of an eight-manifold and hence
could emerge in dimensional reduction of $D=11$ supergravity.
Furthermore, in contrast to e.g. $N=8$, $D=4$, even for all the
noncompact gauge groups, there is a vacuum of maximal supersymmetry
where all scalar VEVs vanish, and the corresponding background
supergroups also encompass the exceptional supergroups $G(3)$ and
$F(4)$. Turning on scalar fields, the landscape of the corresponding
potentials of the various models on the coset $\Eee/SO(16)$ are
probably the most intricate analytical potentials encountered so far
in supergravity and beyond. Indeed, from the computational aspect,
there is good evidence that, while conceptually simple, it is probably
entirely impossible to write down an explicit analytic expression for
the potential on the whole $128$-dimensional manifold of scalars,
since the number of terms produced would exceed the number of
particles available in the accessible universe by many orders of
magnitude!

All this makes the study of the extremal structure of these models an
interesting and challenging subject, despite, or maybe even because,
at present there is only a fragmentary understanding how they fit into
the bigger scheme of things.

This work is organized as follows: the present introductory chapter is
intended to provide background information and lay out the conventions
for subsequent chapters and was written with the intention to also
give non-experts in the field of supergravity/superstring theory who
are interested in this work due to the group-theoretical tools
developed here at least a conceptual overview over the underlying
physical motivation. In chapter $\ref{ch3dSUGRA}$, we recite some
important steps of the constructions of ungauged and gauged maximal
three-dimensional supergravity models. Chapter $\ref{ch3dCompact}$ is
dedicated to the study of the compact semi-simple gauge group
$SO(8)\times SO(8)$, while noncompact gaugings are investigated in
chapter $\ref{ch3dNoncompact}$. In order to demonstrate the generality
of our tools, we present a single instructive example from
four-dimensional supergravity in chapter $\ref{chapter4d}$ which
increases the level of complexity to which the corresponding
supergravity scalar potential has been studied by three orders of
magnitude. The novel approaches to symbolic algebra employed by the
tools developed to make such a deep investigation possible are
discussed in chapter $\ref{LambdaTensorCh}$.  Chapter
$\ref{chConclusion}$ concludes. Restricted potentials which would
have been too lengthy to be given in the main text have been moved to
appendix $\ref{chPotentials}$; since numerous new results on the
vacuum structure of supergravity theories are claimed by this work
which might appear quite bold, being well out of reach of previously
existing technology, appendix $\ref{chLISPcode}$ was added giving
explicit LISP code which should enable the reader (in conjunction with
the \LambdaTensor{} software package) to redo the calculations leading
to the main results of this work and serve as a starting point for
further investigations.

\section{General Conventions}

Interpretation of conventions plays a key role for this work for two
reasons: first, among our main results there are `charts' where to
find nontrivial vacua of gauged maximal supergravity theories that are
given as elements of exceptional Lie groups (not algebras); since it
is difficult to develop an intuitive understanding of these groups, it
is much easier to overlook mistakes that can be traced back to a
mis-understanding of conventions than in many other branches of
supergravity/string theory where enough structure is visible to give
additional hints at correct interpretations of formulae. Second, due
to the richness of structure in three-dimensional supergravity
theories, it is attractive to employ computer aid in the quest for a
systematic and exhaustive treatment, which naturally favours a
low-level presentation and interpretation of formulae, since all
conventions eventually have to be spelled out in detail for the
machine. For the sake of reproducibility of the results presented, a
separate appendix is devoted to \LambdaTensor{} definitions
corresponding to key formulae given in the main text.

Basically, one could -- at least in principle -- for the sake
of definiteness resort to abandoning all conventions, including the
Einstein summation convention, explicitly spelling out all summations,
symmetrization factors, and subgroup embeddings. However, it is easy
to convince oneself that doing so would clutter many formulae so badly
that they become totally unpalatable. Hence, finding the right level
of abbreviation hidden in conventions is a nontrivial issue for the
presentation of results.

The set of rules given in the following should provide a complete and
unambiguous definition how to interpret every new result in the
main text. Although this may seem unusual and maybe even unnecessary
at first, in particular since many of these rules are widely used and
well known, it is nevertheless hard to draw a line between those for
which there is general consensus and those for which there is
not.

\begin{itemize}

\item {\em (Einstein summation convention)} When in a product,
an index name appears twice, the product is to be prefixed with a sum
of this index over all possible values of the index with this
name. Index counting starts at 1\footnote{Here, we follow usual
conventions although it would be more natural to start counting at
zero in the context of index splitting.} (Note that we do not require
this index to show up once as upper and once as lower; for example, in
a sum like $P^j Q^j$ over $SU(8)$ indices, applying these rules will
yield a non-$SU(8)$-covariant (but $SO(8)$ covariant) quantity.)
Indices which are explicitly bound otherwise, e.g. by a summation sign,
are of course exempt from this summation convention.

\item {\em (Index raising and lowering)} When a tensor
that was defined with an upper (lower) index $I$ is used with the
corresponding index as lower (upper), the corresponding group metric
(or its inverse) is used implicitly to lower (raise) this index.

\item {\em (Index interpretation and index splitting)}
Index names are (potentially composed) glyphs from various
alphabets. The assignment of index names to various group
representations is explained in the text; whenever (with one sole
exception given below) a tensor that is defined with some index $J$
corresponding to transformation under the group $G$ occurs in a
formula where in place of this index $J$, an index $A$ of a different
set of glyphs corresponding to a representation of a subgroup
$H\subset G$ shows up, and if there is a decomposition of the
$G$-representation labeled by indices of the type of $J$ with respect
to the subgroup $H$ such that the range of index values for $A$
directly corresponds to a sub-range of index values of the index $J$,
then the $A$ index is implicitly promoted to a $J$ index by
offsetting; that is, one implicitly inserts an extra embedding tensor
factor $H_{JA}$ which contains entries $1$ for
$J=A+\langle\mbox{\it offset}\rangle$,
zero for all other values of $J$ and $A$.

\item {\em (One-to-many index splitting)} If a $G$-index decomposes to
  a group of $H$-indices in such a way that there is a one-to-one
  correspondence between every possible combination for this group of
  $H$-indices and a corresponding $G$-index, and a group of
  $H$-indices appears on a tensor in a place where this tensor was
  defined with the corresponding $G$ index, then this group of indices
  is implicitly promoted to the corresponding $G$ index by the
  insertion of an appropriate extra embedding tensor factor
  $H_{g\;h_1h_2\ldots h_n}$ whose entries are all from the set
  $\{0;1\}$.\footnote{Note that these index splitting rules may
    generate ambiguities if one is not careful in the choice of
    alphabets. This has to be avoided. Furthermore, index groups
    obtained by implicit splitting will usually be offset a bit from
    other indices not belonging to the group to aid visual
    distinction. Note that this extra convention is not well suited
    for recursive application.}
  
\item {\em ($\epsilon$ and $\delta$)} The fully antisymmetric rank-$N$
  tensor is denoted by $\epsilon_{i_1 i_2\ldots i_N}$.  Its entries
  are $+1$ for even index permutations, $-1$ for odd permutations, and
  $0$ otherwise. The tensors $\delta^{i_1}_{j_1}$,
  $\delta^{i_1i_2}_{j_1j_2}$, $\delta^{i_1i_2\ldots i_k}_{j_1j_2\ldots
    j_k}$, etc. are completely antisymmetric in their upper, resp.
  lower indices and have entries $1/k!$, resp. $-1/k!$ if the sequence
  of indices $i_1\ldots i_k$ can be obtained from $j_1\ldots j_k$ by
  an even, resp. odd, index permutation, and $0$ otherwise. Hence,
  $\delta_{i_i i_2\ldots i_k}^{j_1 j_2\ldots j_k}\delta_{j_1 j_2\ldots
    j_k}^{m_i m_2\ldots m_k}=\delta^{m_i m_2\ldots m_k}_{i_i i_2\ldots
    i_k}$.

\item {\em (Index symmetrization)} Wherever a pair
of indices is enclosed by antisymmetrizing brackets $[{\bf ab}]$,\footnote{${\bf a}$ and ${\bf b}$ are not indices, but names for indices, and hence typeset in boldface here} they are to be substituted by otherwise unbound indices ${\bf cd}$ and an extra factor $\delta^{\bf ab}_{\bf cd}$ is to be introduced into the corresponding summand. Likewise for antisymmetrizing in more than two indices. (Since the translation of index symmetrization between typographical and machine representation is a little bit awkward, and hence error-prone if lots of formulae have to be translated manually, we frequently do not make use of it even if it could be applied.)

\item {\em ($SO(N)$ adjoint indices)} If $g_1, g_2$ are glyphs
from the alphabet used to designate $SO(N)$ vector indices, then
composed glyphs of the form $\underline{[g_1g_2]}$ will designate
$SO(N)$ adjoint indices. These $N(N-1)/2$ glyphs are consecutively
labeled either by $1,2,\ldots N(N-1)/2$ or equivalently by $\underline{[12]}$,
$\underline{[13]}$,$\ldots$,$\underline{[1N]}$,$\underline{[23]}$,$\ldots$,
$\underline{[(N-1)\;N]}$.

\item {\em (Index mapping)} Whenever index embedding has to be
  performed that cannot be achieved by simple index-range splitting
  (like, for example, the split of an $SO(N)$ adjoint index into a
  pair of $SO(N)$ vector indices), then an explicit embedding tensor
  is given. Conventionally, all embedding tensors are named $H$ and
  distinguished by the types of indices they are carrying. (For
  example, if indices $I,J$ designate $SO(16)$ vector indices, and an
  index $\underline{[IJ]}$ a SO(16) adjoint index in the range
  $1\ldots120$, then $H^{\underline{[IJ]}}_{IJ}$ is the corresponding
  embedding tensor.) Hence, such embedding tensors $H$ themselves are
  exempt from the implicit index promotion rules given above. Unless
  stated otherwise, embedding tensors $H$ are defined in such a way
  that when they are used to map a collection of index values to a
  unique other collection of index values, the corresponding entry is
  $\pm1$; for related entries, the lexicographically first one will be
  $+1$. (For example, an $SO(16)$ adjoint index corresponds to two
  pairs of different $SO(16)$ vector indices, but a pair of different
  $SO(16)$ vector indices is mapped to a single adjoint index, so the
  corresponding $H^{\underline{[IJ]}}_{IJ}$ has
  $H^{\underline{[12]}}_{12}=-H^{\underline{[12]}}_{21}=1$.

\item {\em (Gamma matrices)} $SO(16)$ resp. $SO(8)$ Gamma matrices (to
  be defined in the next section) are are denoted by $\Gamma$, resp.
  $\gamma$. For both types, $\gamma^{i_1\ldots
    i_n}_{\alpha\beta}=\gamma^{j_1}_{\alpha\dot\gamma_1}
  \gamma^{j_2}_{\gamma_2\dot\gamma_1}
  \gamma^{j_3}_{\gamma_2\dot\gamma_3} \ldots
  \gamma^{j_n}_{\beta\dot\gamma_{(n-1)}} \delta_{j_1\ldots
    j_n}^{i_1\ldots i_n}$, and likewise, $\gamma^{i_1\ldots
    i_n}_{\dot\alpha\dot\beta}=\gamma^{j_1}_{\gamma_1\dot\alpha}\ldots
  \gamma^{j_n}_{\gamma_{(n-1)}\dot\beta}\delta_{j_1\ldots
    j_n}^{i_1\ldots i_n}$, $\gamma^{i_1\ldots
    i_n}_{\alpha\dot\beta}=\gamma^{j_1}_{\alpha\dot\gamma_1}\ldots
  \gamma^{j_n}_{\gamma_{(n-1)}\dot\beta}\delta_{j_1\ldots
    j_n}^{i_1\ldots i_n}$.  Furthermore, we define
  $\gamma^{\alpha\beta}_{\dot\gamma\dot\delta}:=
  \delta_{\rho\lambda}^{\alpha\beta}\gamma^i_{\rho\dot\alpha}\gamma^i_{\lambda\dot\beta}$,
  as well as
  $\gamma^{\alpha\beta\gamma\delta}_{\dot\alpha\dot\beta}:=
  \delta_{\alpha'\beta'\gamma'\delta'}^{\alpha\beta\gamma\delta}
  \gamma^i_{\alpha'\dot\alpha} \gamma^i_{\beta'\dot\lambda}
  \gamma^k_{\gamma'\dot\lambda} \gamma^k_{\delta'\dot\beta}$ and
  $\gamma^{ij}_{\alpha\beta\gamma\delta}:=
  \delta_{\alpha'\beta'\gamma'\delta'}^{\alpha\beta\gamma\delta}
  \gamma^i_{\alpha'\dot\alpha} \gamma^k_{\beta'\dot\alpha}
  \gamma^k_{\gamma'\dot\beta} \gamma^j_{\delta'\dot\beta}$; obviously,
  since $\gamma^{\alpha\beta}_{\dot\alpha\dot\beta}$ and
  $\gamma^{ij}_{\dot\alpha\dot\beta}$ are then discerned by the types
  of label only, we must not use ambiguous expressions like
  $\gamma^{78}_{\dot\alpha\dot\beta}$, but rather write either
  $\gamma^{ij}_{\dot\alpha\dot\beta}\delta_i^7\delta_j^8$ or
  $\gamma^{\alpha\beta}_{\dot\alpha\dot\beta}\delta_\alpha^7\delta_\beta^8$.

\end{itemize}

\section{$\Eee$ conventions}
\label{secEeight}

The Lie algebra $\Eee$ plays a key role for this work. Hence we spell
out all the conventions in full detail. Following \cite{GrScWi87}, we define
\begin{equation}
\begin{array}{ll}
\sigma_1=\left(\begin{array}{rr}1&0\\0&1\end{array}\right)&
\sigma_x=\left(\begin{array}{rr}0&1\\1&0\end{array}\right)\\
\sigma_z=\left(\begin{array}{rr}1&0\\0&-1\end{array}\right)&
\sigma_e=\left(\begin{array}{rr}0&1\\-1&0\end{array}\right)
\end{array}
\end{equation}
from which we obtain $SO(8)$ $\gamma$-matrices using the tensor
$G_{i\lambda\mu\rho}$ implementing the $2\times2\times2\rightarrow 8$
mapping\footnote{Note that this matrix tensoring convention,
which seems to be more widespread,
accidentally is just the opposite of that implicitly used in \cite{Fischbacher:2002hg}}
\begin{equation}
\begin{array}{lclclclclclclcl}
G_{1111}=1&\quad&G_{2112}=1&\quad&G_{3121}=1&\quad&G_{4122}=1\\
G_{5211}=1&\quad&G_{6212}=1&\quad&G_{7221}=1&\quad&G_{8222}=1
\end{array}
\end{equation}
as well as the abbreviation
\begin{equation}
Z_{\alpha\dot\beta}(\sigma_{(A)};\sigma_{(B)};\sigma_{(C)})=\sigma_{(A)\alpha_1\dot\beta_1}\,\sigma_{(B)\alpha_2\dot\beta_2}\sigma_{(C)\alpha_3\dot\beta_3}G_{\alpha\alpha_1\alpha_2\alpha_3}G_{\dot\beta\dot\beta_1\dot\beta_2\dot\beta_3}
\end{equation}
via
\begin{equation}\label{sigma8}
\begin{array}{lcl c lcl}
\gamma^1&=&Z(\sigma_e;\sigma_e;\sigma_e)&\quad&\gamma^2&=&Z(\sigma_1;\sigma_z;\sigma_e)\\
\gamma^3&=&Z(\sigma_e;\sigma_1;\sigma_z)&\quad&\gamma^4&=&Z(\sigma_z;\sigma_e;\sigma_1)\\
\gamma^5&=&Z(\sigma_1;\sigma_x;\sigma_e)&\quad&\gamma^6&=&Z(\sigma_e;\sigma_1;\sigma_x)\\
\gamma^7&=&Z(\sigma_x;\sigma_e;\sigma_1)&\quad&\gamma^8&=&Z(\sigma_1;\sigma_1;\sigma_1)\\
\end{array}
\end{equation}
from which we form $SO(16)$ $\Gamma$-matrices using to the splitting
$J\rightarrow(j,\bar k)$ of SO(16) vector and
$A\rightarrow(\alpha\beta,\dot\gamma\dot\delta)$, $\dot
A\rightarrow(\alpha\dot\beta,\dot\gamma\delta)$ of Majorana-Weyl spinor and
co-spinor indices by
\begin{equation}
\begin{array}{lcl}
\Gamma^{I}_{A\dot A}&=&\phantom+H^I_i H^A_{\alpha\beta} H^{\dot A}_{\gamma\dot\delta}\delta_{\alpha\gamma}\gamma^i_{\beta\dot\delta}
+H^I_i H^A_{\dot\alpha\dot\beta} H^{\dot A}_{\dot\gamma\delta}\delta_{\dot\alpha\dot\gamma} \gamma^i_{\delta\dot\beta}\\
&&+H^I_{\bar i} H^A_{\alpha\beta} H^{\dot A}_{\dot\gamma\delta}\delta_{\beta\delta}\gamma^{\bar i}_{\alpha\dot\gamma}
-H^I_{\bar i} H^A_{\dot\alpha\dot\beta} H^{\dot A}_{\gamma\dot\delta}\delta_{\dot\beta\dot\delta}\gamma^{\bar i}_{\gamma\dot\alpha},\\
\Gamma^{IJ}_{AB}&=&\delta^{IJ}_{KL}\Gamma^{K}_{A\dot C}\Gamma^{L}_{B\dot C}
\end{array}
\end{equation}

If we denote $SO(16)$ adjoint indices running from $1$ to $120$ by
$\underline{[IJ]}$, which naturally decompose into $SO(16)$ vector
indices $I,J$ and split $E_{8(8)}$ adjoint indices
$\mathcal{A}\rightarrow(A,\underline{[IJ]})$, then $E_{8(8)}$
structure constants are given by
\begin{equation}
\begin{array}{lcl}
f_{\cA\cB}{}^\cC&=&-H^\cA_{IJ}H^\cB_{KL}H_\cC^{MN}\delta^{IJ}_{I'J'}\delta^{KL}_{K'L'}\delta^{I'K'}\delta_{L'J'}^{MN}\\
&&+\frac{1}{4}\Gamma^{IJ}_{AB}\left(H^\cA_{IJ}H^\cB_AH_\cC^B+H^\cA_BH^\cB_{IJ}H_\cC^A-H^\cA_A H^\cB_B H_\cC^{IJ}\right).
\end{array}
\end{equation}

In the literature, the intermediate $SO(16)$ adjoint index in the
splitting chain $\mathcal{A}\rightarrow\underline{[IJ]}\rightarrow IJ$
frequently is not displayed explicitly. Instead, a modification of the
usual Einstein summation convention is introduced where one has to
include an extra factor $1/2$ whenever a sum is performed over a pair
of antisymmetric indices, correcting $IJ, JI$ double-counting. Since
it is not entirely clear that this rule will not leave room for
interpretation in some subtle cases (like the definition of $\Eee$
tensors in index-split notation), we will try to avoid it by
explicitly including index splitting projection tensors in our
formulae just as above when presenting new results.

The conventionally normalized Cartan-Killing metric for $\Eee$
\beq
\eta_{\cA\cB}=\frac{1}{2g}f_{\cA\cP}{}^{\cQ}f_{\cB\cQ}{}^{\cP}
\eeq
where $g$ is the dual Coxeter number of the Lie algebra ($30$ for $E_8$)
is then given by
\beq
\eta_{AB}=-\delta_{AB},\quad \eta_{\IJ\,\KL}=\delta_{\IJ\,\KL}.
\eeq

$\Eee$ is the only simple Lie algebra for which the fundamental and
adjoint representation are the same and this property plays an
important role in the construction of $N=16$ $D=3$ supergravity. It can
be expressed by the relation
\beq
\cV^{-1}t^\cM\cV=\cV^\cM{}_\cA t^\cA\Leftrightarrow \cV^\cM{}_\cA=\frac{1}{60}\tr\left(t^\cM\cV t_\cA \cV^{-1}\right).
\eeq

Furthermore, we define compact and noncompact $\Eee$ generators $X_\IJ, Y_A$ by
\beq
\left(X_\IJ\right)^\cC{}_\cB=f_{\IJ\cB}{}^\cC,\qquad \left(Y_A\right)^\cC{}_\cB=f_{A\cB}{}^\cC.
\eeq

It is convenient to take as a Cartan subalgebra the compact generators
$X_{\underline{[1\,2]}},X_{\underline{[3\,4]}},\ldots,X_{\underline{[15\,16]}}$;
this gives the conventional choice for the set of $E_8$ root vectors as 120 vectors
of the form $\{\pm e_i\pm e_j|i,j\in\{1,\ldots,8\}\}$ plus 128 vectors of the form
$\{\frac{1}{2}\left(\pm e_1\pm e_2\ldots \pm e_8\right)\}$ where the total number
of minus signs is even. Most explicitly, the ladder operators corresponding to the
simple roots\footnote{w.r.t. lexicographical ordering} are then given by ($\left(t_\cA\right)^\cC{}_\cB:=f_{\cA\cB}{}^\cC$):
\bea\label{e8_simple_roots}
T_{+\frac{1}{2}-\frac{1}{2}-\frac{1}{2}-\frac{1}{2}-\frac{1}{2}-\frac{1}{2}-\frac{1}{2}+\frac{1}{2}}&=&+t_{84}+i\,t_{86}-i\,t_{100}+t_{102}\nonumber\\
T_{\phnu+1-1\phnu\phnu\phnu\phnu\phnu}&=&+t_{159}+i\,t_{160}-i\,t_{171}+t_{172}\nonumber\\
T_{\phnu\phnu+1-1\phnu\phnu\phnu\phnu}&=&+t_{184}+i\,t_{185}-i\,t_{194}+t_{195}\nonumber\\
T_{\phnu\phnu\phnu+1-1\phnu\phnu\phnu}&=&+t_{205}+i\,t_{206}-i\,t_{213}+t_{214}\nonumber\\
T_{\phnu\phnu\phnu\phnu+1-1\phnu\phnu}&=&+t_{222}+i\,t_{223}-i\,t_{228}+t_{229}\\
T_{\phnu\phnu\phnu\phnu\phnu+1-1\phnu}&=&+t_{235}+i\,t_{236}-i\,t_{239}+t_{240}\nonumber\\
T_{\phnu\phnu\phnu\phnu\phnu\phnu+1+1}&=&-t_{244}+i\,t_{245}+i\,t_{246}+t_{247}\nonumber\\
T_{\phnu\phnu\phnu\phnu\phnu\phnu+1-1}&=&+t_{244}+i\,t_{245}-i\,t_{246}+t_{247}\nonumber
\eea
where our complexity conventions are such that for these compact generators, we have e.g.
\beq
[X_{\underline{[1\,2]}},T_{+1+1\phnu\phnu\phnu\phnu\phnu\phnu}]=+1\cdot i\,T_{+1+1\phnu\phnu\phnu\phnu\phnu\phnu}.
\eeq

Note that if we substitute the ladder operator of the first simple root by
\bea
T_{+1-1\phnu\phnu\phnu\phnu\phnu\phnu}&=&+t_{130}+i\,t_{131}-i\,t_{144}+t_{145}
\eea
then the new set will spawn the maximal compact subalgebra $SO(16)$.
The ladder operator for the lowest $E_8$ root is
\bea
T_{-1-1\phnu\phnu\phnu\phnu\phnu\phnu}&=&-t_{130}-i\,t_{131}-i\,t_{144}+t_{145}.
\eea

From here on, we will use an abbreviating notation of $E_8$ roots; for
the spinorial roots like
$\left(+\frac{1}{2},-\frac{1}{2},-\frac{1}{2},-\frac{1}{2},-\frac{1}{2},-\frac{1}{2},-\frac{1}{2},+\frac{1}{2}\right)$,
we just write a sequence of eight signs, hence `$+------+$' in this example, while for roots of the form
$\pm e_i\pm e_j$, we write $\pm i\pm j$, e.g. $+3-4$ for $(0,0,+1,-1,0,0,0,0)$.

Sometimes, as in the construction of the $E_{6(+6)}\times SL(3)$
embedding tensor, a noncompact Cartan subalgebra is more useful.
Here, we may take the generators
$Y_{\alpha\beta}\delta_\alpha^n\delta_\beta^n, n\in{1,\ldots 8}$. With
respect to this Cartan subalgebra, simple roots are given by
\bea\label{e8_simple_roots_nc}
T_{+-------+}&=&-t_{72}-t_{79}-t_{86}+t_{93}-t_{100}+t_{107}+t_{114}+t_{121}\nonumber\\
&&-t_{143}+t_{156}+t_{167}+t_{180}-t_{188}-t_{199}-t_{206}+t_{213}\nonumber\\
T_{+2-3}&=&2\,t_{11}+2\,t_{18}+t_{130}-t_{145}+t_{185}+t_{194}+t_{222}-t_{229}+t_{245}+t_{246}\nonumber\\
T_{+3-4}&=&-2\,t_{20}-2\,t_{27}+t_{134}-t_{149}+t_{160}+t_{171}+t_{226}-t_{233}+t_{236}+t_{239}\nonumber\\
T_{+4-5}&=&-2\,t_{29}-2\,t_{36}+t_{135}+t_{148}-t_{159}+t_{172}+t_{227}+t_{232}-t_{235}+t_{240}\nonumber\\
T_{+5-6}&=&-2\,t_{38}-2\,t_{45}-t_{134}+t_{149}-t_{160}+t_{171}-t_{226}+t_{233}-t_{236}+t_{239}\nonumber\\
T_{+6-7}&=&-2\,t_{47}-2\,t_{54}-t_{130}-t_{145}-t_{185}+t_{194}-t_{222}+t_{229}-t_{245}+t_{246}\nonumber\\
T_{+7-8}&=&+2\,t_{56}+2\,t_{63}+t_{134}+t_{149}-t_{160}+t_{171}+t_{226}+t_{233}-t_{236}+t_{239}\nonumber\\
T_{+7+8}&=&-2\,t_{56}+2\,t_{63}-t_{134}-t_{149}+t_{160}-t_{171}+t_{226}+t_{233}-t_{236}+t_{239}\nonumber
\eea
and complexity conventions are such that
\beq
[Y_{11},T_{+1+1\phnu\phnu\phnu\phnu\phnu\phnu}]=+1\,T_{+1+1\phnu\phnu\phnu\phnu\phnu\phnu}\;.
\eeq

\begin{figure}
\label{e8-dynkin}
\begin{center}
\includegraphics{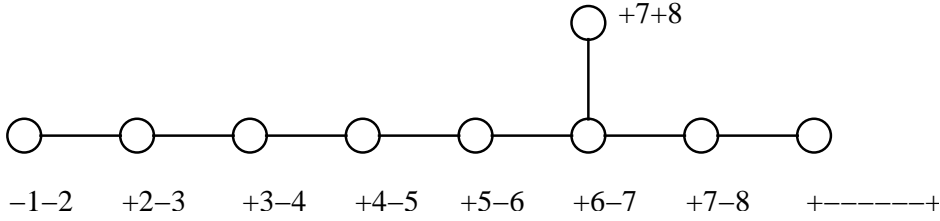}
\end{center}
\caption{The extended Dynkin diagram of $E_8$}
\end{figure}

\chapter[Maximal $D=3$ SUGRA]{Maximal three-dimensional supergravity}
\label{ch3dSUGRA}

\section{Ungauged $N=16$ $D=3$ SUGRA}

In the following presentation, we closely follow \cite{Nicolai:2001sv}.
It is convenient to introduce gauged maximal three-dimensional
supergravity theories via incremental definitions that proceed through
ungauged maximal three-dimensional supergravity, which was first
constructed in $\cite{Marcus:1983hb}$; the physical fields of this
theory form an irreducible supermultiplet of 128 bosons and 128
fermions which transform as spinors and co-spinors of $SO(16)$. In
three dimensions, the dreibein $e_\mu{}^\alpha$ as well as the
gravitini $\psi^I_\mu$ do not carry propagating degrees of
freedom. Due to the hidden invariance of the ungauged theory under
global $\Eee$ and local $SO(16)$ transformations \cite{Cremmer:1997ct},
scalar fields can be described by an element $\cV$ of the non-compact
coset manifold $\Eee/SO(16)$. Using the 248-dimensional
fundamental $\Eee$ representation, $\cV$ transforms as
\beq
\label{ungaugedN=16}
\cV(x)\rightarrow g\cV(x) h^{-1}(x),\qquad g\in\Eee, h(x)\in SO(16).
\eeq

Following the conventions of \cite{Marcus:1983hb} and \cite{Nicolai:2001sv,Nicolai:2000sc},
the scalar fields couple to the fermions via the currents
\beq
\cV^{-1}\partial_\mu\cV=\frac{1}{2}Q_\mu^{IJ} X^{IJ}+P_\mu^A Y^A.
\eeq
Defining the $SO(16)$ field strength $Q_{\mu\nu}^{IJ}$ obtained from the connection $Q_\mu^{IJ}$ via
\beq
Q_{\mu\nu}^{IJ}:=2\left(\partial_{[\mu} Q_{\nu]}^{IJ}+Q_\mu^{K[I}Q_\nu^{J]K}\right),
\eeq
as well as the covariant derivative $D_\mu$
\bea
D_\mu\psi^I_\nu&:=&\partial_\mu\psi^I_\nu+\frac{1}{4}\omega_\mu^{ab}\gamma_{ab}\psi^I_\nu+Q_\mu^{IJ}\psi_\nu^J\\
D_\mu\chi^\dA&:=&\partial_\mu\chi^A+\frac{1}{4}\omega_\mu^{ab}\gamma_{ab}\chi^\dA+\frac{1}{4}Q_\mu^{IJ}\Gamma^{IJ}_{\dA\dB}\chi^\dB
\eea
this implies the integrability relations
\beq\label{IntegrabilityEq}
Q^{IJ}_{\mu\nu}+\frac{1}{2}\Gamma^{IJ}_{AB}P^A_\mu P^B_\nu=0,\qquad D_{[\mu}P_{\nu]}^A=0.
\eeq

If we define the supercovariant current
\beq
\hat P^A_\mu:=P^A_\mu-\bar\psi^I_\mu\chi^\dA\Gamma^I_{A\dA},
\eeq
then the Lagrangian
\beq
\begin{array}{lcl}\label{UngaugedLagrangian}
\mathcal{L}_0&=&-\frac{1}{4}eR+\frac{1}{4}eP^{\mu A}P_\mu^A+\frac{1}{2}e^{\lambda\mu\nu}\bar\psi^I_\lambda D_\mu \psi^I_\nu\\
&&-\frac{i}{2}e\bar\chi^\dA\gamma^\mu D_\mu\chi^\dA-\frac{1}{2}e\bar\chi^\dA\gamma^\mu\gamma^\nu\psi^I_\mu\Gamma^I_{A\dA}P^A_\nu\\
&&-\frac{1}{8}e\left(\bar\chi\gamma_\rho\Gamma^{IJ}\chi\left(\bar\psi^I_\mu\gamma^{\mu\nu\rho}\psi^J_\nu-\bar\psi^I_\mu\gamma^\rho\psi^{\mu J}\right)+\bar\chi\chi\bar\psi^I_\mu\gamma^\nu\gamma^\mu\psi^I_\nu\right)\\
&&+e\left(\frac{1}{8}\left(\bar\chi\chi\right)\left(\bar\chi\chi\right)-\frac{1}{96}\bar\chi\gamma^\mu\Gamma^{IJ}\chi\bar\chi\gamma_\mu\Gamma^{IJ}\chi\right)
\end{array}
\eeq
is invariant under the supersymmetry variations
\beq
\begin{array}{lcl}
        \delta e_\mu{}^\alpha&=&i\bar\epsilon^I\gamma^\alpha\psi^I_\mu\\
        \cV^{-1}\delta\cV&=&\Gamma^I_{A\dA}\bar\chi^\dA\epsilon^I Y^A\\
        \delta\psi^I_\mu&=&D_\mu\epsilon^I-\frac{i}{4}\gamma^\mu\epsilon^J\bar\chi\Gamma^{IJ}\gamma_{\mu\nu}\chi\\
        \delta\chi^\dA&=&\frac{i}{2}\gamma^\mu\epsilon^I\Gamma^I_{A\dA}\hat P^A_\mu.
\end{array}
\eeq

The equation of motion for scalars obtained from this Lagrangian is
\beq
\begin{array}{l}
D_\mu\left(e\left(P^{\mu A}-\bar\psi^I_\nu\gamma^\mu\gamma^\nu\chi^\dA\Gamma^I_{A\dA}\right)\right)\\
=\frac{1}{2}\epsilon^{\mu\nu\rho}\bar\psi^I_\mu\psi^J_\nu\Gamma^{IJ}_{AB}P^B_\rho+\frac{1}{8}ie\bar\chi\gamma^\mu\Gamma^{IJ}\chi\Gamma^{IJ}_{AB}P^B_\mu
\end{array}
\eeq
which can be written in the form (by using the Rarita-Schwinger and Dirac equation for the gravitini and fermions)
\beq
\partial^\mu\left(e\cJ_\mu^\cM\right)=0
\eeq
with $\cJ^\cM_\mu$ the conserved Noether current corresponding to the global $\Eee$ symmetry:
\beq
\begin{array}{lcl}
e\cJ^{\mu\cM}&=&2\cV^\cM{}_\cB\hat P^{\mu B}-\frac{i}{2}\cV^\cM_{IJ}\bar\chi\gamma^\mu\Gamma^{IJ}\chi\\&&
-2e^{-1}\epsilon^{\mu\nu\rho}\left(\cV^\cM_{IJ}\bar\psi^I_\nu\psi^J_\rho-i\Gamma^I_{A\dA}\cV^\cM{}_A\bar\psi^I_\nu\gamma_\rho\chi^\dA\right).
\end{array}
\eeq
Since this current is conserved, we can introduce 248 abelian vector
fields $B_\mu{}^\cM$, defined up to gauge transformations, whose field
strength $B_{\mu\nu}{}^\cM=2\partial_{[\mu}B_{\nu]}{}^\cM$ obeys
$\epsilon^{\mu\nu\rho}B_{\nu\rho}{}^\cM=e\cJ^{\mu\cM}$.  In order to
find supersymmetry transformation rules for these vectors, one can
generalize the supersymmetry transformations of the 36 vector fields
that are obtained by direct dimensional reduction of $D=11$
supergravity
\beq\label{VectorVariations}
\delta B_\mu{}^\cM=-2\cV^\cM{}_{IJ}\bar\epsilon^I\psi^J_\mu+i\Gamma^I_{A\dA}\cV^\cM_A\bar\epsilon^I\gamma_\mu\chi^\dA
\eeq
which of course has to be compatible with the duality relation for the vectors and the Noether current,
which can be re-written in supercovariant form by using the supercovariant field strength
\beq
\hat B_{\mu\nu}{}^\cM:=B_{\mu\nu}{}^\cM+2\cV^\cM{}_{IJ}\bar\psi^I_\mu\psi^J_\nu-2i\Gamma^I_{A\dA}\cV^\cM{}_A\bar\psi^I_{[\mu}\gamma_{\nu]}\chi^\dA
\eeq
so that it is given by
\beq\label{DualityEq}
\epsilon^{\mu\nu\rho}\hat B_{\nu\rho}{}^\cM=2e\cV^\cM{}_A\hat P^{\mu A}-\frac{i}{2}e\cV^\cM{}_{IJ}\bar\chi\gamma^\mu\Gamma^{IJ}\chi.
\eeq

This equation defines 248 vector fields as nonlocal, nonlinear
functions of the original 128 physical scalars $+120$ $SO(16)$ gauge
degrees of freedom (as long as the equations of motion are obeyed,
i.e. the Noether current is divergence-free). By using the
integrability equation $(\ref{IntegrabilityEq})$, we can obtain the
vector equation of motion from the derivative of the duality equation
$(\ref{DualityEq})$:
\beq
\partial_\nu B^{\mu\nu\cM}=-\frac{1}{2}\epsilon^{\mu\nu\rho}\cV^\cM_{IJ} Q^{IJ}_{\nu\rho} + \langle\mbox{fermionic terms($\cV$)}\rangle.
\eeq

\section{Gauged $N=16$ $D=3$ SUGRA}

Part of the $\Eee$ symmetry of the theory described in the previous
section can be made local; let $G_0\subset\Eee$ be a subgroup of the
global symmetry group $\Eee$ that can be promoted to a gauge group (we
will see later that there is a simple group-theoretical restriction
for which groups this is possible). Then, $(\ref{ungaugedN=16})$ is
replaced by
\beq
\label{gaugedN=16}
\cV(x)\rightarrow g_0(x)\cV(x) h^{-1}(x),\qquad g_0(x)\in G_0, h(x)\in SO(16).
\eeq

The subgroup $G_0$ of $\Eee$ that will be promoted to a gauge group is
characterized by its embedding tensor $\Theta_{\cM\cN}$, which is the
restriction of the Cartan-Killing metric $\eta_{\cM\cN}$ to the
algebra corresponding to $G_0$, and thus is given as a linear
combination of projectors onto the simple factors of $G_0$ (where the
relative coefficients in this linear combination turn out to be fixed
by group theory, so that only a single gauge coupling parameter
survives).

Labeling $G_0$ adjoint indices by $m, n, \ldots$, there is an
embedding tensor $\Omega^\cA{}_m$ that will map these indices to $E_8$
indices. By choice of an appropriate basis for the $E_8$ algebra, we
can make these $\nu={\rm dim}\;G_0$ indices the first block of indices in a
split $248=\nu+\ldots$. As usual with such index splitting, we would
then use the convention that use of a $G_0$ index $m, n,\ldots$ in
place of an $E_8$ index $\cA, \cB, \ldots$ corresponds to the silent
omission of some embedding tensor $\Omega$. Although this way of
splitting indices will in general not be compatible with the index
split suitable for the $SO(16)$ decomposition of $\Eee$ explained in
section \ref{secEeight}, we can nevertheless keep the convention that
usage of a $m, n,\ldots$ indices corresponds to the silent omission of
such embedding tensors. In particular, our convention shall be that
silent indices are to be interpreted as in $B_\mu{}^m t_m\equiv
B_\mu{}^\cA\Theta_{\cA\cB}t^\cB$.

From now on, let $g$ be the gauge coupling constant; we first
gauge-covariantize derivatives
\beq
\begin{array}{lcl}
\cV^{-1}\partial_\mu\cV&=&\frac{1}{2}Q_\mu{IJ} X^{IJ}+P_\mu^A Y^A\\
\Rightarrow \cV^{-1}\cD_\mu\cV&=&\frac{1}{2}\cQ_\mu^{IJ} X^{IJ}+\cP_\mu^A Y^A\\
&=&\cV^{-1}\partial_\mu\cV+gB_\mu{}^m\cV^{-1}t_m\cV
\end{array}.
\eeq

The non-abelian field strength associated with this connection is
\beq
\cB_{\mu\nu}{}^m=2\partial_{[\mu}B_{\nu]}{}^m+g f^m{}_{np}B_\mu{}^n B_\nu{}^p
\eeq
and integrability conditions $(\ref{IntegrabilityEq})$ become
\beq
\begin{array}{lcl}
\cQ^{IJ}_{\mu\nu}+\frac{1}{2}\Gamma^{IJ}_{AB}\cP^A_\mu P^B_\nu&=&g\cB_{\mu\nu}{}^m \cV_{mIJ}\\
2D_{[\mu}\cP_{\nu]}^A&=&g\cB_{\mu\nu}{}^m \cV_{mA}
\end{array}.
\eeq
Supersymmetry variations for the vectors still are given by
$(\ref{VectorVariations})$. For the modified currents, they are
\beq
\begin{array}{lcl}
\delta\cQ_\mu^{IJ}&=&\frac{1}{2}\left(\Gamma^IJ\Gamma^K\right)_{A\dA}\cP_\mu^A\bar\chi^\dA\epsilon^K+g\left(\delta B_\mu{}^m\right)\cV_{mIJ}\\
\delta\cP_\mu^A&=&\Gamma^I_{A\dA}D_\mu\left(\bar\chi^\dA\epsilon^I\right)+g\left(\delta B_\mu{}^m\right)\cV_{mA}.
\end{array}
\eeq
(Note that the new currents also depend on $g$.) The modifications
introduced here violate the supersymmetry of the Lagrangian
$(\ref{UngaugedLagrangian})$ with currents replaced by the
covariantized definitions. Restoring local supersymmetry will
therefore require additional modifications to the Lagrangian as well
as the supersymmetry variations. The extra terms that have to be added
to the Lagrangian can be obtained by the Noether procedure and
turn out to be (at first order in $g$) Chern-Simons couplings for the
vectors and scalar-fermion couplings of Yukawa type as well as (at
second order in $g$) a scalar field potential.
These couplings as well as the scalar potential are functions of
tensors formed from the scalar matrix $\cV^\cM{}_\cA$ as well as the
gauge group embedding tensor $\Theta_{\cM\cN}$; one finds that these
are $SO(16)$ representations that can be obtained via suitable
linear projections from the tensor
\beq\label{T-tensor}
T_{\cA\cB}:=\cV^\cM{}_\cA \cV^\cN{}_\cB\Theta_{\cM\cN}
\eeq
which is analogous to the T-tensors given in \cite{deWit:1982ig} and
\cite{Gunaydin:1984qu},
but (as a consequence of the equivalence of the $E_8$
fundamental and adjoint representation) is quadratic and not cubic in
$\cV$. From $T_{\cA\cB}$, we form the
tensors\footnote{The index splitting is given in most explicit form
here, since these quantities are of prime importance for all
subsequent investigations}
\bea\label{A-tensors}
\theta&=&\frac{1}{248}\eta^{\cM\cN}\Theta_{\cM\cN}\\
A_1^{IJ}&=&\frac{8}{7}\theta\delta_{IJ}+\frac{1}{7}T_{\cA\cB} H^\cA_{\underline{[IK]}}H^\cB_{\underline{[JK]}} H^{\underline{[IK]}}_{IK} H^{\underline{[JK]}}_{JK}\\
A_2^{I\dot A}&=&-\frac{1}{7} T_{\cA\cB} H^\cA_{\underline{[IJ]}} H^{\underline{[IJ]}}_{IJ} H^\cB_{A}\Gamma^J_{A\dot A}\\
A_3^{\dot A\dot B}&=&2\theta\delta_{\dot A\dot B}+\frac{1}{48}\Gamma^{IJKL}_{\dot A\dot B}T_{\cA\cB} H^\cA_{\underline{[IJ]}}H^\cB_{\underline{[KL]}} H^{\underline{[IJ]}}_{IJ} H^{\underline{[KL]}}_{KL}
\eea
The full Lagrangian of gauged $N=16$ $D=3$ supergravity is then
\beq
\begin{array}{lcl}
\cL^{(CS)}&=&\frac{1}{4}g\epsilon^{\mu\nu\rho}B_\mu{}^m\left(\partial_\nu B_{\rho m}+\frac{1}{3}gf_{mnp}B_\nu{}^n B_\rho{}^p\right)\\
\cL^{(Y)}&=&\frac{1}{2}geA_1^{IJ}\bar\psi^I_\mu\gamma^{\mu\nu}\psi^J_\nu+igeA_2^{I\dA}\bar\chi^\dA\gamma^\mu\psi^I_\mu+\frac{1}{2}geA_3^{\dA\dB}\bar\chi^\dA\chi^\dB\\
\cL^{(V)}&=&\frac{1}{8}g^2e\left(A_1^{IJ}A_1^{IJ}-\frac{1}{2}A_2^{I\dA}A_2^{I\dA}\right)\\
\cL&=&\cL_0+\cL^{(CS)}+\cL^{(Y)}+\cL^{(V)}
\end{array}
\eeq
where $\cL_0$ is the Lagrangian of the ungauged theory, but with modified covariantized currents,
$\cL^{(CS)}$ are the Chern-Simons couplings of the vectors,
$\cL^{(Y)}$ Yang-Mills couplings between scalars and fermions, and
$\cL^{(V)}$ is the scalar potential term. Supersymmetry variations
of the fermions must be further modified by terms
\beq
\delta_g\psi^I_\mu=igA_1^{IJ}\gamma_\mu\epsilon^J,\qquad \delta_g\chi^\dA=gA_2^{I\dA}\epsilon^I.
\eeq
Explicit calculation then shows that this Lagrangian is supersymmetric
under the given transformations and the superalgebra closes if and
only if the $T$-tensor satisfies a series of linear relations which
state that it is entirely determined by $A_1, A_2, A_3$, as well as a
set of differential and quadratic identities on $A_1, A_2, A_3$.
Remarkably, all these constraints can be recast into a single
group-theoretic constraint on the gauge group embedding tensor $\Theta$
which states that under the decomposition into $E_8$ irreps
${\bf 248}\times {\bf 248}_{\tt sym}$=${\bf 1}+{\bf 3875}+{\bf 27000}$
the ${\bf 27000}$ component vanishes.

Under the maximal compact subgroup $SO(16)$, these $E_8$ irreps decompose into
($SO(16)$ Dynkin labels given in parentheses):
\bea
{\bf 3875}&\Rightarrow&{\bf 135}_{(20000000)}+{\bf 1920}_{(10000001)}+{\bf 1820}_{(00010000)}\nonumber\\
{\bf 27000}&\Rightarrow&{\bf 5304}_{(02000000)}+{\bf 13312}_{(01000010)}+{\bf 6435}_{(00000020)}\nonumber\\
&&+{\bf 1820}_{(00010000)}+{\bf 128}_{(00000010)}+{\bf 1}_{(00000000)}
\eea


The main subject of this work is the extremal structure of the scalar potential term

\beq\label{ThePotential}
\fbox{$V:=-\frac{1}{8}g^2\left(A_1^{IJ}A_1^{IJ}-\frac{1}{2}A_2^{I\dA}A_2^{I\dA}\right)$}
\eeq
which determines possible vacua of these gauged maximal three-dimensional supergravity theories.

Of the quadratic constraints on the $T$-tensor which eventually are
subsumed under the gauge group embedding tensor projection condition,
Eq. $(3.25)$ of \cite{Nicolai:2001sv}
\beq
A_1^{IK}A_1^{KJ}-\frac{1}{2}A_2^{I\dot A}A_2^{J\dot A}=\frac{1}{16}\delta^{IJ}\left(A_1^{KL}A_1^{KL}-\frac{1}{2}A_2^{K\dot A}A_2^{K\dot A}\right)
\eeq
deserves special attention for this work,
since it provides a nontrivial consistency check that the relative
factor between the $A_1$ and $A_2$ contributions to the scalar
potential has been chosen correctly in the machine code transliteration of
$(\ref{T-tensor}), (\ref{A-tensors}), (\ref{ThePotential})$.

Taking derivatives of the $A$-tensors with respect to an invariant vector field,
$\delta\cV^\cM{}_\cB/\delta\Sigma^A=f_\cB{}^{\cC A}\cV^\cM{}_\cC$, we
get \cite{Nicolai:2001sv}
\bea
\frac{\delta A_1^{IJ}}{\delta\Sigma^A}&=&\frac{1}{14}\,\left(\Gamma^{IK}_{AB}T_{KJ\,B}\Gamma^{JK}_{AB}T_{KI\,B}\right)\\
\frac{\delta A_2^{I\dot A}}{\delta\Sigma^A}&=&\frac{1}{14}\,\Gamma^J_{B\dot A}\left(\Gamma^{IJ}_{AC}T_{BC}+\frac{1}{2}\,\Gamma^{MN}_{AB} T_{IJ\,MN}\right)\\
\frac{\delta A_3^{\dot A\dot B}}{\delta\Sigma^A}&=&-\frac{1}{48}\,\gamma^{IJKL}_{\dot A\dot B}\Gamma^{KL}_{AB} T_{IJ\,B}
\eea
from which we obtain by re-writing projections of the $T$-tensor as
the corresponding $A$-tensors the scalar mass matrix of second
derivatives for an arbitrary vacuum \cite{Fischbacher:2002fx}
\bea
-4\,g^{-2}\cM_{AB}&=&-8g^{-2}\frac{\delta^2 V}{\delta\Sigma^A\delta\Sigma^B}\nonumber\\
&=&\phantom+\frac{3}{4}\left(\Gamma^I_{A\dot A}A_2^{J\dot A}A_2^{J\dot B}\Gamma^I_{B\dot B}+\Gamma^I_{A\dot A}A_2^{J\dot A}A_2^{I\dot B}\Gamma^J_{B\dot B}\right)\nonumber\\
&&+\frac{3}{4}\,A_1^{IJ}A_1^{IJ}\delta_{AB}-\frac{3}{4}\,A_1^{II} T_{A\,B}\nonumber\\
&&+\frac{1}{2}\,\Gamma^I_{A\dot A}A_1^{IJ}A_3^{\dot A\dot B}\Gamma^J_{B\dot B}-\frac{1}{4}\,\Gamma^I_{A\dot A}A_3^{\dot A\dot C}A_3^{\dot C\dot B}\Gamma^I_{B\dot B}\nonumber\\
&&+\frac{1}{4}\,\Gamma^I_{A\dot A}A_3^{\dot A\dot C}\Gamma^I_{C\dot C}T_{C\,B}
\eea
while the scalar kinetic term is uniformly normalized as
\beq
\cL_{\rm kin}=\frac{1}{4}\,e\partial^\mu\Sigma^A\partial_\mu\Sigma^A+\ldots,
\eeq
independent of the particular vacuum.

While the vectors do not carry propagating degrees of freedom when
mass terms are absent, the vacuum will spontaneously break a gauge
group $G_0$ to a compact subgroup $H_0$, causing the vectors
associated with the broken generators to absorb the corresponding
Goldstone bosons and hence become massive by a topological
three-dimensional variant of the Brout-Englert-Higgs effect. As
explained in \cite{Fischbacher:2002fx}, the vector mass matrix is
obtained by restricting $\cV^M{}_A\cV^\cL{}_A\Theta_{\cN\cL}$ to
$G_0$, and this information may be extracted directly from
\beq
\cM^{\rm vec}_{AB}=g T_{A\,B}\;.
\eeq

The fermionic analogue of this transfer of degrees of freedom from
matter fields to previously nonpropagating gauge fields is realized by
some of the gravitini absorbing the Goldstinos produced by
supersymmetry breaking via the super-Higgs effect \cite{Deser:uq}.

\section{Supergravity in $AdS_3$}

In three dimensions, it is possible to define a dual spin connection
\beq
A_\mu^\alpha:=-\frac{1}{2}\epsilon^{abc}\omega_{\mu bc}
\eeq
whose field strength
\beq
F^a_{\mu\nu}:=2\partial_{[\mu}A^a_{\nu]}+\epsilon^a{}_{bc}A^b_\mu A^c_\nu
\eeq
can be used to re-write the Einstein-Hilbert term in the action in Chern-Simons form
\beq
-\frac{1}{4}eR=\frac{1}{4}\epsilon^{\mu\nu\rho}e_\mu{}^a F_{\nu\rho a}.
\eeq

Generically, stationary points of the supergravity potentials
considered here are not true minima, but either maxima (as is e.g. the
origin of the $D=4$ $N=8$ potential) or saddle points.  Nevertheless,
as has been shown by Breitenlohner and
Freedman\cite{Breitenlohner:jf}, such stationary points can
nevertheless possess at least perturbative stability for negative
cosmological constant if none of the second derivatives get too large
with respect to the corresponding AdS radius; vacua with remaining
supersymmetry will always be stable, since the positive energy
argument of Poincare supersymmetry can be generalized to the
corresponding supergroups. For nonsupersymmetric vacua, an expansion
to second order in gravitational and scalar perturbations shows that,
since allowed fluctuations are required to have finite energy and
hence fall off to infinity sufficiently fast in AdS, the corresponding
positive kinetic energy term can overcompensate a negative second
derivative of the potential as long as it is not too large. In $d$
dimensions, one finds the bound for the scalar mass eigenvalues $m$
\cite{Mezincescu:ev}
\beq
4\,m^2L^2\ge (d-1)^2
\eeq
hence, in our case, $m^2 L^2\ge -1$. The AdS scale $L$ is given by
\beq
L^{-2}=-2\,V_0
\eeq
where $V_0$ is the value of the potential at the vacuum. The Ricci tensor is given by
\beq
R_{\mu\nu}=4\;V_0 g_{\mu\nu} = \Lambda g_{\mu\nu}\;
\eeq
and the corresponding $AdS_3$ covariant
derivative is given by
\beq
\cD^\pm_\mu:=\partial_\mu+\frac{i}{2}\gamma_a\left(A_\mu{}^a\pm L^{-1} e_\mu{}^a\right)
\eeq
with commutator
\beq
\left[\cD^\pm_\mu,\cD^\pm_\nu\right]=\frac{1}{2}\,i\gamma_a\left(F_{\mu\nu}{}^a+L^{-2}\epsilon^{abc}e_{\mu b} e_{\nu c}\right)
\eeq

Concerning the supersymmetries of a given vacuum,
it has been shown \cite{Nicolai:2001sv} by using arguments from
\cite{Gunaydin:1984qu} that the number of unbroken supersymmetries is the
number of eigenvalues $\alpha$ of $A_1$ for which
\beq
16\alpha^2=A_1^{IJ}A_1^{IJ}-\frac{1}{2}A_2^{I\dot A}A_2^{I\dot A}=\frac{4}{g^2L^2}
\eeq
with $L$ being the AdS scale, given by the value of the potential
at a stationary point via%
\beq
 4\,g^{-2}L^{-2}=-8\,g^{-2}V_0.
\eeq
(Note that unbroken supersymmetry cannot be realized with positive
cosmological constant; maximal supersymmetry is equivalent to the
vanishing of $A_2$.)

\chapter[$SO(8)\times SO(8)$ gauged $N=16$ $D=3$ SUGRA]{$N=16$ $D=3$ Supergravity with compact gauge group}
\label{ch3dCompact}
\section{On compact gauge groups}

The maximal compact subgroup of $\Eee$ is $SO(16)$.  For the obvious
embedding of $SO(16)$ into $\Eee$, we have $\Theta_{\IJ A}=0$ and
$\Theta_{AB}=0$, hence the projection condition $P^{27000}\Theta=0$
for subgroups of this $SO(16)$ reduces to the condition that $\Theta$
may only carry the ${\bf 135}$ representation of $SO(16)$.\footnote{We
  may of course apply a boost to rotate (conjugate) this $SO(16)$
  inside $\Eee$, rendering the $\Theta_{\IJ A}=0$ and $\Theta_{AB}=0$
  components nonzero. This observation plays an important role for the
  construction of nonsemisimple gaugings \cite{FNS}.}
(Of the other possible pieces, the ${\bf 5304}$
is part of the ${\bf 27000}$ of $E_8$, and the ${\bf 1}$ and ${\bf 1820}$ have to coincide
with the corresponding vanishing pieces of $\Theta_{AB}$.) Hence, due to tracelessness,
there is no simple compact gauge group. One can show (cf. \cite{Nicolai:2001sv}) that of the maximal
subgroups of $SO(16)$, only for $SO(8)\times SO(8)$ there is a choice of the relative gauge
coupling constants for which $P^{27000}\Theta=0$ (namely $g_2/g_1=-1$). If we split $SO(16)$
vector indices via $I\rightarrow(i,\bar{i})$, the corresponding embedding tensor is given
by\footnote{that is, its $(129,129)$, $(130,130)$, \ldots, $(135,135)$; $(143,143)$,
\ldots, $(149,149)$; \ldots ; $(204,204)$ components are $2$, while its $(221,221),\ldots(248,248)$
components are $-2$. Note that with our splitting conventions, $H_\cA^\IJ H_\IJ^{IJ}=H_\cA^{IJ}$.}
\begin{equation}
\Theta_{\cA\cB}=H_\cA^{IJ} H_\cB^{KL}\left(H^I_i H^J_j H^K_k H^L_l\delta^{ij}_{kl}
-H^I_{\bar{i}}H^J_{\bar{j}}H^K_{\bar{k}}H^L_{\bar{l}}\delta^{\bar{i}\bar{j}}_{\bar{k}\bar{l}}\right).
\end{equation}

The question whether a group embedding tensor $\Theta$ with vanishing
{\bf 27000} component can be constructed from a nonmaximal semisimple
subgroup of $SO(16)$ has not been answered yet.

\section{Techniques for finding nontrivial vacua}


The most fruitful technique for a study of the extremal structure of
these potentials known so far appears to be that introduced in
\cite{Warner:vz}: first, choose a subgroup $H$ of the gauge group~$G$
($SO(8)$ in forementioned analysis of $N=8, D=4$ gauged supergravity,
$SO(8)\times SO(8)$ for the case considered here); then, determine
a parametrization of the submanifold $M$ of $H$-singlets of the
manifold of physical scalars $P$. Every point of this submanifold for
which all derivatives within $M$ vanish must also have vanishing
derivatives within $P$. The reason is that, with the potential $V$
being invariant under $G$ and hence also under $H$, the power series
expansion of a variation $\delta z$ of $V$ around a stationary point
$z_0$ in $M$ where $\delta z$ points out of the submanifold $M$ of
$H$-singlets can not have a $\mathcal{O}(\delta z)$ term, since each
term of this expansion must be invariant under $H$ and it is not
possible to form a $H$-singlet from just one $H$-nonsinglet. All the
stationary points found that way will break the gauge group down to a
symmetry group that contains $H$.

The general tendency is that, with $H$ getting smaller, the number of
$H$-singlets among the supergravity scalars will increase. For
$H$-singlet spaces of low dimension, it easily happens that the scalar
potential does not feature any nontrivial stationary points at all,
while for higher-dimensional singlet spaces, the potential soon
becomes intractably complicated. For the maximal gauged $N=8, D=4$
model, using the embedding of $SU(3)\subset SO(8)$ under which the
scalars, vectors and co-vectors of $SO(8)$ decompose into ${\bf
3}+\bar {\bf 3}+{\bf 1}+{\bf 1}$ gives a case of manageable complexity
with six-dimensional scalar manifold for which five nontrivial extrema
were given in a complete analysis in
\cite{Warner:vz}. (It seems reasonable to expect further
yet undiscovered extrema breaking $SO(8)$ down to groups smaller than
$SU(3)$; cf. chapter $\ref{chapter4d}$.)

\section{The potential on submanifolds}

\subsection{Partial results for residual symmetry of $SU(3)_{\rm diag}$}

Since it is interesting to see how the extremal structure of $N=8,
D=4$ gauged $SO(8)$ supergravity is related to $N=16, D=3$
$SO(8)\times SO(8)$ gauged supergravity, especially since common lore
tells that stationary points in higher dimensions have corresponding
counterparts in lower dimensions, it is reasonable to try to lift the
construction given in \cite{Warner:vz} to this case via an embedding
of $E_{7(+7)}$ in $E_{8(+8)}$, which works as follows: under the
$SO(8)_L\times SO(8)_R$ gauge group considered here, $\Eee$ decomposes
into ${\bf 248}\rightarrow ({\bf 28},{\bf 1}) + ({\bf 1},{\bf 28}) +
({\bf 8}_v,{\bf 8}_v) + ({\bf 8}_s,{\bf 8}_s) + ({\bf 8}_c,{\bf 8}_c)$
(where we could apply further triality rotations to each of the
$SO(8)$ that generate permutations of the left and right $v,s,c$
labels).  Note that the compact generators from $({\bf 8}_v,{\bf
8}_v)$ extend $SO(8)\times SO(8)$ to the maximal compact subgroup
$SO(16)$ of $\Eee$, while the noncompact ones form a ${\bf 128}$ of
$SO(16)$.\footnote{This decomposition is particularly easy to
understand starting from the set of roots of $E_8$ given in
conventional notation by $120$ $\pm e_i\pm e_j$ and $128$ $\sum_k \pm
1/2 e_k$ with an even number of minus signs.  This decomposition just
corresponds to the split of $8$-dimensional root vectors into pairs of
four-dimensional root vectors, where the Cartan subalgebra of $\Eee$
is taken to be the sum of the conventional Cartan subalgebras of both
$SO(8)$, and for both $SO(16)$ and $SO(8)$, spinors are those vectors
with entries $\pm 1/2$ and an even number of minus signs (cospinors
odd).}

When taking the diagonal of both $SO(8)$ rotations in this scheme,
each of the $({\bf 8},{\bf 8})$ decompose into ${\bf 1}+{\bf 28}+{\bf
35}_{v,s,c}$; the three singlets (two noncompact, one compact) form an
$SL(2)$, while the compact generators from ${\bf 35}_v$ may be used to
extend the diagonal $SO(8)$ to $SU(8)$ which is further extended by
the ${\bf 35}_s$ and ${\bf 35}_c$ to $E_{7(+7)}$ which commutes with
forementioned $SL(2)$. Hence, under this $E_{7(+7)}\times SL(2)$,
${\bf 248}\rightarrow ({\bf 133},{\bf 1}) + ({\bf 1},{\bf 3}) + ({\bf
56},{\bf 2})$.

Re-identifying the $E_{8(+8)}$ generators corresponding to the $SU(3)$
singlets, resp. the $SU(8)$ rotations used to parametrize the singlet
manifold given in \cite{Warner:vz} is straightforward; exponentiating
them, however, is not. Looking closely at explicit $248\times248$
matrix representations of these generators reveals that, after
suitable re-ordering of coordinates, they decompose into blocks of
maximal size $8\times8$ and are (by using a computer) sufficiently
easy to diagonalize.\footnote{One advantage of this brute-force
approach in comparison to more sophisticated considerations involving
the octonionic structure of $\Eee$ is that no extra work is necessary
to handle the same problem for $E_{7(+7)}$ and $E_{6(+6)}$.}

Considerable simplification of the task of computing explicit analytic
expressions for the scalar potential by making use of as much group
theoretical structure as possible is expected, but nowadays computers
are powerful enough to not only allow a head-on approach using
explicit 248-dimensional component notation and symbolic algebra on
sparsely occupied tensors, but also make this the preferable route
when the aim is to investigate high-dimensional singlet spaces. This
is explained in detail in chapter $\ref{LambdaTensorCh}$.

One important complication arises from the fact that the manifold $M$
of $SU(3)\subset SU(4)\subset SO(8)_{\rm diag}\subset SO(8)\times
SO(8)$ singlets from the ${\bf 128}$ is not six-dimensional, as in the
$D=4$ $E_{7(+7)}$ case, but twelve-dimensional, since there are two
additional singlets from the noncompact directions of $SL(2)$ as well
as four more from the $({\bf 56},{\bf 2})$. While explicit analytic
calculation of the potential on submanifolds of $M$ reveals that the
full 12-dimensional potential is\footnote{by now probably just} out of
reach of a complete analysis using standard techniques, it is
nevertheless possible to make progress by making educated guesses at
the possible locations of extrema; for example, one notes that for
four of the five stationary points given in \cite{Warner:vz}, the
angular parameters are just such that the sines and cosines appearing
in the potential all have values $\{-1;0;+1\}$. Hence it seems
reasonable to try to search for stationary points by letting these
compact coordinates of this particular parametrization run through a
discrete set of special values only, thereby reducing the number of
coordinates.

The immediate problem with the investigation of only proper
submanifolds $M'$ of the full manifold $M$ of singlets is that, aside
from the fundamental inability to prove the exhaustiveness of the list
of stationary points with remaining symmetry of at least $H$ thus
obtained, the vanishing of derivatives within $M'$ does not guarantee
to have a stationary point of the full potential. A sieve for true
solutions is given by the stationarity condition (4.12) in
\cite{Nicolai:2001sv}
\begin{equation}
\label{SolutionSieve}
3\,A_1^{IM}A_2^{M\dot A}=A_2^{I\dot B}A_3^{\dot A\dot B}.
\end{equation}
which has to be checked anyway, since coordinate singularities may
also create `fake' stationary points.\footnote{For the purpose of
illustration, imagine a 2-sphere conventionally parametrized by
$\theta, \phi$ and a scalar field $\Phi$ on the sphere whose gradient
at the north pole is parallel to the $\phi=\pi/2$ great circle. Then
for the scalar field given in $\theta,\phi$ coordinates,
$\partial_\theta\Phi_{|N}=\partial_\phi\Phi_{|N}=0$, but this point is
not a stationary point.}

By doing the calculation, one finds that on the six-dimensional
$SU(3)\subset SO(8)$ invariant subspace of $E_{7(+7)}/SU(8)$
(which one can identify as the space
$(SU(2,1)/(SU(2)\times U(1)))\times (SL(2)/U(1))$) given in
\cite{Warner:vz}, the $E_8$ supergravity potential does not feature
any true nontrivial stationary points, hence the plan here is to
extend the calculation to an eight-dimensional subspace of the full
singlet manifold by also parametrizing the noncompact directions of
$SL(2)$. Explicitly, if we define an embedding of the $su(8)$ algebra
(here with indices ${\tt i}, {\tt j}$ into the $so(16)$ algebra via
\begin{equation}
H^{IJ}{}_{\tt i}{}^{\tt j} = \left\{\begin{array}{llll}
+1/2&\;\mbox{if}\;&I={\tt i}&J={\tt j}\\
+1/2&\;\mbox{if}\;&I={\tt i}+8,&J={\tt j}+8\\
+1/2&\;\mbox{if}\;&I={\tt i}+8,&J={\tt j}\\
-1/2&\;\mbox{if}\;&I={\tt i},&J={\tt j}+8\\
0&&\mbox{otherwise}
\end{array}\right.
\end{equation}
then this parametrization is given by\footnote{Note that $G_1^+$ and $G_2^+$ commute.}
\begin{eqnarray}\label{phi-so8xso8-map}
G_1^+{}^\cC{}_\cB&:=&\frac{1}{4}f_{\alpha\beta\,\cB}{}^\cC\left(\gamma^{1234}_{\alpha\beta}+\gamma^{1256}_{\alpha\beta}+\gamma^{1278}_{\alpha\beta}\right)\nonumber\\
G_2^+{}^\cC{}_\cB&:=&\frac{1}{4}f_{\alpha\beta\,\cB}{}^\cC\left(-\gamma^{1357}_{\alpha\beta}+\gamma^{1368}_{\alpha\beta}+\gamma^{1458}_{\alpha\beta}+\gamma^{1467}_{\alpha\beta}\right)\nonumber\\
X^{(a)}{}^\cC{}_\cB&:=&f_{\IJ\,\cB}{}^\cC H^\IJ_{IJ} H^{IJ}{}_{\tt i}{}^{\tt j}\left(\delta^{\tt i}_{\tt j}-4\,\delta^{\tt i}_7\delta_{\tt j}^7-4\,\delta^{\tt i}_8\delta_{\tt j}^8\right)\nonumber\\
X^{(\phi)}{}^\cC{}_\cB&:=&f_{\IJ\,\cB}{}^\cC H^\IJ_{IJ} H^{IJ}{}_{\tt i}{}^{\tt j}\left(i\,\delta^{\tt i}_7\delta_{\tt j}^7-i\,\delta^{\tt i}_8\delta_{\tt j}^8\right)\nonumber\\
X^{(\vartheta)}{}^\cC{}_\cB&:=&f_{\IJ\,\cB}{}^\cC H^\IJ_{IJ} H^{IJ}{}_{\tt i}{}^{\tt j}\left(\delta^{\tt i}_8\delta_{\tt j}^7-\,\delta^{\tt i}_7\delta_{\tt j}^8\right)\nonumber\\
X^{(\psi)}{}^\cC{}_\cB&:=&X^{(\vartheta)}{}^\cC{}_\cB\\
W^\cC{}_\cB&:=&\frac{1}{4}f_{\IJ\,\cB}{}^\cC H^\IJ_{IJ} H^{I}_{i} H^{J}_{\bar j} \delta^{i\bar j}\nonumber\\
Z^\cC{}_\cB&:=&\frac{1}{4}f_{\alpha\beta\,\cB}{}^\cC \delta^{\alpha\beta}\nonumber\\
\cV&=&\exp(wW)\exp(zZ)\exp(-wW)\nonumber\\
&&\exp(aX^{(a)})\exp(\phi X^{(\phi)})\exp(\vartheta X^{(\vartheta)})\exp(\psi X^{(\psi)})\nonumber\\
&&\exp\left(\lambda_1 G_1^+ +\lambda_2 G_2^+\right)\nonumber\\
&&\exp(-\psi X^{(\psi)})\exp(-\vartheta X^{(\vartheta)})\exp(-\phi X^{(\phi)})\exp(-aX^{(a)})\nonumber
\end{eqnarray}
and the corresponding analytic form of the potential
reads\footnote{Due to right $SO(16)$ invariance, this calculation can
be greatly simplified by omitting the $\exp(-\ldots)$ factors in this
expression. The computation of this potential took slightly more than
two hours of CPU time on a 1.7 GHz Pentium-IV.}
{\small
\setlongtables
\begin{longtable}{lll}\label{potential_su3diag}
$-8g^{-2}V$&=&\hfill
\\
&&$\frac{189}{8}+\frac{3}{8}\,\cos(4\,\phi)+3\,\cos(2\,a)\,\cos(2\,\phi)+\frac{7}{2}\,\cosh(2\,\lambda_2)$\\
&&$+\frac{7}{8}\,\cosh(4\,\lambda_2)-\frac{1}{2}\,\cosh(2\,\lambda_2)\,\cos(4\,\phi)$\\
&&$+\frac{1}{8}\,\cosh(4\,\lambda_2)\,\cos(4\,\phi)-3\,\cosh(2\,\lambda_2)\,\cos(2\,a)\,\cos(2\,\phi)$\\
&&$+3\,\cosh(2\,\lambda_1)-3\,\cosh(2\,\lambda_1)\,\cos(2\,a)\,\cos(2\,\phi)+9\,\cosh(2\,\lambda_1)\,\cosh(2\,\lambda_2)$\\
&&$+3\,\cosh(2\,\lambda_1)\,\cosh(2\,\lambda_2)\,\cos(2\,a)\,\cos(2\,\phi)+\frac{405}{32}\,\cosh(z)\,\cosh(\lambda_1)$\\
&&$+\frac{23}{32}\,\cosh(z)\,\cosh(3\,\lambda_1)-\frac{9}{32}\,\cosh(z)\,\cosh(\lambda_1)\,\cos(4\,\phi)$\\
&&$-\frac{3}{32}\,\cosh(z)\,\cosh(3\,\lambda_1)\,\cos(4\,\phi)$\\
&&$-\frac{3}{8}\,\cosh(z)\,\cosh(\lambda_1)\,\cos(2\,a)\,\cos(2\,\phi)$\\
&&$+\frac{3}{8}\,\cosh(z)\,\cosh(3\,\lambda_1)\,\cos(2\,a)\,\cos(2\,\phi)$\\
&&$+\frac{93}{8}\,\cosh(z)\,\cosh(\lambda_1)\,\cosh(2\,\lambda_2)$\\
&&$-\frac{9}{32}\,\cosh(z)\,\cosh(\lambda_1)\,\cosh(4\,\lambda_2)$\\
&&$-\frac{1}{8}\,\cosh(z)\,\cosh(3\,\lambda_1)\,\cosh(2\,\lambda_2)$\\
&&$-\frac{19}{32}\,\cosh(z)\,\cosh(3\,\lambda_1)\,\cosh(4\,\lambda_2)$\\
&&$+\frac{3}{8}\,\cosh(z)\,\cosh(\lambda_1)\,\cosh(2\,\lambda_2)\,\cos(4\,\phi)$\\
&&$-\frac{3}{32}\,\cosh(z)\,\cosh(\lambda_1)\,\cosh(4\,\lambda_2)\,\cos(4\,\phi)$\\
&&$+\frac{1}{8}\,\cosh(z)\,\cosh(3\,\lambda_1)\,\cosh(2\,\lambda_2)\,\cos(4\,\phi)$\\
&&$-\frac{1}{32}\,\cosh(z)\,\cosh(3\,\lambda_1)\,\cosh(4\,\lambda_2)\,\cos(4\,\phi)$\\
&&$+\frac{3}{8}\,\cosh(z)\,\cosh(\lambda_1)\,\cosh(4\,\lambda_2)\,\cos(2\,a)\,\cos(2\,\phi)$\\
&&$-\frac{3}{8}\,\cosh(z)\,\cosh(3\,\lambda_1)\,\cosh(4\,\lambda_2)\,\cos(2\,a)\,\cos(2\,\phi)$\\
&&$+\frac{15}{32}\,\cos(w-3\,a)\,\sinh(\lambda_1)\,\sinh(z)+\frac{99}{16}\,\cos(w+a)\,\sinh(\lambda_1)\,\sinh(z)$\\
&&$-\frac{5}{32}\,\cos(w-3\,a)\,\sinh(3\,\lambda_1)\,\sinh(z)-\frac{9}{16}\,\cos(w+a)\,\sinh(3\,\lambda_1)\,\sinh(z)$\\
&&$-\frac{9}{32}\,\cos(w-3\,a)\,\cos(4\,\phi)\,\sinh(\lambda_1)\,\sinh(z)$\\
&&$+\frac{45}{8}\,\cos(w-a)\,\cos(2\,\phi)\,\sinh(\lambda_1)\,\sinh(z)$\\
&&$+\frac{3}{32}\,\cos(w-3\,a)\,\cos(4\,\phi)\,\sinh(3\,\lambda_1)\,\sinh(z)$\\
&&$-\frac{3}{8}\,\cos(w-a)\,\cos(2\,\phi)\,\sinh(3\,\lambda_1)\,\sinh(z)$\\
&&$-\frac{3}{8}\,\cos(w-3\,a)\,\cosh(2\,\lambda_2)\,\sinh(\lambda_1)\,\sinh(z)$\\
&&$-6\,\cos(w+a)\,\cosh(2\,\lambda_2)\,\sinh(\lambda_1)\,\sinh(z)$\\
&&$-\frac{3}{32}\,\cos(w-3\,a)\,\cosh(4\,\lambda_2)\,\sinh(\lambda_1)\,\sinh(z)$\\
&&$-\frac{3}{16}\,\cos(w+a)\,\cosh(4\,\lambda_2)\,\sinh(\lambda_1)\,\sinh(z)$\\
&&$+\frac{1}{8}\,\cos(w-3\,a)\,\cosh(2\,\lambda_2)\,\sinh(3\,\lambda_1)\,\sinh(z)$\\
&&$+\frac{1}{32}\,\cos(w-3\,a)\,\cosh(4\,\lambda_2)\,\sinh(3\,\lambda_1)\,\sinh(z)$\\
&&$+\frac{9}{16}\,\cos(w+a)\,\cosh(4\,\lambda_2)\,\sinh(3\,\lambda_1)\,\sinh(z)$\\
&&$+\frac{3}{8}\,\cos(w-3\,a)\,\cosh(2\,\lambda_2)\,\cos(4\,\phi)\,\sinh(\lambda_1)\,\sinh(z)$\\
&&$-6\,\cos(w-a)\,\cosh(2\,\lambda_2)\,\cos(2\,\phi)\,\sinh(\lambda_1)\,\sinh(z)$\\
&&$-\frac{3}{32}\,\cos(w-3\,a)\,\cosh(4\,\lambda_2)\,\cos(4\,\phi)\,\sinh(\lambda_1)\,\sinh(z)$\\
&&$+\frac{3}{8}\,\cos(w-a)\,\cosh(4\,\lambda_2)\,\cos(2\,\phi)\,\sinh(\lambda_1)\,\sinh(z)$\\
&&$-\frac{1}{8}\,\cos(w-3\,a)\,\cosh(2\,\lambda_2)\,\cos(4\,\phi)\,\sinh(3\,\lambda_1)\,\sinh(z)$\\
&&$+\frac{1}{32}\,\cos(w-3\,a)\,\cosh(4\,\lambda_2)\,\cos(4\,\phi)\,\sinh(3\,\lambda_1)\,\sinh(z)$\\
&&$+\frac{3}{8}\,\cos(w-a)\,\cosh(4\,\lambda_2)\,\cos(2\,\phi)\,\sinh(3\,\lambda_1)\,\sinh($z)
\end{longtable}
}

Note that again, just as in four dimensions, the dependency on the
angular coordinates $\vartheta,\psi$ drops out. Parametrizing six
dimensions, this potential is slightly more general than the one given
in \cite{Fischbacher:2002hg} (which furthermore uses a different
normalization of the embedding tensor and hence has to be re-scaled by
a factor $1/4$ to compare it with this result); at first, it looks a
bit surprising that the extra $SL(2)$ angle $w$ combines so nicely
with the $SU(8)$ angle $a$, where one a priori might have expected
this extra angle to at least double the complexity of the potential.

\subsection{Residual symmetry of $G_{2\,{\rm diag}}$}

Before we turn to the discussion of stationary points of the potential
of $N=16$ $D=3$ $SO(8)\times SO(8)$ gauged SUGRA, let us look at
parametrizations of some other particularly interesting submanifolds.
The subgroup of $SO(8)_{\rm diag}$ that leaves both the vector and the
spinor whose coordinates are $(0,0,\ldots,0,1)$ invariant is $G_2$
which in our conventions $(\ref{sigma8})$ also leaves the co-spinor
$(0,0,\ldots,0,1)$ invariant. Hence, we get one $G_2$ singlet from
each of the $({\bf 8},{\bf 8})$ $SO(8)\times SO(8)$ representations,
and the group commuting with $G_{2\,{\rm diag}}$ is $SL(2)\times
SL(2)$, yielding a four-parameter submanifold of $\Eee/SO(16)$ on
which the potential can be calculated without further truncation, so
one can hope do deduce a statement about {\em all} stationary points
with at least a remaining symmetry of $G_{2\,{\rm diag}}$.

The simplest case featuring a nontrivial stationary point is obtained
by restriction to the submanifold of $SO(7)$ singlets where $SO(7)$ is
chosen in such a way that either ${\bf 8}_s\rightarrow {\bf 7}+{\bf
1}$ or (here, equivalently) ${\bf 8}_c\rightarrow {\bf 7}+{\bf 1}$. When
using a suggestive parametrization for the $G_{2\,{\rm diag}}$ case,
this simpler case follows by omitting one angular parameter.

The manifold of $G_{2\,{\rm diag}}$ singlet scalars is parametrized by
\begin{eqnarray}\label{g2singlets}
V^\cC{}_\cB&:=&f_{\IJ\cB}{}^\cC H^\IJ_{IJ}H^I_i H^J_{\bar j}\left(2\delta^i_8\delta^{\bar j}_{8}-\frac{1}{4}\delta^{i\bar j}\right)\nonumber\\
S^\cC{}_\cB&:=&\frac{1}{4}f_{\alpha\beta\,\cB}{}^\cC\left(8\delta^\alpha_8\delta^\beta_8-\delta^{\alpha\beta}\right)\nonumber\\
W^\cC{}_\cB, Z^\cC{}_\cB&&\mbox{as in $(\ref{phi-so8xso8-map})$}\\
\cV&=&\exp(wW)\exp(zZ)\exp(-wW)\nonumber\\
&&\exp(vV)\exp(sS)\exp(-vV)\nonumber
\end{eqnarray}
yielding the potential
\begin{equation}
\begin{array}{lcl}
-8g^{-2}V&=&\frac{243}{8}+\frac{7}{2}\,\cosh(2\,s)+\frac{49}{8}\,\cosh(4\,s)+\frac{21}{8}\,\cos(4\,v)\\
&&-\frac{7}{2}\,\cos(4\,v)\,\cosh(2\,s)+\frac{7}{8}\,\cos(4\,v)\,\cosh(4\,s)\\
&&+\frac{1141}{64}\,\cosh(z)\,\cosh(s)+\frac{427}{64}\,\cosh(z)\,\cosh(3\,s)\\
&&-\frac{7}{64}\,\cosh(z)\,\cosh(5\,s)-\frac{25}{64}\,\cosh(z)\,\cosh(7\,s)\\
&&-\frac{21}{64}\,\cosh(z)\,\cos(4\,v)\,\cosh(s)\\
&&+\frac{21}{64}\,\cosh(z)\,\cos(4\,v)\,\cosh(3\,s)\\
&&+\frac{7}{64}\,\cosh(z)\,\cos(4\,v)\,\cosh(5\,s)\\
&&-\frac{7}{64}\,\cosh(z)\,\cos(4\,v)\,\cosh(7\,s)\\
&&-\frac{1645}{128}\,\cos(w-v)\,\sinh(s)\,\sinh(z)\\
&&+\frac{651}{128}\,\cos(w-v)\,\sinh(3\,s)\,\sinh(z)\\
&&+\frac{7}{128}\,\cos(w-v)\,\sinh(5\,s)\,\sinh(z)\\
&&-\frac{49}{128}\,\cos(w-v)\,\sinh(7\,s)\,\sinh(z)\\
&&-\frac{315}{64}\,\cos(w+3\,v)\,\sinh(s)\,\sinh(z)\\
&&+\frac{133}{64}\,\cos(w+3\,v)\,\sinh(3\,s)\,\sinh(z)\\
&&-\frac{7}{64}\,\cos(w+3\,v)\,\sinh(5\,s)\,\sinh(z)\\
&&-\frac{7}{64}\,\cos(w+3\,v)\,\sinh(7\,s)\,\sinh(z)\\
&&+\frac{35}{128}\,\cos(w+7\,v)\,\sinh(s)\,\sinh(z)\\
&&-\frac{21}{128}\,\cos(w+7\,v)\,\sinh(3\,s)\,\sinh(z)\\
&&+\frac{7}{128}\,\cos(w+7\,v)\,\sinh(5\,s)\,\sinh(z)\\
&&-\frac{1}{128}\,\cos(w+7\,v)\,\sinh(7\,s)\,\sinh(z)
\end{array}
\end{equation}
which reduces to the $SO(7)$ potential by setting $v=0$, giving
\begin{equation}
\begin{array}{lcl}
-8g^{-2}V&=&33+7\cosh(4\,s)+\frac{35}{2}\,\cosh(z)\,\cosh(s)+7\,\cosh(z)\,\cosh(3\,s)\\
&&-\frac{1}{2}\,\cosh(z)\,\cosh(7\,s)-\frac{35}{2}\,\cos(w)\,\sinh(s)\,\sinh(z)\\
&&+7\,\cos(w)\,\sinh(3\,s)\,\sinh(z)-\frac{1}{2}\,\cos(w)\,\sinh(7\,s)\,\sinh(z)
\end{array}
\end{equation}
.

\subsection{Residual symmetry of $SO(6)_{\rm diag}$}
\label{so6diag}
For noncompact gauge groups (to be discussed in the next chapter), any
vacuum will break the gauge group down to a subgroup of its maximal
compact subgroup, thus when studying gaugings like $SO(6,2)\times
SO(6,2)$, it is natural to consider breaking to groups like
$H=SO(6)_{\rm diag}$. Once a suitable parametrization of the
submanifold of $H$-singlets is obtained, this can often be re-cycled
for other gauge groups by re-doing the calculation with a different
embedding tensor $\Theta$. Hence, of all the possible ways to embed
$SO(6)$ into $SO(8)\times SO(8)$, we will consider here only the case
of the diagonal $SO(6)$ of the $SO(6,2)\times SO(6,2)$ subgroup of
$SO(8,8)$ which itself is formed from the $({\bf 1},{\bf 28})+({\bf 28},{\bf 1})+({\bf 8}_s, {\bf 8}_s)$ of the
$E_8\rightarrow SO(8)\times SO(8)$ decomposition
${\bf 248}=({\bf 1},{\bf 28})+({\bf 28},{\bf 1})+({\bf 8}_v, {\bf 8}_v)+({\bf 8}_s, {\bf 8}_s)+({\bf 8}_c, {\bf 8}_c)$.
The particular choice we make is that our $SO(6)_{\rm diag}$
shall leave fixed the last two spinor coordinates.
Using such a diagonal embedding, there are five singlets
under $SO(6)\times SO(2)$, seven singlets under $SO(6)$,
and twelve singlets under $SU(3)$, so from the point
of complexity, so it is reasonable to try to calculate
the potential on the $SO(6)$ invariant seven-manifold, which
consists of the noncompact directions of the $SL(3)\times SL(2)$
group commuting with $SO(6)_{\rm diag}$. The generators of this
$SL(3)\times SL(2)$ are given as follows, in the same notation
as in \cite{Fischbacher:2002fx}:
{\small
\beq\label{so6-sl3xsl2}
\begin{array}{lclclcl}
p_1{}^{\mathcal{C}}{}_{\mathcal{B}}&=&\multicolumn{5}{l}{\frac{1}{2}\left(\delta^i_1\delta^j_7+\delta^i_2\delta^j_8-\delta^i_3\delta^j_6+\delta^i_4\delta^j_5\right)\left(\delta_i^I\delta_j^{J-8}-\delta_i^{J-8}\delta_j^{I}+\delta_i^{I-8}\delta_j^J-\delta_i^{J}\delta_j^{I-8}\right)f_{\underline{[IJ]}\mathcal{B}}{}^{\mathcal{C}}}\\
p_2{}^{\mathcal{C}}{}_{\mathcal{B}}&=&\multicolumn{5}{l}{\frac{1}{2}\delta^{ij}\left(\delta_i^{I-8}\delta_j^J-\delta_i^I\delta_j^{J-8}\right)f_{\underline{[IJ]}\mathcal{B}}{}^{\mathcal{C}}}\\
p_3{}^{\mathcal{C}}{}_{\mathcal{B}}&=&\multicolumn{5}{l}{-\frac{1}{2}\left(\delta^i_1\delta^j_7+\delta^i_2\delta^j_8-\delta^i_3\delta^j_6+\delta^i_4\delta^j_5\right)\left(\delta_i^I\delta_j^J-\delta_i^J\delta_j^I-\delta_i^{I-8}\delta_j^{J-8}+\delta_i^{J-8}\delta_j^{I-8}\right)f_{\underline{[IJ]}\mathcal{B}}{}^{\mathcal{C}}}\\
p_4{}^{\mathcal{C}}{}_{\mathcal{B}}&=&\multicolumn{5}{l}{\frac{1}{2}\left(\delta^{\dot\gamma}_1\delta^{\dot\delta}_2-\delta^{\dot\gamma}_3\delta^{\dot\delta}_4+\delta^{\dot\gamma}_5\delta^{\dot\delta}_6+\delta^{\dot\gamma}_7\delta^{\dot\delta}_8\right)\left(\delta^{\dot\alpha}_{\dot\gamma}\delta^{\dot\beta}_{\dot\delta}-\delta^{\dot\beta}_{\dot\gamma}\delta^{\dot\alpha}_{\dot\delta}\right)f_{\dot\alpha\dot\beta\mathcal{B}}{}^{\mathcal{C}}}\\
p_5{}^{\mathcal{C}}{}_{\mathcal{B}}&=&\frac{1}{2}\,\delta^{\dot\alpha\dot\beta}\,f_{\dot\alpha\dot\beta\mathcal{B}}{}^{\mathcal{C}}&\qquad&
p_6{}^{\mathcal{C}}{}_{\mathcal{B}}&=&\left(\delta^\alpha_7\delta^\beta_8-\delta^\alpha_8\delta^\beta_7\right)f_{\alpha\beta\mathcal{B}}{}^{\mathcal{C}}\\
p_7{}^{\mathcal{C}}{}_{\mathcal{B}}&=&\frac{1}{2}\,\delta^{\alpha\beta}\,f_{\alpha\beta\mathcal{B}}{}^{\mathcal{C}}&\qquad&
p_8{}^{\mathcal{C}}{}_{\mathcal{B}}&=&p_7{}^{\mathcal{C}}{}_{\mathcal{B}}-2\left(\delta^\alpha_7\delta^\beta_7+\delta^\alpha_8\delta^\beta_8\right)f_{\alpha\beta\mathcal{B}}{}^{\mathcal{C}}\\
q_1{}^{\mathcal{C}}{}_{\mathcal{B}}&=&\multicolumn{5}{l}{\frac{1}{4}\left(\delta^i_1\delta^j_7+\delta^i_2\delta^j_8-\delta^i_3\delta^j_6+\delta^i_4\delta^j_5\right)\left(\delta_i^I\delta_j^J-\delta_i^J\delta_j^I+\delta_i^{I-8}\delta_j^{J-8}-\delta_i^{J-8}\delta_j^{I-8}\right)f_{\underline{[IJ]}\mathcal{B}}{}^{\mathcal{C}}}\\
q_2{}^{\mathcal{C}}{}_{\mathcal{B}}&=&\frac{1}{2}\left(\delta^\alpha_7\delta^\beta_7-\delta^\alpha_8\delta^\beta_8\right)f_{\alpha\beta\mathcal{B}}{}^{\mathcal{C}}&\qquad&
q_3{}^{\mathcal{C}}{}_{\mathcal{B}}&=&\frac{1}{2}\left(\delta^\alpha_7\delta^\beta_8+\delta^\alpha_8\delta^\beta_7\right)f_{\alpha\beta\mathcal{B}}{}^{\mathcal{C}}.\\
&&
\end{array}
\eeq
}

The $p$-, resp. $q$-generators stand in one-to-one correspondence to
the following $SL(3)$ resp. $SL(2)$ generator matrices in the defining
representations that satisfy the same commutation relations (e.g.
$[p_8,p_6]=3\,p_3$ and $[\tilde p_8,\tilde p_6]=3\,\tilde p_3$) and
hence are much better suited to read off such commutation relations
than above $E_8$ generator definitions.

{\small
\beq\label{sl3map}
\begin{array}{lclclclclcl}
\tilde p_1&=&\left(\begin{array}{ccc}0&0&0\\0&0&1\\0&-1&0\end{array}\right)&\quad&\tilde p_2&=&\left(\begin{array}{ccc}0&1&0\\-1&0&0\\0&0&0\end{array}\right)&\quad&\tilde p_3&=&\left(\begin{array}{ccc}0&0&-1\\0&0&0\\1&0&0\end{array}\right)\\
\tilde p_4&=&\left(\begin{array}{ccc}0&0&0\\0&0&1\\0&1&0\end{array}\right)&\quad&\tilde p_5&=&\left(\begin{array}{ccc}0&1&0\\1&0&0\\0&0&0\end{array}\right)&\quad&\tilde p_6&=&\left(\begin{array}{ccc}0&0&-1\\0&0&0\\-1&0&0\end{array}\right)\\
\tilde p_7&=&\left(\begin{array}{ccc}-1&0&0\\0&1&0\\0&0&0\end{array}\right)&\quad&\tilde p_8&=&\left(\begin{array}{ccc}1&0&0\\0&1&0\\0&0&-2\end{array}\right)&&&&\\
\tilde q_1&=&\left(\begin{array}{cc}0&1/2\\-1/2&0\end{array}\right)&\quad&
\tilde q_2&=&\left(\begin{array}{cc}1/2&0\\0&-1/2\end{array}\right)&\quad&
\tilde q_3&=&\left(\begin{array}{cc}0&1/2\\1/2&0\end{array}\right)\\
&&&&
\end{array}
\eeq
}

The noncompact directions of $SL(3)$ are hence equivalent to the
symmetric traceless $3\times3$ matrices, and since the axes of any
tensor ellipsoid can be aligned with the coordinate axes by a suitable
rotation, we parametrize the scalars from $SL(3)$ by applying all
$SO(3)$ rotations (which we parametrize by $yzx$ Euler angles, since
such a system is better behaved in terms of coordinate singularities
than a $zxz$ system\footnote{Hence, the yaw, pitch, roll system used
  in aviation is also of $zxy$ type.}) to all linear combinations of
two linear independent commuting diagonal matrices. Thus a convenient
coordinate parametrization of our seven-dimensional singlet space is
given by
\beq\label{so6diagparam}
\begin{array}{lcl}
\mathcal{V}&=&\exp\left({r_1\,p_1}\right)\,\exp\left({r_2\,p_3}\right)\,\exp\left({r_3\,p_2}\right)\\
&&\,\exp\left({zp_8-sp_7}\right)\,\exp\left({-r_3\,p_2}\right)\\
&&\exp\left({-r_2\,p_3}\right)\,\exp\left({-r_1\,p_1}\right)\\
&&\exp\left({r_5\,q_1}\right)\,\exp\left({v\,q_2}\right)\,\exp\left({-r_5\,q_1}\right)\;
\end{array}
\eeq
and the potential we obtain reads
{\small
\setlongtables
\begin{longtable}{lll}\label{potential_so6diag}
$-8g^{-2}V$&$=$&\hfill
\\
&&$27+3\,\cosh(4\,z)+3\,\cosh(4\,z)\,\cos(2\,r_2)-3\,\cosh(4\,z)\,\cos(2\,r_1)$\\
&&$-3\,\cosh(4\,z)\,\cos(2\,r_1)\,\cos(2\,r_2)+\frac{1}{4}\,\cosh(4\,s)$\\
&&$+\frac{1}{4}\,\cosh(4\,s)\,\cos(2\,r_2)-\frac{1}{4}\,\cosh(4\,s)\,\cos(2\,r_1)$\\
&&$-\frac{1}{4}\,\cosh(4\,s)\,\cos(2\,r_1)\,\cos(2\,r_2)+9\,\cosh(2\,s)\,\cosh(2\,z)$\\
&&$+\frac{3}{4}\,\cosh(2\,s)\,\cosh(6\,z)-3\,\cos(2\,r_3)\,\sinh(2\,z)\,\sinh(2\,s)$\\
&&$+\frac{1}{4}\,\cos(2\,r_3)\,\sinh(6\,z)\,\sinh(2\,s)$\\
&&$-3\,\cosh(2\,s)\,\cosh(2\,z)\,\cos(2\,r_2)-\frac{1}{4}\,\cosh(2\,s)\,\cosh(6\,z)\,\cos(2\,r_2)$\\
&&$-3\,\cos(2\,r_2)\,\cos(2\,r_3)\,\sinh(2\,z)\,\sinh(2\,s)$\\
&&$+\frac{1}{4}\,\cos(2\,r_2)\,\cos(2\,r_3)\,\sinh(6\,z)\,\sinh(2\,s)$\\
&&$+3\,\cosh(2\,s)\,\cosh(2\,z)\,\cos(2\,r_1)+\frac{1}{4}\,\cosh(2\,s)\,\cosh(6\,z)\,\cos(2\,r_1)$\\
&&$-9\,\cos(2\,r_1)\,\cos(2\,r_3)\,\sinh(2\,z)\,\sinh(2\,s)$\\
&&$+\frac{3}{4}\,\cos(2\,r_1)\,\cos(2\,r_3)\,\sinh(6\,z)\,\sinh(2\,s)$\\
&&$+3\,\cosh(2\,s)\,\cosh(2\,z)\,\cos(2\,r_1)\,\cos(2\,r_2)$\\
&&$+\frac{1}{4}\,\cosh(2\,s)\,\cosh(6\,z)\,\cos(2\,r_1)\,\cos(2\,r_2)$\\
&&$-12\,\sin(2\,r_3)\,\sin(r_2)\,\sin(2\,r_1)\,\sinh(2\,z)\,\sinh(2\,s)$\\
&&$+\sin(2\,r_3)\,\sin(r_2)\,\sin(2\,r_1)\,\sinh(6\,z)\,\sinh(2\,s)$\\
&&$+3\,\cos(2\,r_1)\,\cos(2\,r_2)\,\cos(2\,r_3)\,\sinh(2\,z)\,\sinh(2\,s)$\\
&&$-\frac{1}{4}\,\cos(2\,r_1)\,\cos(2\,r_2)\,\cos(2\,r_3)\,\sinh(6\,z)\,\sinh(2\,s)$\\
&&$+\cosh(2\,v)+9\,\cosh(v)\,\cosh(4\,z)$\\
&&$-3\,\cosh(v)\,\cosh(4\,z)\,\cos(2\,r_2)$\\
&&$+3\,\cosh(v)\,\cosh(4\,z)\,\cos(2\,r_1)$\\
&&$+3\,\cosh(v)\,\cosh(4\,z)\,\cos(2\,r_1)\,\cos(2\,r_2)$\\
&&$-\frac{1}{4}\,\cosh(2\,v)\,\cosh(4\,s)-\frac{1}{4}\,\cosh(2\,v)\,\cosh(4\,s)\,\cos(2\,r_2)$\\
&&$+\frac{1}{4}\,\cosh(2\,v)\,\cosh(4\,s)\,\cos(2\,r_1)$\\
&&$+\frac{1}{4}\,\cosh(2\,v)\,\cosh(4\,s)\,\cos(2\,r_1)\,\cos(2\,r_2)$\\
&&$+15\,\cosh(v)\,\cosh(2\,s)\,\cosh(2\,z)-\frac{3}{4}\,\cosh(2\,v)\,\cosh(2\,s)\,\cosh(6\,z)$\\
&&$+3\,\cosh(v)\,\cos(2\,r_3)\,\sinh(2\,z)\,\sinh(2\,s)$\\
&&$-\frac{1}{4}\,\cosh(2\,v)\,\cos(2\,r_3)\,\sinh(6\,z)\,\sinh(2\,s)$\\
&&$+3\,\cosh(v)\,\cosh(2\,s)\,\cosh(2\,z)\,\cos(2\,r_2)$\\
&&$+\frac{1}{4}\,\cosh(2\,v)\,\cosh(2\,s)\,\cosh(6\,z)\,\cos(2\,r_2)$\\
&&$+3\,\cosh(v)\,\cos(2\,r_2)\,\cos(2\,r_3)\,\sinh(2\,z)\,\sinh(2\,s)$\\
&&$-\frac{1}{4}\,\cosh(2\,v)\,\cos(2\,r_2)\,\cos(2\,r_3)\,\sinh(6\,z)\,\sinh(2\,s)$\\
&&$-3\,\cosh(v)\,\cosh(2\,s)\,\cosh(2\,z)\,\cos(2\,r_1)$\\
&&$-\frac{1}{4}\,\cosh(2\,v)\,\cosh(2\,s)\,\cosh(6\,z)\,\cos(2\,r_1)$\\
&&$+9\,\cosh(v)\,\cos(2\,r_1)\,\cos(2\,r_3)\,\sinh(2\,z)\,\sinh(2\,s)$\\
&&$-\frac{3}{4}\,\cosh(2\,v)\,\cos(2\,r_1)\,\cos(2\,r_3)\,\sinh(6\,z)\,\sinh(2\,s)$\\
&&$-3\,\cosh(v)\,\cosh(2\,s)\,\cosh(2\,z)\,\cos(2\,r_1)\,\cos(2\,r_2)$\\
&&$-\frac{1}{4}\,\cosh(2\,v)\,\cosh(2\,s)\,\cosh(6\,z)\,\cos(2\,r_1)\,\cos(2\,r_2)$\\
&&$+12\,\cosh(v)\,\sin(2\,r_3)\,\sin(r_2)\,\sin(2\,r_1)\,\sinh(2\,z)\,\sinh(2\,s)$\\
&&$-\cosh(2\,v)\,\sin(2\,r_3)\,\sin(r_2)\,\sin(2\,r_1)\,\sinh(6\,z)\,\sinh(2\,s)$\\
&&$-3\,\cosh(v)\,\cos(2\,r_1)\,\cos(2\,r_2)\,\cos(2\,r_3)\,\sinh(2\,z)\,\sinh(2\,s)$\\
&&$+\frac{1}{4}\,\cosh(2\,v)\,\cos(2\,r_1)\,\cos(2\,r_2)\,\cos(2\,r_3)\,\sinh(6\,z)\,\sinh(2\,s)$.
\end{longtable}
}

\subsection{Residual symmetry of $SO(5)_{\rm diag}$}

For an embedding of $SO(5)_{\rm diag}$ analogous to the $SO(6)$
embedding studied in the previous section (i.e. $SO(5)$ leaves fixed
the spinor indices $6,7,8$), the manifold of invariant scalars is
$14$-dimensional, and therefore too big to justify its use for this
approach. However, a particularly interesting subgroup to consider is
an analogously formed $\left(SO(5)\times SO(3)\right)_{\rm diag}$,
since there is a known de Sitter vacuum of $SO(5,3)$ gauged $N=8$
$D=4$ supergravity. Employing the techniques of the last section, it
is very tempting to go even a bit further and delete the $6,8$ and
$7,8$-rotations from $SO(3)$ to break to $SO(5)\times SO(2)$, since
the invariant manifold then is the six-dimensional space
$\mathbb{R}\times SL(3)/SO(3)$.

We first want to consider the simpler $\left(SO(5)\times
  SO(3)\right)_{\rm diag}$ case whose three singlets are the two
noncompact directions of the $SL(2)$ commuting with $E_{7(7)}$ which
we again parametrize as in $\ref{phi-so8xso8-map}$, plus an extra
singlet given by
\beq\label{so5xso3param}
M^{\cC}{}_\cB=\frac{1}{4}\left(3\sum_{j=1}^5\delta_\alpha^j\delta_\beta^j-5\sum_{j=6}^8\delta_\alpha^j\delta_\beta^j\right)\,f_{\alpha\beta\cB}{}^\cC,
\eeq
and thus the total parametrization is
\beq
\mathcal{V}=\exp(s M)\;\exp(w W)\,\exp(z Z)\,\exp(-w W)
\eeq
from which we obtain the potential
{\small
\bea
-8g^{-2}V&=&25+15\cosh(4\,s)+15\,\cosh(s)\,\cosh(z)+15/2\,\cosh(3\,s)\,\cosh(z)\nonumber\\
&&+3/2\,\cosh(5\,s)\,\cosh(z)-15\,\cos(w)\,\sinh(z)\,\sinh(s)\\
&&+15/2\,\cos(w)\,\sinh(z)\,\sinh(3\,s)-3/2\,\cos(w)\,\sinh(z)\,\sinh(5\,s).\nonumber
\eea
}

The eight generators of $SL(3)$ commuting with $SO(5)\times SO(2)$ are
{\small
\beq
\begin{array}{lcl}
h_{\alpha\beta}&=&\delta_\alpha^5\delta_\beta^6-\delta_\alpha^6\delta_\beta^5
\\p_1{}^{\mathcal{C}}{}_{\mathcal{B}}&=&\frac{1}{2}\,f_{\underline{[IJ]}\cB}{}^\cC\,H^{\underline{[IJ]}}_{IJ}H^I_{i}H^J_{\bar j}\delta^{i\bar j}\\
p_2{}^{\mathcal{C}}{}_{\mathcal{B}}&=&-\frac{1}{8}\,f_{\underline{[IJ]}\cB}{}^\cC\left(H^I_iH^J_j\gamma^{ij}_{\alpha\beta}-H^I_{\bar i}H^J_{\bar j}\gamma^{\bar i\bar j}_{\alpha\beta}\right)h_{\alpha\beta}\\
p_3{}^{\mathcal{C}}{}_{\mathcal{B}}&=&-\frac{1}{4}\,f_{\underline{[IJ]}\cB}{}^\cC H^I_iH^J_{\bar j}\gamma^{i\bar j}_{\alpha\beta}h_{\alpha\beta}\\
p_4{}^{\mathcal{C}}{}_{\mathcal{B}}&=&\frac{1}{2}\,f_{\dot\alpha\dot\beta\,\cB}{}^\cC \delta^{\dot\alpha\dot\beta}\\
p_5{}^{\mathcal{C}}{}_{\mathcal{B}}&=&-f_{\alpha\beta\,\cB}{}^\cC h^{\alpha\beta}\\
p_6{}^{\mathcal{C}}{}_{\mathcal{B}}&=&f_{\dot\alpha\dot\beta\,\cB}{}^\cC \delta^{\dot\alpha\dot\beta}_{\dot\gamma\dot\delta}\left(
\delta^{\dot\gamma}_{1}\delta^{\dot\delta}_{4}
+\delta^{\dot\gamma}_{3}\delta^{\dot\delta}_{2}
+\delta^{\dot\gamma}_{5}\delta^{\dot\delta}_{8}
+\delta^{\dot\gamma}_{6}\delta^{\dot\delta}_{7}
\right)\\
p_7{}^{\mathcal{C}}{}_{\mathcal{B}}&=&f_{\alpha\beta\,\cB}{}^\cC\left(\delta^\alpha_6\delta^\beta_6+\delta^\alpha_7\delta^\beta_7\right)\\
p_8{}^{\mathcal{C}}{}_{\mathcal{B}}&=&f_{\alpha\beta\,\cB}{}^\cC\delta^{\alpha\beta}-p_7{}^{\mathcal{C}}{}_{\mathcal{B}}
\end{array}
\eeq
}
which again are given in the proper order to share the commutation
relations of the $\tilde p$ from $(\ref{sl3map})$, hence we can again
use the parametrization
\beq\label{so5xso2param}
\begin{array}{lcl}
\mathcal{V}&=&\exp\left({r_1\,p_1}\right)\,\exp\left({r_2\,p_3}\right)\,\exp\left({r_3\,p_2}\right)\\
&&\exp\left({zp_8-sp_7}\right)\,\exp\left({-r_3\,p_2}\right)\\
&&\exp\left({-r_2\,p_3}\right)\,\exp\left({-r_1\,p_1}\right)
\end{array}
\eeq
and obtain the potential given in the appendix.

\subsection{Residual symmetry of $\left(SO(4)\times SO(4)\right)_{\rm diag}$}

Of the many different ways to embed $SO(4)\times SO(4)$ into
$SO(8)\times SO(8)$, we only consider the one analogous to the
previous $SO(5)_{\rm diag}$ and $SO(6)_{\rm diag}$ constructions: if
we label our gauge group as $SO(8)_L\times SO(8)_R$ and perform the
split ${\bf 8}_{L/R, s}\rightarrow({\bf 4}_{L/R, 1},{\bf 4}_{L/R,
  2})$, then we consider the diagonal
$SO(4)_{{\rm diag}(L1,R1)}\times SO(4)_{{\rm diag}(L2,R2)}$.
There are four singlets under this $SO(4)\times SO(4)$,
parametrizing the space $\left(SL(2)/U(1)\right)^2$,
five singlets under an analogous $SO(4)\times SO(3)$,
eleven singlets under $SO(4)\times SO(2)$,
and ten singlets under an $SO(3)\times SO(3)$.

If we parametrize the $SO(4)\times SO(3)$ singlet space via
\bea\label{so43map}
S_1^{\mathcal C}{}_{\mathcal B}&=&\frac{1}{4}\left(\sum_{j=1}^4\delta_\alpha^j\delta_\beta^j-\sum_{j=5}^8\delta_\alpha^j\delta_\beta^j\right)\,f_{\alpha\beta\,\cB}{}^\cC\nonumber\\
S_2^{\mathcal C}{}_{\mathcal B}&=&\frac{1}{4}\left(\sum_{j=1}^4\delta_{\dot \alpha}^j\delta_{\dot \beta}^j-\sum_{j=5}^8\delta_{\dot \alpha}^j\delta_{\dot \beta}^j\right)\,f_{\dot\alpha\dot\beta\,\cB}{}^\cC\nonumber\\
X^{\mathcal C}{}_{\mathcal B}&=&\left(\sum_{j=5}^7\delta_{\alpha}^j\delta_{\beta}^j-3\delta_\alpha^8\delta_\beta^8\right)\,f_{\alpha\beta\,\cB}{}^\cC\\
\mathcal{V}&=&\exp(wW)\,\exp(zZ)\,\exp(-wW)\nonumber\\
&&\exp(v[S_1,S_2])\,\exp(S_1)\,\exp(-v[S_1,S_2])\,\exp(xX),\nonumber
\eea
we obtain the potential
{\small
\setlongtables
\begin{longtable}{lll}\label{potential_so43diag}
$-8g^{-2}V$&=&\hfill
\\
&&$21+3\,\cosh(8\,x)+12\,\cosh(z)\,\cosh(2\,x)+4\,\cosh(z)\,\cosh(6\,x)$\\
&&$+12\,\cos(w)\,\sinh(2\,x)\,\sinh(z)-4\,\cos(w)\,\sinh(6\,x)\,\sinh(z)$\\
&&$+12\,\cosh(s)\,\cosh(2\,x)+4\,\cosh(s)\,\cosh(6\,x)+4\,\cosh(s)\,\cosh(z)$\\
&&$+\frac{9}{2}\,\cosh(s)\,\cosh(z)\,\cosh(4\,x)-\frac{1}{2}\,\cosh(s)\,\cosh(z)\,\cosh(12\,x)$\\
&&$-\frac{3}{2}\,\cosh(s)\,\cos(w)\,\sinh(4\,x)\,\sinh(z)$\\
&&$+\frac{1}{2}\,\cosh(s)\,\cos(w)\,\sinh(12\,x)\,\sinh(z)-12\,\cos(v)\,\sinh(2\,x)\,\sinh(s)$\\
&&$+4\,\cos(v)\,\sinh(6\,x)\,\sinh(s)+\frac{3}{2}\,\cos(v)\,\cosh(z)\,\sinh(4\,x)\,\sinh(s)$\\
&&$-\frac{1}{2}\,\cos(v)\,\cosh(z)\,\sinh(12\,x)\,\sinh(s)$\\
&&$+4\,\cos(v)\,\cos(w)\,\sinh(z)\,\sinh(s)$\\
&&$-\frac{9}{2}\,\cos(v)\,\cos(w)\,\cosh(4\,x)\,\sinh(z)\,\sinh(s)$\\
&&$+\frac{1}{2}\,\cos(v)\,\cos(w)\,\cosh(12\,x)\,\sinh(z)\,\sinh(s)$
\end{longtable}
}
which reduces to the potential on the $SO(4)\times SO(4)$ singlet space
\beq
-8g^{-2}V=24+16\,\cosh(z)+16\,\cosh(s)+8\,\cosh(s)\,\cosh(z)
\eeq
by setting $x=0$.

\section{Vacua}

For many submanifolds for which these restricted potentials can be
analytically calculated, they turn out to be too complicated for a
fully analytic determination of their extremal structure (using
presently available technology); one way to proceed from here is to
make educated guesses and further restrict the analysis to directions
in the potential with special properties by letting the angular
parameters run through a set of special values. While this technique
was used in \cite{Fischbacher:2002hg} to identify analoga of all
stationary points of the $SO(8)\times SO(8)$ potential that correspond
to the known extrema of $SO(8)$ gauged $N=8$ $D=4$ supergravity with
at least remaining $SU(3)$ symmetry (except the vacuum with $G_2$
symmetry which was identified in \cite{Fischbacher:2002fx}, a more
promising approach seems to be to use these analytic
expressions\footnote{whose evaluation is much faster than a full
numerical exponentiation of $E_8$ generator matrices -- which
nevertheless is also available and useful to cross-check results} in a
numerical search for vacua\footnote{Since stationary points of this
potential usually are saddle points, one has to minimize the length of
the gradient.} whose results then are subjected to educated inspection
to give conjectures about exact locations of stationary points.
Typically, numerical search will produce values for angular
coordinates very close to rational multiples of $\pi$, or simple
relations between some of the hyperbolic angular coordinates (for
example, one being close to the negative value of another one). By
substituting these conjectured properties back into the analytic
potential, the problem typically is simplified far enough to produce a
complete set of coordinates which then are subjected to
$(\ref{SolutionSieve})$ to filter out 'fake' solutions this process
may have generated.\footnote{This procedure is documented in full
detail in one of the examples provided in the \LambdaTensor{}
package.}

For the more accessible vacua, we collect eigenvalues of the vector
and scalar mass matrices as well as the $A_1$ and $A_3$ fermion and
gravitino mass tensors in tables like the following corresponding to
the trivial vacuum at the origin $\cV^\cC{}_\cB=\delta^\cC_\cB$ with
$(n_L,n_R)=(8,8)$ supersymmetry:
\begin{eqnarray}
\begin{tabular}{|l|l|}\hline
$\Lambda/(2g^2)$&$-16$\\\hline
$\mathcal{M}/g^2$&$-12_{(\times 128)}$\\\hline
$\mathcal{M}^{\rm vec}/g$&$0_{(\times 128)}$\\\hline
$A_1$&$2_{(\times 8)},\;-2_{(\times 8)}$\\\hline
$A_3$&$0_{(\times 128)}$\\\hline
\end{tabular}
\end{eqnarray}
Multiplicities are given as subscripts in these tables; the Goldstone
modes are contained in the $m^2=0$ $\mathcal{M}$-eigenvalues and
goldstino modes (identified by projection with $A_2$) will be marked
with an asterisk.

\subsection{The $G_2\times G_2$ vacuum}

Finding vacua is a highly nontrivial task; checking the existence of a
vacuum (and its properties) once it has been identified is -- at least
in principle -- amenable to a manual calculation. We demonstrate this
by presenting explicit intermediate quantities for the particular case
of the vacuum with unbroken gauge symmetry $G_2\times G_2$ and
$(n_L,n_R)=(1,1)$ supersymmetry.

This vacuum is located at 
\beq
\begin{array}{lcl}
M^\cA&:=&\sqrt{2}\left(H^\cA_{\alpha\beta}\delta^\alpha_8\delta^\beta_8-H^\cA_{\dot\alpha\dot\beta}\delta^{\dot\alpha}_8\delta^{\dot\beta}_8\right)\\
m&:=&\frac{1}{2}\log\left(\frac{7}{3}+\frac{2}{3}\sqrt{10}\right)\\
\cV{}^\cC{}_\cB&=&\exp\left(mM^\cA f_{\cA\cB}{}^\cC\right)
\end{array}
\eeq

Even and odd powers of the generator
$N^\cC{}_\cB:=M^\cA f_{\cA\cB}{}^\cC$
(from which one can directly read off its exponential in terms of $\sin$ and $\cos$ contributions)
are given by
{\small
\bea
P^{(7v)}_{ij}&:=&\delta_{ij}-\delta_i^8\delta_j^8,\quad
P^{(7s)}_{\alpha\beta}:=\delta_{\alpha\beta}-\delta_\alpha^8\delta_\beta^8,\quad
P^{(7c)}_{\dot\alpha\dot\beta}:=\delta_{\dot\alpha\dot\beta}-\delta_{\dot\alpha}^8\delta_{\dot\beta}^8\nonumber\\
D^{(7v)}_{ij}&:=&\delta_{ij}-8\,\delta_i^8\delta_j^8,\quad
D^{(7s)}_{\alpha\beta}:=\delta_{\alpha\beta}-8\,\delta_\alpha^8\delta_\beta^8,\quad
D^{(7c)}_{\dot\alpha\dot\beta}:=\delta_{\alpha\beta}-8\,\delta_\alpha^8\delta_\beta^8\nonumber\\
E^{(7)}_{\alpha\dot\beta}&:=&\gamma^i_{\alpha\dot\beta}\delta^i_8-8\delta_\alpha^8\delta_{\dot\beta}^8\nonumber\\
A_{ijkl}&:=&\delta^{mn}_{kl} P^{(7v)}_{im} P^{(7v)}_{jn}\nonumber\\
B_{ijkl}&:=&\delta^{ij}_{kl}-A_{ijkl}\nonumber\\
Q^{(7c)}_{ij\dot\alpha\dot\beta}&:=&\delta_{\dot\beta}^8\delta_{\dot\gamma}^8 P^{(7c)}_{\dot\delta\dot\alpha}P^{(7v)}_{ik}P^{(7v)}_{jl}\gamma^{kl}_{\dot\delta\dot\gamma}\nonumber\\
Q^{(c)}_{ij\dot\alpha\dot\beta}&:=&\delta_{\dot\beta}^8\delta_{\dot\gamma}^8\gamma^{ij}_{\dot\alpha\dot\gamma}-Q^{(7c)}_{ij\dot\alpha\dot\beta}\nonumber\\
Q^{(7s)}_{ij\alpha\beta}&:=&\delta_{\alpha}^8\delta_{\gamma}^8 P^{(7s)}_{\delta\beta}P^{(7v)}_{ik}P^{(7v)}_{jl}\gamma^{kl}_{\beta\gamma}\nonumber\\
Q^{(s)}_{ij\alpha\beta}&:=&\delta_{\alpha}^8\delta_{\gamma}^8\gamma^{ij}_{\beta\gamma}-Q^{(7s)}_{ij\alpha\beta}\nonumber\\
R^{(7c)}_{ij\dot\alpha\dot\beta}&:=&
\delta_\alpha^8\delta_i^8\delta_{\dot\beta}^8\gamma^j_{\alpha\dot\alpha}
+\delta_\beta^8\delta_j^8\delta_{\dot\alpha}^8\gamma^i_{\beta\dot\beta}
-2\delta^i_8\delta^j_8\delta^{\dot\alpha}_8\delta^{\dot\beta}_8\nonumber\\
R^{(c)}_{ij\dot\alpha\dot\beta}&:=&\gamma^i_{\gamma\dot\beta}\gamma^j_{\alpha\dot\alpha}\delta_\gamma^8\delta_\alpha^8-R^{(7c)}_{ij\dot\alpha\dot\beta}\nonumber\\
R^{(7s)}_{ij\alpha\beta}&:=&
\delta_\beta^8\delta_i^8\delta_{\dot\alpha}^8\gamma^j_{\alpha\dot\alpha}
+\delta_\alpha^8\delta_j^8\delta_{\dot\beta}^8\gamma^i_{\beta\dot\beta}
-2\delta^i_8\delta^j_8\delta^{\alpha}_8\delta^{\beta}_8\nonumber\\
R^{(s)}_{ij\alpha\beta}&:=&\gamma^i_{\beta\dot\gamma}\gamma^j_{\alpha\dot\alpha}\delta_{\dot\gamma}^8\delta_{\dot\alpha}^8-R^{(7s)}_{ij\alpha\beta}\nonumber\\
\left(N^\cC{}_\cB\right)^{2n+1}\!/\sqrt2&=&
-\frac{1}{2}H^\cC_{IJ}H_\cB^{\dot\alpha\dot\beta}H^I_i H^J_j\left(Q^{(c)}_{ij\dot\beta\dot\alpha}+2^{2n}Q^{(7c)}_{ij\dot\beta\dot\alpha}\right)\nonumber\\
&&+\frac{1}{2}H^\cC_{IJ}H_\cB^{\alpha\beta}H^I_i H^J_j\left(Q^{(s)}_{ij\alpha\beta}+2^{2n}Q^{(7s)}_{ij\alpha\beta}\right)\nonumber\\
&&+\frac{1}{2}H^\cC_{IJ}H_\cB^{\alpha\beta}H^I_{\bar i} H^J_{\bar j}\left(Q^{(s)}_{\bar i\bar j\beta\alpha}+2^{2n}Q^{(7s)}_{\bar i\bar j\beta\alpha}\right)\nonumber\\
&&-\frac{1}{2}H^\cC_{IJ}H_\cB^{\dot\alpha\dot\beta}H^I_{\bar i} H^J_{\bar j}\left(Q^{(c)}_{\bar i\bar j\dot\alpha\dot\beta}+2^{2n}Q^{(7c)}_{\bar i\bar j\dot\alpha\dot\beta}\right)\nonumber\\
&&-\frac{1}{2}H^\cC_{\dot\alpha\dot\beta}H_\cB^{IJ}H^I_{\bar i} H^J_{\bar j}\left(Q^{(c)}_{ij\dot\beta\dot\alpha}2^{2n}Q^{(7c)}_{ij\dot\beta\dot\alpha}\right)\nonumber\\
&&+\frac{1}{2}H^\cC_{\alpha\beta}H_\cB^{IJ}H^I_{i} H^J_{j}\left(Q^{(s)}_{ij\alpha\beta}2^{2n}Q^{(7s)}_{ij\alpha\beta}\right)\nonumber\\
&&+\frac{1}{2}H^\cC_{\alpha\beta}H_\cB^{IJ}H^I_{\bar i} H^J_{\bar j}\left(Q^{(s)}_{\bar i\bar j\beta\alpha}+2^{2n}Q^{(7s)}_{\bar i\bar j\beta\alpha}\right)\nonumber\\
&&-\frac{1}{2}H^\cC_{\dot\alpha\dot\beta}H_\cB^{IJ}H^I_{\bar i} H^J_{\bar j}\left(Q^{(s)}_{\bar i\bar j\dot\alpha\dot\beta}+2^{2n}Q^{(7s)}_{\bar i\bar j\dot\alpha\dot\beta}\right)\nonumber\\
&&+H^\cC_{IJ}H_\cB^{\alpha\beta}H^I_{i} H^J_{\bar j}\left(R^{(s)}_{i\bar j\alpha\beta}+2^{2n}R^{(7s)}_{i\bar j\alpha\beta}\right)\nonumber\\
&&+H^\cC_{IJ}H_\cB^{\dot\alpha\dot\beta}H^I_{i} H^J_{\bar j}\left(R^{(c)}_{i\bar j\dot\alpha\dot\beta}+2^{2n}R^{(7c)}_{i\bar j\dot\alpha\dot\beta}\right)\nonumber\\
&&+H^\cC_{\alpha\beta}H_\cB^{IJ}H^I_{i} H^J_{\bar j}\left(R^{(s)}_{i\bar j\alpha\beta}+2^{2n}R^{(7s)}_{i\bar j\alpha\beta}\right)\nonumber\\
&&+H^\cC_{\dot\alpha\dot\beta}H_\cB^{IJ}H^I_{i} H^J_{\bar j}\left(R^{(c)}_{i\bar j\dot\alpha\dot\beta}+2^{2n}R^{(7c)}_{i\bar j\dot\alpha\dot\beta}\right)\nonumber\\
\left(N^\cC{}_\cB\right)^{2n}&=&
\phantom+H^\cC_{\dot\alpha\dot\beta}H_\cB^{\dot\gamma\dot\delta}
\left(\frac{1}{2}\delta_{\dot\alpha\dot\gamma}\delta_{\dot\beta\dot\delta}
+2^{2n-1}\delta_{\dot\alpha\dot\gamma}\delta_{\dot\beta}^8\delta_{\dot\delta}^8\right.\nonumber\\&&\left.
+2^{2n-1}\delta_{\dot\beta\dot\delta}\delta_{\dot\alpha}^8\delta_{\dot\gamma}^8
-2^{2n}\delta_{\dot\alpha}^8\delta_{\dot\beta}^8\delta_{\dot\gamma}^8\delta_{\dot\delta}^8
\right)\nonumber\\
&&+H^\cC_{\alpha\beta}H_\cB^{\gamma\delta}
\left(\frac{1}{2}\delta_{\alpha\gamma}\delta_{\beta\delta}
+2^{2n-1}\delta_{\alpha\gamma}\delta_{\beta}^8\delta_{\delta}^8\right.\nonumber\\&&\left.
+2^{2n-1}\delta_{\beta\delta}\delta_{\alpha}^8\delta_{\gamma}^8
-2^{2n}\delta_{\alpha}^8\delta_{\beta}^8\delta_{\gamma}^8\delta_{\delta}^8
\right)\nonumber\\
&&+H^\cC_{\alpha\beta}H_\cB^{\dot\gamma\dot\delta}
\left(\frac{1}{2}\gamma^i_{\alpha\dot\gamma}\gamma^j_{\beta\dot\delta}\delta_i^8\delta_j^8
-2^{2n-1}\gamma^i_{\alpha\dot\gamma}\delta_i^8\delta_{\beta}^8\delta_{\dot\delta}^8\right.\nonumber\\&&\left.
-2^{2n-1}\gamma^i_{\beta\dot\delta}\delta_i^8\delta_{\alpha}^8\delta_{\dot\gamma}^8
+2^{2n}\delta_{\alpha}^8\delta_{\beta}^8\delta_{\dot\gamma}^8\delta_{\dot\delta}^8
\right)\nonumber\\
&&+H^\cC_{\dot\alpha\dot\beta}H_\cB^{\gamma\delta}
\left(\frac{1}{2}\gamma^i_{\gamma\dot\alpha}\gamma^j_{\delta\dot\beta}\delta_i^8\delta_j^8
-2^{2n}\gamma^i_{\gamma\dot\alpha}\delta_i^8\delta_{\dot\beta}^8\delta_{\delta}^8\right.\nonumber\\&&\left.
-2^{2n}\gamma^i_{\beta\delta}\delta_i^8\delta_{\dot\alpha}^8\delta_{\gamma}^8
+2^{2n+1}\delta_{\dot\alpha}^8\delta_{\dot\beta}^8\delta_{\dot\gamma}^8\delta_{\dot\delta}^8
\right)\nonumber\\
&&+2^{2n-6} H^\cC_{IJ}H_\cB^{KL} H^I_{\bar i} H^J_j H^K_k H^L_l \delta_{\bar i}^8 E^{(7)}_{\alpha\dot\gamma}\gamma^{jkl}_{\alpha\dot\beta}\nonumber\\
&&+2^{2n-6} H^\cC_{IJ}H_\cB^{KL} H^I_i H^J_j H^K_k H^L_{\bar l} \delta_{\bar l}^8 E^{(7)}_{\alpha\dot\gamma}\gamma^{ijk}_{\alpha\dot\beta}\nonumber\\
&&+2^{2n-6} H^\cC_{IJ}H_\cB^{KL} H^I_i H^J_{\bar j} H^K_k H^L_l \delta_{\bar j}^8 E^{(7)}_{\alpha\dot\gamma}\gamma^{ikl}_{\alpha\dot\beta}\nonumber\\
&&+2^{2n-6} H^\cC_{IJ}H_\cB^{KL} H^I_i H^J_j H^K_{\bar k} H^L_l \delta_{\bar k}^8 E^{(7)}_{\alpha\dot\gamma}\gamma^{ijl}_{\alpha\dot\beta}\nonumber\\
&&H^\cC_{IJ}H_\cB^{KL} H^I_i H^J_j H^K_k H^L_l
\left(2^{2n-3}A_{ijkl}-\frac{1}{16}\left(1+7\cdot2^{2n-2}\right)B_{ijkl}\right.\nonumber\\&&\left.
+2^{2n-5}\gamma^{ijkl}_{\alpha\beta}P^{(7s)}_{\alpha\beta}
+2^{2n-5}\gamma^{ijkl}_{\dot\alpha\dot\beta}P^{(7c)}_{\dot\alpha\dot\beta}
\right)\nonumber\\
&&H^\cC_{IJ}H_\cB^{KL} H^I_{\bar i} H^J_{\bar j} H^K_{\bar k} H^L_{\bar l}
\left(2^{2n-3}A_{\bar i\bar j\bar k\bar l}-\frac{1}{16}\left(1+7\cdot2^{2n-2}\right)B_{\bar i\bar j\bar k\bar l}\right.\nonumber\\&&\left.
+2^{2n-5}\gamma^{\bar i\bar j\bar k\bar l}_{\alpha\beta}P^{(7s)}_{\alpha\beta}
+2^{2n-5}\gamma^{\bar i\bar j\bar k\bar l}_{\dot\alpha\dot\beta}P^{(7c)}_{\dot\alpha\dot\beta}
\right)\nonumber\\
&&H^\cC_{IJ}H_\cB^{KL} H^I_{i} H^J_{\bar j} H^K_{k} H^L_{\bar l}
\left(\delta_{ik}\delta_{\bar j\bar l}
+\left(2^{2n-2}-1\right)\delta_{ik}\delta_{\bar j}^8\delta_{\bar l}^8\right.\nonumber\\&&\left.
+\left(2^{2n-2}-1\right)\delta_{\bar j\bar l}\delta_{i}^8\delta_{k}^8
-\left(2^{2n-1}-2\right)\delta_{i}^8\delta_{\bar j}^8\delta_{k}^8\delta_{\bar l}^8
\right)
\eea
}

The $T$-tensor for this vacuum is then given by
\bea
K^{(c)}_{\dot\alpha\dot\beta kl}&:=&
\frac{2}{3}\sqrt{5}\left(2\,\gamma^{kl}_{\dot\alpha\dot\gamma}\delta_{\dot\gamma}^8\delta_{\dot\beta}^8
-\gamma^{il}_{\dot\alpha\dot\gamma}\delta_k^8\delta_{\dot\gamma}^8\delta_i^8\delta_{\dot\beta}^8
-\gamma^{ki}_{\dot\alpha\dot\gamma}\delta_l^8\delta_{\dot\gamma}^8\delta_i^8\delta_{\dot\beta}^8\right)\nonumber\\
K^{(s)}_{\alpha\beta kl}&:=&
\frac{2}{3}\sqrt{5}\left(2\,\gamma^{kl}_{\alpha\gamma}\delta_\gamma^8\delta_{\beta}^8
-\gamma^{il}_{\alpha\gamma}\delta_k^8\delta_{\gamma}^8\delta_i^8\delta_{\beta}^8
-\gamma^{ki}_{\alpha\gamma}\delta_l^8\delta_{\gamma}^8\delta_i^8\delta_{\beta}^8\right)\nonumber\\
K^{(v)}_{ijkl}&:=&-\frac{47}{6}\delta^{ij}_{kl}-\frac{1}{6}\delta^{mn}_{kl}D^{(7v)}_{im}D^{(7v)}_{jn}
-\frac{1}{8}D^{(7s)}_{\alpha\beta}\gamma^{ijkl}_{\alpha\beta}
-\frac{1}{8}D^{(7c)}_{\dot\alpha\dot\beta}\gamma^{ijkl}_{\dot\alpha\dot\beta}\nonumber\\
T_{\cA\cB}&=&
\phantom+\frac{1}{2}\left(H_\cA^{\dot\alpha\dot\beta}H_\cB^{IJ}+H_\cB^{\dot\alpha\dot\beta}H_\cA^{IJ}\right) H_I^{\bar i}H_J^{\bar j}K^{(c)}_{\dot\alpha\dot\beta\bar i\bar j}\nonumber\\
&&+\frac{1}{2}\left(H_\cA^{\alpha\beta}H_\cB^{IJ}+H_\cB^{\alpha\beta}H_\cA^{IJ}\right) H_I^{\bar i}H_J^{\bar j}K^{(s)}_{\beta\alpha\bar i\bar j}\nonumber\\
&&-\frac{2}{3}\sqrt{5}\left(H_\cA^{\alpha\beta}H_\cB^{IJ}+H_\cB^{\alpha\beta}H_\cA^{IJ}\right) H_I^{i}H_J^{\bar j}\left(\delta_\beta^8\delta_i^8\delta_{\dot \gamma}^8\gamma^{\bar j}_{\alpha\dot\gamma}-\delta_\alpha^8\delta_{\bar j}^8\delta_{\dot\gamma}^8\gamma^i_{\beta\dot\gamma}\right)\nonumber\\
&&-\frac{2}{3}\sqrt{5}\left(H_\cA^{\dot\alpha\dot\beta}H_\cB^{IJ}+H_\cB^{\dot\alpha\dot\beta}H_\cA^{IJ}\right) H_I^{i}H_J^{\bar j}\left(\delta_{\dot\beta}^8\delta_i^8\delta_{\gamma}^8\gamma^{\bar j}_{\gamma\dot\alpha}-\delta_{\dot\alpha}^8\delta_{\bar j}^8\delta_{\gamma}^8\gamma^i_{\gamma\dot\beta}\right)\nonumber\\
&&+\frac{8}{3}\sqrt{5}\left(H_\cA^{\alpha\beta}H_\cB^{\dot\gamma\dot\delta}+H_\cB^{\alpha\beta}H_\cA^{\dot\gamma\dot\delta}\right)
\left(\delta_i^8\delta_\beta^8\delta_{\dot\delta}^8\gamma^i_{\alpha\dot\gamma}
-\delta_i^8\delta_\alpha^8\delta_{\dot\gamma}^8\gamma^i_{\beta\dot\delta}
\right)\nonumber\\
&&-\frac{1}{12}\left(H_\cA^{IJ}H_\cB^{KL}+H_\cB^{IJ}H_\cA^{KL}\right)H_I^i H_J^{\bar j}H_K^k H_L^l\delta_{\bar j}^8\delta_m^8\gamma^{mikl}_{\alpha\beta}D^{(7s)}_{\alpha\beta}\nonumber\\
&&+\frac{1}{12}\left(H_\cA^{IJ}H_\cB^{KL}+H_\cB^{IJ}H_\cA^{KL}\right)H_I^{\bar i} H_J^{\bar j}H_K^k H_L^{\bar l}\delta_{k}^8\delta_m^8\gamma^{m\bar i\bar k\bar l}_{\alpha\beta}D^{(7s)}_{\alpha\beta}\nonumber\\
&&-\frac{1}{2}\left(H_\cA^{\alpha\beta}H_\cB^{IJ}+H_\cB^{\alpha\beta}H_\cA^{IJ}\right)H_I^{\bar i} H_J^{\bar j}K^{(s)}_{\alpha\beta\bar i\bar j}\\
&&-4H_\cA^{\alpha\beta}H_\cB^{\gamma\delta}\left(\delta_\beta^8\delta_\delta^8\delta_{\alpha\gamma}-\delta_\alpha^8\delta_\gamma^8\delta_{\beta\delta}\right)\nonumber\\
&&-4H_\cA^{\dot\alpha\dot\beta}H_\cB^{\dot\gamma\dot\delta}\left(\delta_{\dot\beta}^8\delta_{\dot\delta}^8\delta_{\dot\alpha\dot\gamma}-\delta_{\dot\alpha}^8\delta_{\dot\gamma}^8\delta_{\dot\beta\dot\delta}\right)\nonumber\\
&&-\frac{1}{2}\left(H_\cA^{\dot\alpha\dot\beta}H_\cB^{IJ}+H_\cB^{\dot\alpha\dot\beta}H_\cA^{IJ}\right) H_I^{i}H_J^{j}K^{(c)}_{\dot\beta\dot\alpha ij}\nonumber\\
&&+\frac{1}{4}H_\cA^{IJ}H_\cB^{KL}H_I^{\bar i}H_J^{\bar j}H_K^{\bar k}H_L^{\bar l}K^{(v)}_{\bar i\bar j\bar k\bar l}\nonumber\\
&&-\frac{1}{4}H_\cA^{IJ}H_\cB^{KL}H_I^{i}H_J^{j}H_K^{k}H_L^{l}K^{(v)}_{ijkl}\nonumber\\
&&+\frac{2}{3}H_\cA^{IJ}H_\cB^{KL}H_I^{i}H_J^{\bar j}H_K^{k}H_L^{\bar l}
\left(\delta_{\bar j}^8\delta_{\bar l}^8P^{(7)}_{ik}
-\delta_{i}^8\delta_{k}^8P^{(7)}_{\bar j\bar l}
\right)\nonumber
\eea
and the $A_{1,2,3}$ tensors are
\bea
A_{1}^{IJ}&=&H^I_i H^J_j\left(4\,P^{(7v)}_{ij}+\frac{8}{3}\delta_i^8\delta_j^8\right)-H^I_{\bar i} H^J_{\bar j}\left(4\,P^{(7v)}_{\bar i\bar j}+\frac{8}{3}\delta_{\bar i}^8\delta_{\bar j}^8\right)\nonumber\\
A_{2}^{I\dot A}&=&\frac{4}{3}\sqrt{5}\left(H^I_i H^{\dot A}_{\alpha\dot\beta}\delta_\alpha^8\delta_{\gamma}^8P^{(7v)}_{ij}\gamma^j_{\gamma\dot\beta}
-H^I_{\bar i} H^{\dot A}_{\alpha\dot\beta}\delta_{\dot\beta}^8\delta_{\dot\gamma}^8P^{(7v)}_{\bar i\bar j}\gamma^{\bar j}_{\alpha\dot\gamma}\right.\nonumber\\
&&\left.
-H^I_i H^{\dot A}_{\dot\alpha\beta}\delta_{\dot\alpha}^8\delta_{\dot\gamma}^8P^{(7v)}_{ij}\gamma^j_{\beta\dot\gamma}
-H^I_{\bar i} H^{\dot A}_{\dot\alpha\beta}\delta_{\beta}^8\delta_{\gamma}^8P^{(7v)}_{\bar i\bar j}\gamma^{\bar j}_{\gamma\dot\alpha}
\right)\\
A_{3}^{\dot A\dot B}&=&-8\,H^{\dot A}_{\alpha\dot\beta}H^{\dot B}_{\gamma\dot\delta}\left(P^{(7s)}_{\alpha\gamma}\delta_{\dot\beta}^8\delta_{\dot\delta}^8-P^{(7c)}_{\dot\beta\dot\delta}\delta_\alpha^8\delta_\gamma^8\right)\nonumber\\
&&
-8\,H^{\dot A}_{\dot\alpha\beta}H^{\dot B}_{\dot\gamma\delta}\left(P^{(7c)}_{\dot\alpha\dot\gamma}\delta_{\beta}^8\delta_{\delta}^8-P^{(7s)}_{\beta\delta}\delta_{\dot\alpha}^8\delta_{\dot\gamma}^8\right)\nonumber\\
&&
-\frac{4}{3}\left(H^{\dot A}_{\dot\alpha\beta}H^{\dot B}_{\gamma\dot\delta}+H^B_{\dot\alpha\beta}H^{\dot A}_{\gamma\dot\delta}\right)\delta_i^8\times\nonumber\\
&&\times\left(\delta_j^8\gamma^i_{\gamma\dot\alpha}\gamma^j_{\beta\dot\delta}
-4\,\gamma^i_{\gamma\dot\alpha}\delta_{\beta}^8\delta_{\dot\delta}^8
-4\,\gamma^i_{\beta\dot\delta}\delta_{\dot\alpha}^8\delta_{\gamma}^8
\right)\nonumber
\eea
And one sees that indeed, the stationarity condition
$3\,A_1^{IJ}A_2^{J\dot A}=A_2^{I\dot B}A_3^{\dot B\dot A}$ is
satisfied.

Thus, we obtain:

\begin{tabular}{|l|l|}\hline
$\Lambda/(2g^2)$&$-256/9$\\\hline
$\mathcal{M}/g^2$&$1040/9_{(\times 1)},\;16_{(\times 1)},\;0_{(\times 28)},\;-112/9_{(\times 49)},\;-80/3_{(\times 49)}$\\\hline
$\mathcal{M}^{\rm vec}/g$&$20/3_{(\times 7)},\;4/3_{(\times 7)},\;0_{(\times 100)},\;-4/3_{(\times 7)},\;-20/3_{(\times 7)}$\\\hline
$A_1$&$4_{(\times 7)},\;8/3_{(\times 1)},\;-8/3_{(\times 1)},\;-4_{(\times 7)}$\\\hline
$A_3$&$12_{(\times 7)^*},\;28/3_{(\times 1)},\;4_{(\times 7)},\;4/3_{(\times 49)},$\\&$-4/3_{(\times 49)},\;-4_{(\times 7)},\;-28/3_{(\times 1)},\;-12_{(\times 7)^*}$\\\hline
\end{tabular}

\subsection{The other vacua}

For the other vacua, we do not present the full calculation but just
list their locations and properties.

The vacuum with largest unbroken symmetry has a remaining symmetry of
$SO(7)^\pm\times SO(7)^\pm$ (both are equivalent here), and is located at
\beq
\cV^\cC{}_\cB=\exp\left((\Arcosh 2)\,\delta^{\alpha}_8\delta^\beta_8 H^\cA_{\alpha\beta} f_{\cA\cB}{}^\cC\right)
\eeq
-- all supersymmetry is broken here, but it is nevertheless stable:
\begin{eqnarray}\label{E-p8-so7}
\begin{tabular}{|l|l|}\hline
$\Lambda/(2g^2)$&$-25$\\\hline
$\mathcal{M}/g^2$&$96_{(\times 1)},\;0_{(\times 14)},\;-9_{(\times
64)},\;-24_{(\times 49)}$\\\hline
$\mathcal{M}^{\rm vec}/g$&$6_{(\times 7)},\;0_{(\times 114)},\;-6_{(\times
7)}$\\\hline
$A_1$&$7/2_{(\times 8)},\;-7/2_{(\times 8)}$\\\hline
$A_3$&$21/2_{(\times 8)^*},\;3/2_{(\times 56)},\;-3/2_{(\times
56)},\;-21/2_{(\times 8)^*}$\\\hline
\end{tabular}
\end{eqnarray}

There is a further vacuum with $(n_,n_R)=(2,2)$ supersymmetry and a remaining symmetry of $SU(3)\times U(1)\times SU(3)\times U(1)$ at
\beq
\cV^\cC{}_\cB=\exp\left(\frac{1}{2}(\Arcosh 3)\,f_{\cA\cB}{}^\cC
\left(H^\cA_{\alpha\beta}\left(\delta^\alpha_3\delta^\beta_3+\delta^\alpha_5\delta^\beta_5\right)
     +H^\cA_{\dot\alpha\dot\beta}\left(\delta^{\dot\alpha}_4\delta^{\dot\beta}_6+\delta^{\dot\alpha}_6\delta^{\dot\beta}_4\right)
\right)\right)
\eeq
with mass spectrum
\begin{eqnarray}\label{E-p8-su3}
\begin{tabular}{|l|l|}\hline
$\Lambda/(2g^2)$&$-36$\\\hline
$\mathcal{M}/g^2$&$160_{(\times 1)},\;28_{(\times 4)},\;0_{(\times 38)},\;-20_{(\times 36)},\;-32_{(\times 49)}$\\\hline
$\mathcal{M}^{\rm vec}/g$&$8_{(\times 7)},\;2_{(\times 12)},\;0_{(\times 90)},\;-2_{(\times 12)},\;-8_{(\times 7)}$\\\hline
$A_1$&$5_{(\times 6)},\;3_{(\times 2)},\;-3_{(\times 2)},\;-5_{(\times 6)}$\\\hline
$A_3$&$15_{(\times 6)^*},\,11_{(\times2)},\,5_{(\times14)},\,1_{(\times42)},$\\
&$-1_{(\times42)},\,-5_{(\times14)},\,-11_{(\times2)},\,-15_{(\times 6)^*}$\\\hline
\end{tabular}
\end{eqnarray}

Another AdS vacuum without any supersymmetry that breaks $SO(8)\times SO(8)$ down to $SU(4)$
and has $\Lambda/(2g^2)=-52$ is located at
\bea
a&:=&\frac{1}{4}\log\left(\frac{5}{2}+\frac{1}{2}\sqrt{21}\right)\nonumber\\
b&:=&\Arcosh 2\nonumber\\
\cV^\cC{}_\cB&=&\exp\left(f_{\cA\cB}{}^\cC
\left(aH^\cA_{\alpha\beta}\delta^{\alpha\beta}
     +bH^\cA_{\dot\alpha\dot\beta}\left(\delta^{\dot\alpha}_3\delta^{\dot\beta}_5+\delta^{\dot\alpha}_5\delta^{\dot\beta}_3\right)
\right)\right)
\eea

The mass spectrum for this vacuum is slightly more difficult to
calculate than for the preceding cases and has not been determined
yet.

A further AdS vacuum that breaks all supersymmetry and has a remaining
symmetry of $SU(3)\times U(1)\times U(1)$ is given by
\bea
A_{\alpha\beta}&:=&\delta_{\alpha\beta}-\delta_\alpha^3\delta_\beta^3-\delta_\alpha^5\delta_\beta^5\nonumber\\
k&:=&\sqrt{78+14\sqrt{33}}\nonumber\\
m&=&\sqrt{6+6\sqrt{33}}\nonumber\\
p&:=&\log\left(\frac{1}{6}\sqrt{18+6\sqrt{33}+6\,m}\right)\approx0.4616649\nonumber\\
q&:=&-\frac{1}{2}\log\left(\frac{7}{2}+\frac{1}{2}\sqrt{33}-\frac{1}{2}k\right)\approx1.2694452\nonumber\\
\cV^\cC{}_\cB&=&\exp\left(f_{\cA\cB}{}^\cC\left(pH^\cA_{\alpha\beta}A^{\alpha\beta}+qH^\cA_{\dot\alpha\dot\beta}\left(\delta^{\dot\alpha}_3\delta^{\dot\beta}_5+\delta^{\dot\alpha}_5\delta^{\dot\beta}_3\right)\right)\right)\\
\Lambda/(2g^2)&=&-3564\left(1453+253\sqrt{33}+116k+20\sqrt{33}\,k\right)\times\nonumber\\
&&\left(3+\sqrt{33}+m\right)^{-2}\left(7+\sqrt{33}+k\right)^{-2}\times\nonumber\\
&&\left(24+6\sqrt{33}-3m-\sqrt{33}m\right)^{-1}\approx-49.82132\nonumber
\eea

Again, the mass spectrum has not been determined analytically yet.

Furthermore, there is strong numerical evidence for a further vacuum with remaining symmetry of only $SU(3)$ at
\bea
p&\approx&0.43045295,\quad q\approx0.03708009,\quad r\approx1.16386200\nonumber\\
A^{(s)}_{\alpha\beta}&=&
\sum_{n\in \{3,5\}} -q\,\delta_\alpha^n\delta_\beta^n
+\sum_{n\in \{1,2,4,6,7,8\}} -p\,\delta_\alpha^n\delta_\beta^n
-r\,\delta_\alpha^3\delta_\beta^5-r\,\delta_\alpha^5\delta_\beta^3\nonumber\\
A^{(c)}_{\alpha\beta}&=&
\sum_{n\in \{3,5\}} q\,\delta_\alpha^n\delta_\beta^n
+\sum_{n\in \{1,2,4,6,7,8\}} p\,\delta_\alpha^n\delta_\beta^n
+r\,\delta_\alpha^3\delta_\beta^5+r\,\delta_\alpha^5\delta_\beta^3\nonumber\\
\cV^\cC{}_\cB&=&\exp\left(f_{\cA\cB}{}^\cC
\left(pH^\cA_{\alpha\beta}A^{(s)}_{\alpha\beta}+qH^\cA_{\dot\alpha\dot\beta}A^{(c)}_{\dot\alpha\dot\beta}\right)\right)\\
\Lambda/(2g^2)&\approx&105.527621\nonumber
\eea

So far, no analytic expression has been found for this vacuum
candidate. The corresponding parameters for $(\ref{phi-so8xso8-map})$
are $\phi=\pi/4$, $a=15\pi/4$, $w=\pi/4$,$\lambda_1\approx-0.5563132$,
$\lambda_2\approx-1.6459494$, $z\approx-1.8786964$.

\chapter[Noncompact gauge groups]{Noncompact gauge groups}
\label{ch3dNoncompact}

\section{On noncompact gauge groups in supergravity}

Standard quantum field theory textbook lore \cite{Weinberg:kr} tells
that gauge groups have to be compact to avoid the appearance of
negative norm states \cite{Glashow:ep}. Nevertheless, there are some
cases of theories which contain scalar fields where a positive
definite metric can be constructed from these scalars that allows an
implementation of noncompact gauge invariance. (See
\cite{deWit:1983xe} and, in particular, \cite{Hull:vg} for an
explanation of the construction of $SO(7,1)$ gauged $N=8$
supergravity.) 

Generally, a necessary condition for a gauge group candidate is that a
maximal subset of the vectors transform in the adjoint representation.
When it is not possible to absorb the extra charged vectors that
transform nontrivially under the gauge group (for example, in $D=5$
the vectors can be dualized into massive self-dual
2-forms \cite{Pernici:ju, Gunaydin:1984qu}), as is the case in four
dimensions, then the whole set of vectors must be used. This gives
quite severe restrictions on possible gaugings in higher dimensions;
in particular, for $D=4$ $N=8$ one basically has just
the $SO(p,8-p)$ noncompact versions of $SO(8)$ and their
contractions \cite{deWit:2002vt}, while for $N=16$, $D=3$ a much bigger
 freedom exists due to the duality between vectors and scalars.
Of the maximal semisimple subgroups
of $\Eee$, the analogous noncompact versions of $SO(8)\times SO(8)$
of the form $SO(p,8-p)\times SO(p,8-p)$, which are contained
in $\Eee$ via their embedding in $SO(8,8)$,
as well as the exceptional cases $\Eee$,
$E_{7(+7)}\times SL(2)$,
$E_{7(-5)}\times SU(2)$,
$E_{6(+6)}\times SL(3)$,
$E_{6(+2)}\times SU(2,1)$,
$E_{6(-14)}\times SU(3)$,
$G_2\times F_{4(-20)}$,
and $G_{2(+2)}\times F_{4(+4)}$
are possible.

\section{$SO(p,8-p)^2$}

Going back to the $E_8\rightarrow SO(8)_L\times SO(8)_R$ decomposition
${\bf 248}=({\bf 1},{\bf 28})+({\bf 28},{\bf 1})+({\bf 8}_v, {\bf
  8}_v)+({\bf 8}_s, {\bf 8}_s)+({\bf 8}_c, {\bf 8}_c)$, we obtain
$SO(8,8)$ from $SO(16)$ by using a triality rotation to exchange the
role of vectors and spinors.  Then, the compact subgroup of the
first factor of $SO(p,8-p)_1\times SO(p,8-p)_2$ is obtained by
splitting each $({\bf 8}_s)\rightarrow({\bf p},{\bf 8-p})$ and
combining the $SO(p)_L$ with the $SO(8-p)_R$ subgroup.  (And vice
versa for the second factor.) As before, the $P^{27000}\Theta=0$
projection condition requires a ratio of gauge coupling constants of
$-1$. Hence, the embedding tensor of $SO(p,8-p)^2$ is given by
\bea
P_{\alpha\beta}&:=&\sum_{j=1}^p \delta_\alpha^p\delta_\beta^p,\quad
Q_{\alpha\beta}:=\delta_{\alpha\beta}-P_{\alpha\beta}\nonumber\\
H_{\cA}^{ij}&:=&H_\cA^{\underline{[IJ]}} H_{\underline{[IJ]}}^{IJ} H_I^i H_J^j,\quad
H_{\cA}^{\bar i\bar j}:=H_\cA^{\underline{[IJ]}} H_{\underline{[IJ]}}^{IJ} H_I^{\bar i} H_J^{\bar j}\nonumber\\
\Theta^{SO(p,8-p)^2}_{\cA\cB}&=&\phantom+\frac{1}{16}H_\cA^{ij}H_\cB^{kl}\gamma^{ij}_{\alpha\beta}\gamma^{kl}_{\gamma\delta}\left(P^{\alpha\gamma}P^{\beta\delta}-Q^{\alpha\gamma}Q^{\beta\delta}\right)\nonumber\\
&&+\frac{1}{16}H_\cA^{\bar i\bar j}H_\cB^{\bar k\bar l}\gamma^{\bar i\bar j}_{\alpha\beta}\gamma^{\bar k\bar l}_{\gamma\delta}\left(Q^{\alpha\gamma}Q^{\beta\delta}-P^{\alpha\gamma}P^{\beta\delta}\right)\nonumber\\
&&+2H_\cA^{\alpha\beta}H_\cB^{\gamma\delta}P_{\alpha\gamma}Q_{\beta\delta}-2H_\cA^{\alpha\beta}H_\cB^{\gamma\delta}Q_{\alpha\gamma}P_{\beta\delta}.
\eea

The subgroups presented in the previous chapter have been chosen in
such a way to maximize the amount of information that can be obtained
on various gaugings of the theory by simply redoing the calculations
using the invariant scalar submanifold parametrizations given there
with other (compatible) embedding tensors.\footnote{Note that the
double ratio\hfill\break
$\frac{\rm Effort(Submanifold\; M,\; hand\; calculation\; Nr.\; 2)}
{\rm Effort(Submanifold\; M,\; hand\; calculation\; Nr.\; 1)}/
\frac{\rm Effort(Submanifold\; M,\; machine\; calculation\; Nr.\; 2)}
{\rm Effort(Submanifold\; M,\; machine\; calculation\; Nr.\; 1)}$
hardly could be larger, since all we have to do
in the machine approach is to change the value $p$
and go for a cup of coffee!} It may happen that in some cases,
some of the singlets under the subgroup in question
of the compact subgroup of the gauge group are part
of the noncompact gauge group, and hence correspond
to trivial flat directions of the potential,
but this does not spoil the analysis.

Using the $G_{2\,{\rm diag}}$ singlet parametrization
$(\ref{g2singlets})$ for $SO(7,1)^2$ gives
{\small
\beq
\begin{array}{lcl}
-8g^{-2}V&=&\frac{909}{32}-\frac{7}{8}\,\cosh(2\,s)-\frac{49}{32}\,\cosh(4\,s)+\frac{6461}{512}\,\cosh(s)\,\cosh(z)\\
&&-\frac{1001}{512}\,\cosh(3\,s)\,\cosh(z)-\frac{203}{512}\,\cosh(5\,s)\,\cosh(z)-\frac{137}{512}\,\cosh(7\,s)\,\cosh(z)\\
&&+\frac{21}{8}\,\cos(2\,v)+\frac{63}{32}\,\cos(4\,v)+\frac{7}{2}\,\cos(2\,v)\,\cosh(2\,s)\\&&
-\frac{49}{8}\,\cos(2\,v)\,\cosh(4\,s)-\frac{21}{8}\,\cos(4\,v)\,\cosh(2\,s)+\frac{21}{32}\,\cos(4\,v)\,\cosh(4\,s)\\
&&+\frac{5145}{1024}\,\cos(2\,v)\,\cosh(s)\,\cosh(z)-\frac{5229}{1024}\,\cos(2\,v)\,\cosh(3\,s)\,\cosh(z)\\
&&+\frac{217}{1024}\,\cos(2\,v)\,\cosh(5\,s)\,\cosh(z)-\frac{133}{1024}\,\cos(2\,v)\,\cosh(7\,s)\,\cosh(z)\\
&&-\frac{21}{512}\,\cos(4\,v)\,\cosh(s)\,\cosh(z)-\frac{63}{512}\,\cos(4\,v)\,\cosh(3\,s)\,\cosh(z)\\
&&+\frac{147}{512}\,\cos(4\,v)\,\cosh(5\,s)\,\cosh(z)-\frac{63}{512}\,\cos(4\,v)\,\cosh(7\,s)\,\cosh(z)\\
&&-\frac{105}{1024}\,\cos(6\,v)\,\cosh(s)\,\cosh(z)+\frac{189}{1024}\,\cos(6\,v)\,\cosh(3\,s)\,\cosh(z)\\
&&-\frac{105}{1024}\,\cos(6\,v)\,\cosh(5\,s)\,\cosh(z)+\frac{21}{1024}\,\cos(6\,v)\,\cosh(7\,s)\,\cosh(z)\\
&&-\frac{28987}{2048}\,\cos(v-w)\,\sinh(z)\,\sinh(s)-\frac{651}{512}\,\cos(v+w)\,\sinh(z)\,\sinh(s)\\
&&-\frac{2835}{2048}\,\cos(v-w)\,\sinh(z)\,\sinh(3\,s)-\frac{3255}{512}\,\cos(v+w)\,\sinh(z)\,\sinh(3\,s)\\
&&-\frac{623}{2048}\,\cos(v-w)\,\sinh(z)\,\sinh(5\,s)+\frac{21}{512}\,\cos(v+w)\,\sinh(z)\,\sinh(5\,s)\\
&&-\frac{343}{2048}\,\cos(v-w)\,\sinh(z)\,\sinh(7\,s)-\frac{63}{512}\,\cos(v+w)\,\sinh(z)\,\sinh(7\,s)\\
&&+\frac{1323}{1024}\,\cos(3\,v-w)\,\sinh(z)\,\sinh(s)-\frac{7875}{2048}\,\cos(3\,v+w)\,\sinh(z)\,\sinh(s)\\
&&-\frac{385}{1024}\,\cos(3\,v-w)\,\sinh(z)\,\sinh(3\,s)+\frac{2541}{2048}\,\cos(3\,v+w)\,\sinh(z)\,\sinh(3\,s)\\
&&+\frac{35}{1024}\,\cos(3\,v-w)\,\sinh(z)\,\sinh(5\,s)+\frac{609}{2048}\,\cos(3\,v+w)\,\sinh(z)\,\sinh(5\,s)\\
&&-\frac{49}{1024}\,\cos(3\,v-w)\,\sinh(z)\,\sinh(7\,s)-\frac{399}{2048}\,\cos(3\,v+w)\,\sinh(z)\,\sinh(7\,s)\\
&&+\frac{35}{2048}\,\cos(5\,v-w)\,\sinh(z)\,\sinh(s)+\frac{315}{1024}\,\cos(5\,v+w)\,\sinh(z)\,\sinh(s)\\
&&-\frac{189}{2048}\,\cos(5\,v-w)\,\sinh(z)\,\sinh(3\,s)+\frac{63}{1024}\,\cos(5\,v+w)\,\sinh(z)\,\sinh(3\,s)\\
&&+\frac{175}{2048}\,\cos(5\,v-w)\,\sinh(z)\,\sinh(5\,s)-\frac{189}{1024}\,\cos(5\,v+w)\,\sinh(z)\,\sinh(5\,s)\\
&&-\frac{49}{2048}\,\cos(5\,v-w)\,\sinh(z)\,\sinh(7\,s)+\frac{63}{1024}\,\cos(5\,v+w)\,\sinh(z)\,\sinh(7\,s)\\
&&+\frac{315}{2048}\,\cos(7\,v+w)\,\sinh(z)\,\sinh(s)-\frac{189}{2048}\,\cos(7\,v+w)\,\sinh(z)\,\sinh(3\,s)\\
&&+\frac{63}{2048}\,\cos(7\,v+w)\,\sinh(z)\,\sinh(5\,s)-\frac{9}{2048}\,\cos(7\,v+w)\,\sinh(z)\,\sinh(7\,s)
\end{array}
\eeq
}
and by setting $v=0$, we again obtain the potential on the manifold of $SO(7)$ singlets
\beq
\begin{array}{lcl}
-8g^{-2}V&=&33-7\cosh(4\,s)+\frac{35}{2}\,\cosh(s)\,\cosh(z)-7\,\cosh(3\,s)\,\cosh(z)\\
&&-\frac{1}{2}\,\cosh(7\,s)\,\cosh(z)-\frac{35}{2}\,\cos(w)\,\sinh(z)\,\sinh(s)\\
&&-7\,\cos(w)\,\sinh(z)\,\sinh(3\,s)-\frac{1}{2}\,\cos(w)\,\sinh(z)\,\sinh(7\,s).
\end{array}
\eeq

The potentials obtained by using the $SO(6)_{\rm diag}$ singlet
parametrization $(\ref{sl3map})$ for the embedding tensors of
$SO(7,1)^2$ and $SO(6,2)^2$ as well as the
$(SO(5)\times SO(2))_{\rm diag}$ potential for
$\left(SO(5)\times SO(3)\right)^2$ are given in appendix {\ref{chPotentials}}.
The potential of the $(SO(5)\times SO(3))^2$ gauged theory on the manifold
of $(SO(5)\times SO(3))_{\rm diag}$ singlets parametrized by $(\ref{so5xso3param})$
reads
{\small
\beq
\begin{array}{lcl}
-8g^{-2}V&=&25-15\,\cosh(4\,s)-15\,\cosh(s)\,\cosh(z)+\frac{15}{2}\,\cosh(3\,s)\,\cosh(z)\\
&&+\frac{3}{2}\,\cosh(5\,s)\,\cosh(z)+15\,\cos(w)\,\sinh(z)\,\sinh(s)\\
&&+\frac{15}{2}\,\cos(w)\,\sinh(z)\,\sinh(3\,s)-\frac{3}{2}\,\cos(w)\,\sinh(z)\,\sinh(5\,s).
\end{array}
\eeq
}

For the gauge group $SO(4,4)^2$, the potential on the manifold of
$(SO(4)\times SO(4))_{\rm diag}$ singlets parametrized by
$(\ref{so43map})$ with $x=0$ is
{\small
\beq
-8g^{-2}V=24-16\,\cosh(z)-16\,\cosh(s)+8\,\cosh(s)\,\cosh(z),
\eeq
}
and including the extra $(SO(4)\times SO(3))_{\rm diag}$ singlet, this reads
{\small
\beq
\begin{array}{lcl}
-8g^{-2}V&=&21+3\,\cosh(8\,x)-12\,\cosh(z)\,\cosh(2\,x)-4\,\cosh(z)\,\cosh(6\,x)\\
&&-12\,\cos(w)\,\sinh(2\,x)\,\sinh(z)+4\,\cos(w)\,\sinh(6\,x)\,\sinh(z)\\
&&-12\,\cosh(s)\,\cosh(2\,x)-4\,\cosh(s)\,\cosh(6\,x)+4\,\cosh(s)\,\cosh(z)\\
&&+\frac{9}{2}\,\cosh(s)\,\cosh(z)\,\cosh(4\,x)\\
&&-\frac{1}{2}\,\cosh(s)\,\cosh(z)\,\cosh(12\,x)-\frac{3}{2}\,\cosh(s)\,\cos(w)\,\sinh(4\,x)\,\sinh(z)\\
&&+\frac{1}{2}\,\cosh(s)\,\cos(w)\,\sinh(12\,x)\,\sinh(z)+12\,\cos(v)\,\sinh(2\,x)\,\sinh(s)\\
&&-4\,\cos(v)\,\sinh(6\,x)\,\sinh(s)+\frac{3}{2}\,\cos(v)\,\cosh(z)\,\sinh(4\,x)\,\sinh(s)\\
&&-\frac{1}{2}\,\cos(v)\,\cosh(z)\,\sinh(12\,x)\,\sinh(s)+4\,\cos(v)\,\cos(w)\,\sinh(z)\,\sinh(s)\\
&&-\frac{9}{2}\,\cos(v)\,\cos(w)\,\cosh(4\,x)\,\sinh(z)\,\sinh(s)\\
&&+\frac{1}{2}\,\cos(v)\,\cos(w)\,\cosh(12\,x)\,\sinh(z)\,\sinh(s).
\end{array}
\eeq
}

If we do the same calculation with the embedding tensor of $SO(7,1)^2$, we obtain

{\small
\beq
\begin{array}{lcl}
-8g^{-2}V&=&21-3\,\cosh(8\,x)+12\,\cosh(z)\,\cosh(2\,x)-4\,\cosh(z)\,\cosh(6\,x)\\
&&+12\,\cos(w)\,\sinh(2\,x)\,\sinh(z)+4\,\cos(w)\,\sinh(6\,x)\,\sinh(z)\\
&&+12\,\cosh(s)\,\cosh(2\,x)-4\,\cosh(s)\,\cosh(6\,x)+4\,\cosh(s)\,\cosh(z)\\
&&-\frac{3}{2}\,\cosh(s)\,\cosh(z)\,\cosh(4\,x)\\
&&-\frac{1}{2}\,\cosh(s)\,\cosh(z)\,\cosh(12\,x)+\frac{9}{2}\,\cosh(s)\,\cos(w)\,\sinh(4\,x)\,\sinh(z)\\
&&+\frac{1}{2}\,\cosh(s)\,\cos(w)\,\sinh(12\,x)\,\sinh(z)-12\,\cos(v)\,\sinh(2\,x)\,\sinh(s)\\
&&-4\,\cos(v)\,\sinh(6\,x)\,\sinh(s)-\frac{9}{2}\,\cos(v)\,\cosh(z)\,\sinh(4\,x)\,\sinh(s)\\
&&-\frac{1}{2}\,\cos(v)\,\cosh(z)\,\sinh(12\,x)\,\sinh(s)+4\,\cos(v)\,\cos(w)\,\sinh(z)\,\sinh(s)\\
&&+\frac{3}{2}\,\cos(v)\,\cos(w)\,\cosh(4\,x)\,\sinh(z)\,\sinh(s)\\
&&+\frac{1}{2}\,\cos(v)\,\cos(w)\,\cosh(12\,x)\,\sinh(z)\,\sinh(s)
\end{array}
\eeq
}

\section{Exceptional gauge groups}

\subsection{$E_{7(+7)}\times SL(2)$}

Since for the gauge group $\Eee$, the potential reduces to just a
cosmological constant, the nontrivial case with the largest number of
noncompact directions in the gauge group is that of $E_{7(+7)}\times SL(2)$.

Here, the embedding tensor is given by
\beq
\begin{array}{lcl}
\Theta^{E_{7(+7)}\times SL(2)}_{\cA\cB}&=&-\frac{1}{2}\,H_\cA^{\underline{[IJ]}}H_\cB^{\underline{[KL]}}H_{\underline{[IJ]}}^{IJ}H_{\underline{[KL]}}^{KL}\\
&&\left(H_I^i H_J^j H_K^k H_L^l \delta^{ij}_{kl}
+H_I^{\bar i} H_J^{\bar j} H_K^{\bar k} H_L^{\bar l} \delta^{\bar i\bar j}_{\bar k\bar l}\right.\\
&&\left.+H_I^{i} H_J^{j} H_K^{\bar k} H_L^{\bar l} \delta^{ij}_{\bar k\bar l}
+H_I^{\bar i} H_J^{\bar j} H_K^{k} H_L^{l} \delta^{\bar i\bar j}_{kl}\right.\\
&&\left.+2\,H_I^{i} H_J^{\bar j} H_K^{k} H_L^{\bar l}
\left(\delta_{ik}\delta_{\bar j\bar l}+\delta_{\bar jk}\delta_{i\bar l}-\delta_{i\bar j}\delta_{k\bar l}\right)
\right)\\
&&+H_\cA^{\alpha\beta}H_\cB^{\gamma\delta}\left(\delta_{\alpha\gamma}\delta_{\beta\delta}+\delta_{\beta\gamma}\delta_{\alpha\delta}-\delta_{\alpha\beta}\delta_{\gamma\delta}\right)\\
&&+H_\cA^{\dot\alpha\dot\beta}H_\cB^{\dot\gamma\dot\delta}\left(\delta_{\dot\alpha\dot\gamma}\delta_{\dot\beta\dot\delta}+\delta_{\dot\beta\dot\gamma}\delta_{\dot\alpha\dot\delta}-\delta_{\dot\alpha\dot\beta}\delta_{\dot\gamma\dot\delta}\right)
\end{array}
\eeq

Using the tools we developed in previous sections, it is
natural to consider the scalar potential on the manifold of
$SO(6)\subset SO(8)$ singlets by using the parametrization
$(\ref{so6diagparam})$. Since the $SL(2)$ parametrized by $v, r_5$ is
part of the gauge group, and hence corresponds to flat directions in
the potential, these two parameters drop out.  Furthermore, three of
the five noncompact directions of the $SO(6)$-invariant $SL(3)$
singlets lie in the gauge group, and the smallest group containing the
remaining two orthogonal directions is $SL(2)$, which we can
parametrize by\footnote{Note that $S^{\cC}{}_\cB$ was incorrectly
  translated from machine form -- where index counting starts at zero
  -- to conventional notation in \cite{Fischbacher:2002fx}.}
\beq
\begin{array}{lcl}
S^\cC{}_\cB&=&\left(\delta^\alpha_7\delta^\beta_8-\delta^\alpha_8\delta^\beta_7\right)H^\cA_{\alpha\beta}f_{\cA\cB}{}^\cC\\
V^\cC{}_\cB&=&\frac{1}{2}\,\delta^{i\bar j}H_i^I H_{\bar j}^J H^\cA_{IJ} f_{\cA\cB}{}^\cC\\
\mathcal{V}&=&\exp\left(v\,V\right)\,\exp\left(s\,S\right)\,\exp\left(-v\,V\right)
\end{array}
\eeq
and obtain the potential
\beq
-8g^{-2}V=22-6\,\cosh(4\,s) \;,
\eeq
which obviously does not have any nontrivial stationary points.

\subsection{$E_{7(-5)}\times SU(2)$}
\label{subsec_e7xsu2}
An embedding of this gauge group into $\Eee$ can be obtained by
deleting the $+2-3$ root from the extended Dynkin diagram
$\ref{e8-dynkin}$. The ladder operators corresponding to the remaining
roots will generate the noncompact real form $E_{7(-5)}\times SU(2)$.
In order to define the embedding tensor, we have to introduce a
further ${\bf 8}\rightarrow {\bf 4}+{\bf 4}$ index split for the
vectors of the left $SO(8)$ of $SO(8)\times SO(8)$;
using superscripts $A,B$ to discern both ${\bf 4}$,
we have $I\rightarrow(i,\bar j)\rightarrow(i^A,i^B,\bar j)$.

The embedding tensor is then given by
\beq
\begin{array}{lcl}
\Theta^{E_{7(-5)}\times SU(2)}_{\cA\cB}&=&\phantom+\frac{1}{2}H_\cA^{\alpha\beta} H_\cB^{\gamma\delta} \delta_{\alpha\gamma}\left(\delta_{\beta\delta}+\gamma^{ijkl}_{\beta\delta}\delta_i^1\delta_j^2\delta_k^3\delta_l^4\right)\\
&&+\frac{1}{2}\,H_\cA^{\dot\alpha\dot\beta} H_\cB^{\dot\gamma\dot\delta} \delta_{\dot\alpha\dot\gamma}\left(\delta_{\dot\beta\dot\delta}+\gamma^{ijkl}_{\dot\beta\dot\delta}\delta_i^1\delta_j^2\delta_k^3\delta_l^4\right)\\
&&+\frac{1}{2}\,H_\cA^{\underline{[IJ]}} H_\cB^{\underline{[KL]}} H_{\underline{[IJ]}}^{IJ} H_{\underline{[KL]}}^{KL} H_I^{i} H_J^{j} H_K^{k} H_L^{l}\\
&&\left(H^{i^A}_i H^{j^A}_j H^{k^A}_k H^{l^A}_l \delta^{i^A j^A}_{k^A l^A}-H^{i^B}_i H^{j^B}_j H^{k^B}_k H^{l^B}_l \delta^{i^B j^B}_{k^B l^B}\right.\\
&&\left.+24\,\delta_{ijkl}^{1234}\right)\\
&&-\frac{1}{2}\,H_\cA^{\underline{[IJ]}} H_\cB^{\underline{[KL]}} H_{\underline{[IJ]}}^{IJ} H_{\underline{[KL]}}^{KL} H_I^{\bar i} H_J^{\bar j} H_K^{\bar k} H_L^{\bar l}\delta^{\bar i\bar j}_{\bar k\bar l}\\
&&-H_\cA^{\underline{[IJ]}} H_\cB^{\underline{[KL]}} H_{\underline{[IJ]}}^{IJ} H_{\underline{[KL]}}^{KL} H_I^{i} H_J^{\bar j} H_K^{k} H_L^{\bar l}\delta_{\bar j\bar l} H^{i^A}_i H^{k^A}_k\delta_{i^A k^A}
\end{array}
\eeq

Shooting into the blue sky, we take the positive roots ${++++++++}$
and ${++++----}$ and parametrize the four noncompact directions of the
corresponding $SL(2)\times SL(2)$ via
\beq
\begin{array}{lcl}
A^{ij}&:=&\delta_i^1\delta_j^2+\delta_i^3\delta_j^4+\delta_i^5\delta_j^6+\delta_i^7\delta_j^8\\
V^\cC{}_\cB&=&\frac{1}{2}\,\delta^{i\bar j} H^I_i H^J_{\bar j} H^\cA_{IJ} f_{\cA\cB}{}^\cC\\
S^\cC{}_\cB&=&\left(-\delta^\alpha_3\delta^\beta_3+\delta^\alpha_5\delta^\beta_5\right)H^\cA_{\alpha\beta} f_{\cA\cB}{}^\cC\\
W^\cC{}_\cB&=&\frac{1}{2}\left(H^I_iH^J_j A^{ij}-H^I_{\bar i}H^J_{\bar j} A^{\bar i\bar j}\right) H^\cA_{IJ} f_{\cA\cB}{}^\cC\\
Z^\cC{}_\cB&=&\left(+\delta^\alpha_3\delta^\beta_3+\delta^\alpha_5\delta^\beta_5\right)H^\cA_{\alpha\beta} f_{\cA\cB}{}^\cC\\
&&\\
\cV&=&\exp\left(vV\right)\exp\left(sS\right)\exp\left(-vV\right)
\exp\left(wW\right)\exp\left(zZ\right)\exp\left(-wW\right)
\end{array}
\eeq
and obtain the potential
\beq
\begin{array}{lcl}
-8g^{-2}V&=&\frac{39}{4}+\frac{1}{4}\,\cosh(4\,z)+\frac{1}{4}\,\cosh(4\,s)\\
&&+6\,\cosh(2\,s)\,\cosh(2\,z)-\frac{1}{4}\,\cosh(4\,s)\,\cosh(4\,z)
\end{array}
\eeq
which indeed does have a stationary point corresponding to a true
vacuum at $s=z=\frac{1}{2}\Arcosh 2$.  The
corresponding singlet generator $\delta^\alpha_5\delta^\beta_5
H_{\alpha\beta}^\cA f_{\cA\cB}{}^\cC \Arcosh 2$ breaks the $\left(SO(12)\times
  SO(3)\right)_{E_7}\times SU(2)_{SU(2)}$ compact subgroup of the
$E_{7(-5)}\times SU(2)$ gauge group in quite an interesting way:
  the $SU(2)$ gauge group factor with positive root ${+1+2}$ forms a
  diagonal $SO(3)$ with an $SO(3)$ factor within the $SO(12)$ of the
  $SO(12)\times SO(3)$ maximal compact subgroup of $E_{7(-5)}$; the
  positive root of this latter $SO(3)$ is ${+3+4}$. The $SO(3)$ factor
  of forementioned $SO(12)\times SO(3)$ (with positive root ${+1-2}$)
  remains unbroken, and a third $SO(3)$ factor with positive root ${+3-4}$
  appears. The rest of $SO(12)$ then forms a $SO(7)$ with an $SU(4)$
  with simple roots ${+5-6}, {+6-7}, {+7-8}$ as simply laced subgroup.
  Hence, the remaining gauge symmetry is
  $SO(7)\times SO(3)\times SO(3)\times SO(3)_{\rm diag}$,
  and this vacuum furthermore features $N=4$ supersymmetry.

\subsection{$E_{6(-14)}\times SU(3)$}

For this gauge group, no attempt has been made so far to find a
nontrivial vacuum; here, we only give the embedding tensor. The subset
${+4-5}$, ${+5-6}$, ${+6-7}$, ${+7-8}$, ${+7+8}$,
${+------+}$ of simple roots from $(\ref{e8_simple_roots})$ with
respect to the compact Cartan subalgebra corresponds to an embedding of
$E_{6(-14)}$ into $\Eee$. $SU(3)$ is likewise obtained from the roots
${-1-2}$, ${+2-3}$. This is sufficient information to complete the
construction.  It gives:
\beq
\begin{array}{lcl}
D^{(46s)}_{\alpha\beta}&:=&\delta_\alpha^4\delta_\beta^4+\delta_\alpha^6\delta_\beta^6,\quad
D^{(46c)}_{\dot\alpha\dot\beta}:=\delta_{\dot\alpha}^4\delta_{\dot\beta}^4+\delta_{\dot\alpha}^6\delta_{\dot\beta}^6\nonumber\\
E_{ij}&:=&\delta_i^7\delta_j^7+\delta_i^8\delta_j^8,\quad F_{ij}:=\delta_{ij}-E_{ij}\nonumber\\
\Theta^{E_{6(-14)}\times SU(3)}_{\cA\cB}&=&
\phantom+H_\cA^{\alpha\beta} H_\cB^{\gamma\delta}\delta^{\alpha\gamma} D^{(46s)}_{\beta\delta}
+H_\cA^{\dot\alpha\dot\beta} H_\cB^{\dot\gamma\dot\delta}\delta^{\dot\alpha\dot\gamma} D^{(46c)}_{\dot\beta\dot\delta}\nonumber\\
&&+H_\cA^\IJ H_\cA^\KL H_\IJ^{IJ} H_\KL^{KL}\nonumber\\
&&\left(-H_I^i H_J^{\bar j} H_K^k H_L^{\bar l} \delta_{\bar j\bar l} E_{ik}
-\frac{1}{2}\,H_I^{\bar i} H_J^{\bar j} H_K^{\bar k} H_L^{\bar l} \delta^{\bar i\bar j}_{\bar k\bar l}\right.\nonumber\\
&&\left.+\frac{1}{2}\,H_I^i H_J^j H_K^k H_L^l
\left(\delta^{mn}_{kl} F_{im} F_{jn}-\delta^{mn}_{kl} E_{im} E_{jn}\right.\right.\nonumber\\
&&\left.\left.-\frac{1}{32}\gamma^{ijkl}_{\alpha\beta}\left(\delta_{\alpha\beta}-4\,D^{(46s)}_{\alpha\beta}\right)
-\frac{1}{32}\gamma^{ijkl}_{\dot\alpha\dot\beta}\left(\delta_{\dot\alpha\dot\beta}-4\,D^{(46c)}_{\dot\alpha\dot\beta}\right)
\right)\right).
\end{array}
\eeq

\subsection{$E_{6(+2)}\times SU(2,1)$}

Applying a ${++++++++}$ Weyl reflection to the previously given set of
simple $E_6$ roots yields the roots ${+1+8}$, ${+4-5}$, ${+5-6}$,
${+6-7}$, ${+7-8}$, and ${------++}$, which correspond to an embedding
of $E_{6(+2)}$ into $\Eee$. In the same way, we obtain the
corresponding $SU(2,1)$ from ${--++++++}$ and ${+2-3}$.
Unfortunately, the corresponding embedding tensor is quite unwieldy,
so we do not give it here, as it perhaps makes sense to try to find a
construction which is more convenient for further calculations.

\subsection{$E_{6(+6)}\times SL(3)$}

One way to obtain the embedding tensor for this group is to take as
$SL(3)$ the subgroup of $\Eee$ that commutes with the compact
$(SO(6)\times SO(2))_{\rm diag}$ subgroup of the diagonal of the gauge
group $SO(6,2)^2$ constructed in sec. $\ref{so6diag}$; the corresponding
generators were explicitly given by $p_{1\ldots8}$ in
$(\ref{so6-sl3xsl2})$.  The group commuting with this $SL(3)$ is just
$E_{6(+6)}$.  With the intention to embed $F_{4(+4)}$ into $E_{6(+6)}$
in mind, a perhaps somewhat more useful construction is given by
starting again from the extended Dynkin diagram, but this time with
roots corresponding to the noncompact Cartan subalgebra, that is,
$(\ref{e8_simple_roots_nc})$.

The embedding tensor thus obtained reads
\beq
\begin{array}{lcl}
D^{(3)}_{\alpha\beta}&:=&\sum_{n=1}^3 \delta_\alpha^n\delta_\beta^n\\
D^{(5)}_{\alpha\beta}&:=&\sum_{n=4}^8 \delta_\alpha^n\delta_\beta^n\\
E_{\alpha\beta\gamma\delta}&:=&D^{(5)}_{\alpha\gamma}D^{(5)}_{\beta\delta}-D^{(3)}_{\alpha\gamma}D^{(3)}_{\beta\delta}\\
F_{\alpha\beta\gamma\delta}&:=&\delta_\alpha^1\delta_\beta^2\delta_\gamma^1\delta_\delta^2+\delta_\alpha^1\delta_\beta^3\delta_\gamma^1\delta_\delta^3-\delta_\alpha^2\delta_\beta^3\delta_\gamma^2\delta_\delta^3\\
G_{\alpha\beta\gamma\delta}&:=&\delta_\alpha^2\delta_\beta^2\delta_\gamma^3\delta_\delta^3+\delta_\alpha^3\delta_\beta^3\delta_\gamma^2\delta_\delta^2-\delta_\alpha^2\delta_\beta^3\delta_\gamma^3\delta_\delta^2-\delta_\alpha^3\delta_\beta^2\delta_\gamma^2\delta_\delta^3\\
\Theta^{E_{6(+6)}\times SL(3)}_{\cA\cB}&=&H_\cA^{\underline{[IJ]}} H_\cB^{\underline{[KL]}}H^{IJ}_{\underline{[IJ]}}H^{KL}_{\underline{[KL]}}\\
&&\left(-\frac{1}{32}H_I^i H_J^j H_K^k H_L^l \gamma^{ij}_{\alpha\beta} \gamma^{kl}_{\gamma\delta} E_{\alpha\beta\gamma\delta}\right.\\
&&-\frac{1}{32}H_I^{\bar i} H_J^{\bar j} H_K^{\bar k} H_L^{\bar l} \gamma^{\bar i\bar j}_{\alpha\beta} \gamma^{\bar k\bar l}_{\gamma\delta} E_{\alpha\beta\gamma\delta}\\
&&-\frac{1}{16}H_I^{\bar i} H_J^{\bar j} H_K^{k} H_L^{l} \gamma^{\bar i\bar j}_{\alpha\beta} \gamma^{kl}_{\gamma\delta} F_{\alpha\beta\gamma\delta}\\
&&-\frac{1}{16}H_I^{i} H_J^{j} H_K^{\bar k} H_L^{\bar l} \gamma^{ij}_{\alpha\beta} \gamma^{\bar k\bar l}_{\gamma\delta} F_{\alpha\beta\gamma\delta}\\
&&-\frac{1}{2}H_I^{i} H_J^{\bar j} H_K^{k} H_L^{\bar l}
\left(\frac{1}{4}\gamma^{i\bar j}_{\alpha\beta} \gamma^{k\bar l}_{\gamma\delta}\left(\delta^\alpha_1\delta^\gamma_1 D^{(5)}_{\beta\delta}+\delta^\alpha_2\delta^\beta_3\delta^\gamma_2\delta^\delta_3\right)\right.\\
&&\left.\left.+\frac{1}{24}\gamma^{i\bar j}_{\alpha_1\alpha_2\alpha_3\alpha_4}\gamma^{k\bar l}_{\beta_1\beta_2\beta_3\beta_4}\delta^{\alpha_1}_1\delta^{\beta_1}_1 D^{(5)}_{\alpha_2\beta_2} D^{(5)}_{\alpha_3\beta_3} D^{(5)}_{\alpha_4\beta_4}\right)\right)\\
&&+H_\cA^{\alpha\beta}H_\cB^{\gamma\delta}\left(D^{(5)}_{\alpha\gamma}D^{(5)}_{\beta\delta}-D^{(3)}_{\alpha\gamma}D^{(3)}_{\beta\delta}+D^{(3)}_{\alpha\delta}D^{(3)}_{\beta\gamma}-D^{(3)}_{\alpha\beta}D^{(3)}_{\gamma\delta}+2G_{\alpha\beta\gamma\delta}\right)\\
&&+\frac{1}{2}H_\cA^{\dot\alpha\dot\beta}H_\cB^{\dot\gamma\dot\delta}
\left(\frac{1}{4}\gamma^{\dot\alpha\dot\beta}_{\alpha\beta} \gamma^{\dot\gamma\dot\delta}_{\gamma\delta}\left(\delta^{\alpha}_1\delta^{\gamma}_1 D^{(5)}_{\beta\delta}+\delta^\alpha_2\delta^\beta_3\delta^\gamma_2\delta^\delta_3\right)\right.\\
&&\left.+\frac{1}{24}\gamma^{\dot\alpha\dot\beta}_{\alpha_1\alpha_2\alpha_3\alpha_4}\gamma^{\dot\gamma\dot\delta}_{\beta_1\beta_2\beta_3\beta_4}\delta^{\alpha_1}_1\delta^{\beta_1}_1 D^{(5)}_{\alpha_2\beta_2} D^{(5)}_{\alpha_3\beta_3} D^{(5)}_{\alpha_4\beta_4}\right)\\
\end{array}
\eeq

\subsection{$F_{4(+4)}\times G_{2(+2)}$}

The $F_{4(+4)}$ subgroup can be obtained from $E_{6(+6)}$ above;
$G_{2(+2)}$ is then the space of all $F_{4(+4)}$ singlets. In
particular, the $F_{4(+4)}$ Lie algebra is generated by
$\left(T_{+7+8}+T_{-7-8}\right)$,
$\left(T_{+7+8}-T_{-7-8}\right)$,
$\left(T_{+6-7}+T_{-6+7}\right)$,
$\left(T_{+6-7}-T_{-6+7}\right)$,
$\left(T_{+------+}+T_{-++++++-} +\right.$ $\left.T_{+4-5}+T_{-4+5}\right)$,
$\left(T_{+------+}-T_{-++++++-} +\right.$ $\left.T_{+4-5}-T_{-4+5}\right)$,
$\left(T_{+5-6}+T_{-6+5} +\right.$ $\left.T_{+7-8}+T_{-7+8}\right)$,
$\left(T_{+5-6}-T_{-6+5} +\right.$ $\left.T_{+7-8}-T_{-7+8}\right)$.
Unfortunately,
this construction mutilates the $SO(8)\times SO(8)$ components of
$\Eee$ so badly that no reasonably simple expression
for the embedding tensor could be found in that language.

\subsection{$G_{2(-14)}\times F_{4(-20)}$}

The subgroup of $SO(8)_L$ that leaves the last vector, spinor, and
co-spinor index fixed is $G_2$; all generators of $E_8$ that commute
with this $G_2$ form a $F_{4(-20)}$ whose compact subgroup is $SO(9)$.
The linear combination of embedding tensors of these two groups with
a relative gauge coupling $g_{G_2}/g_{F_4}=3/2$ satisfies
the $P^{27000}$ projection condition and thus corresponds to a possible gauging.
It is given by
\bea
p_{ij}&:=&\delta_{ij}-2\delta^8_i\delta^8_j\nonumber\\
d_{ijkl}&:=&\delta^{i'j'}_{k'l'}p_{ii'}p_{jj'}p_{kk'}p_{ll'}\nonumber\\
\Theta_{\cA\cB}&=&\phantom+H_\cA^{\alpha\beta}H_\cB^{\gamma\delta}\delta_{\alpha\gamma}\delta_\beta^8\delta_\delta^8\nonumber\\
&&+H_\cA^{\dot\alpha\dot\beta}H_\cB^{\dot\gamma\dot\delta}\delta_{\dot\alpha\dot\gamma}\delta_{\dot \beta}^8\delta_{\dot \delta}^8\nonumber\\
&&-H_\cA^{\underline{[IJ]}}H_\cB^{\underline{[KL]}}H^{IJ}_{\underline{[IJ]}}H^{KL}_{\underline{[KL]}}H_I^{i}H_J^{\bar j}H_K^{k}H_L^{\bar l}\delta_{\bar j\bar l}\delta_i^8\delta_k^8\\
&&-\frac{1}{2}\,H_\cA^{\underline{[IJ]}}H_\cB^{\underline{[KL]}}H^{IJ}_{\underline{[IJ]}}H^{KL}_{\underline{[KL]}}H_I^{\bar i}H_J^{\bar j}H_K^{\bar k}H_L^{\bar l}\delta_{\bar i \bar j}^{\bar k\bar l}\nonumber\\
&&-\frac{1}{16}\,H_\cA^{\underline{[IJ]}}H_\cB^{\underline{[KL]}}H^{IJ}_{\underline{[IJ]}}H^{KL}_{\underline{[KL]}}H_I^{ i}H_J^{ j}H_K^{ k}H_L^{ l}\nonumber\\
&&\left(8\,d_{ijkl}
-\frac{1}{7}\gamma^{ijkl}_{\alpha\beta}\left(\delta_{\alpha\beta}-7\delta^8_\alpha\delta^8_\beta\right)
-\frac{1}{7}\gamma^{ijkl}_{\dot\alpha\dot\beta}\left(\delta_{\dot\alpha\dot\beta}-7\delta^8_{\dot\alpha}\delta^8_{\dot\beta}\right)
\right).\nonumber
\eea

The main problem in this case is to find a suitable
invariance subgroup of the gauge group which is small enough to show nontrivial
structure, but does not give too many singlets. Here, we choose that
particular subgroup $SU(3)\times SU(3)$ of the group $SO(8)_L\times
SO(8)_R$ which stabilizes the vectors $v_1^i=\delta^{i7}$,
$v_2^i=\delta^{i8}$, $v_3^{\bar i}=\delta^{\bar i7}$, $v_4^{\bar
i}=\delta^{\bar i8}$ as well as the spinors
$\psi^{\alpha_L}=\delta^{\alpha_L8}$,
$\psi^{\alpha_R}=\delta^{\alpha_R8}$ (and which therefore is also a subgroup of
$G_2\times F_{4(-20)}$).

This group is stabilized by a subgroup $SU(2,1)\times SU(2,1)$ of
$E_{8(8)}$, hence we have to deal with an eight-dimensional
submanifold of the supergravity scalars here. The intersection of this
eight-dimensional manifold with the gauge group is four-dimensional,
but unfortunately, unlike for the parametrization considered in the
$E_{7(7)}\times SL(2)$ case, the smallest group containing the four
directions orthogonal to the gauge group is the full $SU(2,1)\times
SU(2,1)$, hence we parametrize the full eight-dimensional
manifold. Using the generators $X_{(A,B)}$ of both $SO(3)$ subalgebras as well
as those of two noncompact directions $Y_{(A,B)}$
\beq
\begin{array}{lcl}
Y_{(A)}^{\mathcal{C}}{}_{\mathcal{B}}&=&-\frac{1}{2}\left(\delta^{\dot\alpha}_{2}\delta^{\dot\beta}_{8}-\delta^{\dot\alpha}_{8}\delta^{\dot\beta}_{2}\right)f_{\dot\alpha\dot\beta\mathcal{B}}{}^{\mathcal{C}}\\
X_{(A)1}^{\mathcal{C}}{}_{\mathcal{B}}&=&2\left(\delta^{i}_{7}\delta^{\bar j}_{8}-\delta^{i}_{8}\delta^{\bar j}_{7}\right)f_{\underline{[i\bar j]}\mathcal{B}}{}^{\mathcal{C}}\\
X_{(A)2}^{\mathcal{C}}{}_{\mathcal{B}}&=&2\left(\delta^{i}_{7}\delta^{\bar j}_{7}+\delta^{i}_{8}\delta^{\bar j}_{8}\right)f_{\underline{[i\bar j]}\mathcal{B}}{}^{\mathcal{C}}\\
X_{(A)3}^{\mathcal{C}}{}_{\mathcal{B}}&=&2\left(\delta^{i}_{7}\delta^{j}_{8}\,f_{\underline{[ij]}\mathcal{B}}{}^{\mathcal{C}}-\delta^{\bar i}_{7}\delta^{\bar j}_{8}\,f_{\underline{[\bar i\bar j]}\mathcal{B}}{}^{\mathcal{C}}\right)\\
Y_{B}^{\mathcal{C}}{}_{\mathcal{B}}&=&-\frac{1}{2}\left(\delta^{\dot\alpha}_{2}\delta^{\dot\beta}_{8}+\delta^{\dot\alpha}_{8}\delta^{\dot\beta}_{2}\right)f_{\dot\alpha\dot\beta\mathcal{B}}{}^{\mathcal{C}}\\
X_{(B)1}^{\mathcal{C}}{}_{\mathcal{B}}&=&-2\left(\delta^{i}_{7}\delta^{\bar j}_{8}+\delta^{i}_{8}\delta^{\bar j}_{7}\right)f_{\underline{[i\bar j]}\mathcal{B}}{}^{\mathcal{C}}\\
X_{(B)2}^{\mathcal{C}}{}_{\mathcal{B}}&=&-2\left(\delta^{i}_{7}\delta^{\bar j}_{7}-\delta^{i}_{8}\delta^{\bar j}_{8}\right)f_{\underline{[i\bar j]}\mathcal{B}}{}^{\mathcal{C}}\\
X_{(B)3}^{\mathcal{C}}{}_{\mathcal{B}}&=&-2\left(\delta^{i}_{7}\delta^{j}_{8}\,f_{\underline{[ij]}\mathcal{B}}{}^{\mathcal{C}}+\delta^{\bar i}_{7}\delta^{\bar j}_{8}\,f_{\underline{[\bar i\bar j]}\mathcal{B}}{}^{\mathcal{C}}\right)\;,
\end{array}
\eeq
we parametrize the eight-dimensional singlet manifold as
\beq\label{su3xsu3param}
\begin{array}{lcl}
\mathcal{V}&=&\exp(r_1\,X_{(A)1})\,\exp(r_2\,X_{(A)2})\,
\exp(r_3\,X_{(A)3})\\
&&\exp(r_4\,X_{(B)1})\,\exp(r_5\,X_{(B)2})\,\exp(r_6\,X_{(B)3})\\
&&\,\exp(s\,Y_{(A)})\,\exp(z\,Y_{(B)})\\
&&\exp(-r_6\,X_{(B)3})\,\exp(-r_5\,X_{(B)2})\,\exp(-r_4\,X_{(B)1})\\
&&\exp(-r_3\,X_{(A)3})\,\exp(-r_2\,X_{(A)2})\,\exp(-r_1\,X_{(A)1})
\end{array}
\eeq
and obtain the potential given $(\ref{potential_su3su3_g2xf4})$.

\section{Vacua}

For every model with noncompact gauge group for which we find a
nontrivial vacuum, we also give data for the trivial
$\cV^\cC{}_\cB=\delta^\cC_\cB$ vacuum; of relevance is for example the
square root of the ratio of cosmological constants, since this should
correspond to the ratio of central charges of the boundary CFT.

For the gauge group $SO(7,1)\times SO(7,1)$, the vacuum at the origin has:

\begin{eqnarray}
\begin{tabular}{|l|l|}\hline
$\Lambda/2g^2$&$-9$\\\hline
$\mathcal{M}/g^2$&$16_{(\times 1)},\;0_{(\times 14)},\;-5_{(\times 64)},\;-8_{(\times 49)}$\\\hline
$\mathcal{M}^{\rm vec}/g$&$2_{(\times 7)},\;0_{(\times 114)},\;-2_{(\times 7)}$\\\hline
$A_1$&$3/2_{(\times 8)},\;-3/2_{(\times 8)}$\\\hline
$A_3$&$7/2_{(\times 8)},\;1/2_{(\times 56)},\;-1/2_{(\times 56)},\;-7/2_{(\times 8)}$\\\hline
\end{tabular}
\end{eqnarray}

There is a further stable AdS vacuum with remaining symmetry
$G_2\times G_2$ and no supersymmetry at
\beq
\cV^\cC{}_\cB=\exp\left(f_{\cA\cB}{}^\cC H^\cA_{\dot\alpha\dot\beta}\delta^{\dot\alpha}_8\delta^{\dot\beta}_8\cdot\Arcosh 2\right)
\eeq
with mass spectrum

\begin{eqnarray}\label{E-p7-g2}
\begin{tabular}{|l|l|}\hline
$\Lambda/(2g^2)$&$-211/16$\\\hline
$\mathcal{M}/g^2$&$195/4_{(\times 1)},\;45/2_{(\times 1)},\;0_{(\times
28)},\;-9/2_{(\times 49)},\;-33/4_{(\times 49)}$\\\hline
$\mathcal{M}^{\rm vec}/g$&$9/2_{(\times 7)},\;3_{(\times 7)},\;0_{(\times
100)},\;-3_{(\times 7)},\;-9/2_{(\times 7)}$\\\hline
$A_1$&$35/8_{(\times 1)},\;19/8_{(\times 7)},\;-19/8_{(\times
7)},\;-35/8_{(\times 1)}$\\\hline
$A_3$&$105/8_{(\times 1)^*},\;57/8_{(\times 7)^*},\;33/8_{(\times
7)},\;15/8_{(\times 49)},\;$
\\&$-15/8_{(\times 49)},\;-33/8_{(\times
7)},\;-57/8_{(\times 7)^*},\;-105/8_{(\times 1)^*}$\\\hline
\end{tabular}
\end{eqnarray}

Furthermore, there is strong numerical evidence for another vacuum
with $\Lambda/(2g^2)=-13$ and remaining symmetry $SO(6)\times SO(6)$
whose location is approximately given by
\bea
\cV^\cC{}_\cB&\approx&\exp\left(f_{\cA\cB}{}^\cC H^\cA_{\alpha\beta}\left(-1.1311837\,\delta^\alpha_7\delta^\beta_7+0.3317440\,\delta^\alpha_7\delta^\beta_8\right.\right.\nonumber\\
&&\left.\left.+0.1023203\,\delta^\alpha_8\delta^\beta_7-0.0084405\delta^\alpha_8\delta^\beta_8\right)\right)
\eea
or, in parameters for $(\ref{so6diagparam})$,
$z\approx-s\approx0.2906220$, $v\approx-1.2037291$,
$r_1\approx3\,\pi$, $r_3\approx 5\pi/2$, $r_2\approx6.3825153$,
$r_5\approx 0.3689107$. So far, no corresponding analytic expression
has been obtained for this vacuum.

For the $SO(5,3)\times SO(5,3)$ gauged model, the origin is an AdS
vacuum:

\begin{eqnarray}
\begin{tabular}{|l|l|}\hline
$\Lambda/2g^2$&$-1$\\\hline
$\mathcal{M}/g^2$&$8_{(\times 9)},\;3_{(\times 64)},\;0_{(\times 55)}$\\\hline
$\mathcal{M}^{\rm vec}/g$&$2_{(\times 15)},\;0_{(\times 98)},\;-2_{(\times 15)}$\\\hline
$A_1$&$1/2_{(\times 8)},\;-1/2_{(\times 8)}$\\\hline
$A_3$&$5/2_{(\times 24)},\;3/2_{(\times 40)},\;-3/2_{(\times 40)},\;-5/2_{(\times 24)}$\\\hline
\end{tabular}
\end{eqnarray}

Furthermore, we find a deSitter vacuum with remaining symmetry of
$SO(5)\times SO(5)\times SO(3)_{\rm diag}$ at
\bea
\cV^\cC{}_\cB&=&\exp\left(\frac{1}{2}\Arcosh 5\;f_{\cA\cB}{}^\cC H^\cA_{\alpha\beta}\left(\delta^\alpha_6\delta^\beta_6+\delta^\alpha_7\delta^\beta_7+\delta^\alpha_8\delta^\beta_8\right)\right)
\eea
with mass spectrum
\begin{eqnarray}\label{E-p5-so5}
\begin{tabular}{|l|l|}\hline
$\Lambda/(2g^2)$&$11$\\\hline
$\mathcal{M}/g^2$&$96_{(\times 5)},\;45_{(\times 48)},\;24_{(\times
25)},\;0_{(\times 33)},\;-3_{(\times 16)},\;-48_{(\times 1)}$\\\hline
$\mathcal{M}^{\rm vec}/g$&$6_{(\times 15)},\;0_{(\times 98)},\;-6_{(\times
15)}$\\\hline
$A_1$&$5/2_{(\times 8)},\;-5/2_{(\times 8)}$\\\hline
$A_3$&$15/2_{(\times 8^*,\;\times16)},\;9/2_{(\times 40)},\;-9/2_{(\times
40)},\;-15/2_{(\times 8^*,\;\times16)}$\\\hline
\end{tabular}
\end{eqnarray}

The vacuum at the origin of the $SO(4,4)\times SO(4,4)$ gauged model
is a Minkowski vacuum:
\begin{eqnarray}\label{E-p4-so44}
\begin{tabular}{|l|l|}\hline
$\Lambda/(2g^2)$&$0$\\\hline
$\mathcal{M}/g^2$&$4_{(\times 96)},\;0_{(\times 32)}$\\\hline
$\mathcal{M}^{\rm vec}/g$&
$2_{(\times 16)},\;0_{(\times 96)},\;-2_{(\times 16)}$ \\\hline 
$A_1$&$0_{(\times 16)}$\\\hline
$A_3$&$2_{(\times 64)},\;-2_{(\times 64)}$\\\hline
\end{tabular}
\end{eqnarray}

Here we find a further deSitter vacuum with remaining symmetry $SO(4)\times SO(4)\times SO(4)_{\rm diag}$ at
\bea
\cV^\cC{}_\cB&=&\exp\left(\frac{1}{2}\Arcosh 2\;f_{\cA\cB}{}^\cC H^\cA_{\alpha\beta}\sum_n=1^4 \delta^\alpha_n\delta^\beta_n\right)
\eea
with mass spectrum
\begin{eqnarray}\label{E-p4-so4}
\begin{tabular}{|l|l|}\hline
$\Lambda/(2g^2)$&$2$\\\hline
$\mathcal{M}/g^2$&$12_{(\times 49)},\;9_{(\times 32)},\;0_{(\times
46)},\;-12_{(\times 1)}$\\\hline
$\mathcal{M}^{\rm vec}/g$&$3_{(\times 16)},\;0_{(\times 96)},\;-3_{(\times
16)}$\\\hline
$A_1$&$1_{(\times 8)},\;-1_{(\times 8)}$\\\hline
$A_3$&$3_{(\times 8^*,\;\times56)},\;-3_{(\times 8^*,\;\times56)}$\\\hline
\end{tabular}
\end{eqnarray}

For the $G_{2(-14)}\times F_{4(-20)}$ gauged model, the origin has
$N=(7,9)$ supersymmetry, and the gauge group is broken down to its
maximal compact subgroup $G_2\times SO(9)$:
\begin{eqnarray}\label{E-e1-so9}
\begin{tabular}{|l|l|}\hline
$\Lambda/(2g^2)$&$-4$\\\hline
$\mathcal{M}/g^2$&$0_{(\times 16)},\;-3_{(\times 112)}$\\\hline
$\mathcal{M}^{\rm vec}/g$&$1_{(\times 16)},\;0_{(\times 112)}$\\\hline
$A_1$&$1_{(\times 7)},\;-1_{(\times 9)}$\\\hline
$A_3$&$2_{(\times 16)},\;0_{(\times 112)}$\\\hline
\end{tabular}
\end{eqnarray}
There is a further AdS vacuum with $N=(0,1)$ supersymmetry
which breaks gauge group to $SU(3)\times SO(7)$ at
\bea
\cV^\cC{}_\cB&=&\exp\left(\frac{1}{2}\Arcosh 7\;f_{\cA\cB}{}^\cC H^\cA_{\dot\alpha\dot\beta}\delta^{\dot\alpha}_8\delta^{\dot\beta}_2\right)
\eea
with mass spectrum
\begin{eqnarray}\label{E-e1-so7}
\begin{tabular}{|l|l|}\hline
$\Lambda/(2g^2)$&$-25/4$\\\hline
$\mathcal{M}/g^2$&$24_{(\times 1)},\;0_{(\times 37)},\;-9/4_{(\times
48)},\;-6_{(\times 42)}$\\\hline
$\mathcal{M}^{\rm vec}/g$&$4_{(\times 1)},\;3_{(\times 6)},\;3/2_{(\times
8)},\;1_{(\times 7)},\;0_{(\times 91)},\;-1/2_{(\times 8)},\;-3_{(\times
7)}$\\\hline
$A_1$&$11/4_{(\times 1)},\;7/4_{(\times 6)},\;-5/4_{(\times 1)},\;-7/4_{(\times
8)}$\\\hline
$A_3$&$33/4_{(\times 1)^*},\;21/4_{(\times 6)^*},\;17/4_{(\times
1)},\;11/4_{(\times 8)},\;9/4_{(\times 7)},$\\
&$3/4_{(\times 48)}, -3/4_{(\times 42)},\;-7/4_{(\times 7)},\;-21/4_{(\times 8)^*}$\\\hline
\end{tabular}
\end{eqnarray}

For the $E_{7(-5)}\times SU(2)$ exceptional gauging, the vacuum at the origin breaks the gauge group down to
$SO(12)\times SO(3)\times SU(2)$, with $N=(4,12)$ supersymmetry:
\begin{eqnarray}
\begin{tabular}{|l|l|}\hline
$\Lambda/(2g^2)$&$-4$\\\hline
$\mathcal{M}/g^2$&$0_{(\times 64)},\;-3_{(\times 64)}$\\\hline
$\mathcal{M}^{\rm vec}/g$&$1_{(\times 64)},\;0_{(\times 64)}$\\\hline
$A_1$&$1_{(\times 4)},\;-1_{(\times 12)}$\\\hline
$A_3$&$2_{(\times 64)},\;0_{(\times 64)}$\\\hline
\end{tabular}
\end{eqnarray}

As explained before in sec. \ref{subsec_e7xsu2}, there is a further nontrivial vacuum at
\bea
\cV^\cC{}_\cB&=&\exp\left(\delta^\alpha_5\delta^\beta_5 H_{\alpha\beta}^\cA f_{\cA\cB}{}^\cC \Arcosh 2\right)
\eea
with remaining symmetry $SO(7)\times SO(3)\times SO(3)\times SO(3)_{\rm diag}$, $N=(0,4)$:
\begin{eqnarray}
\begin{tabular}{|l|l|}\hline
$\Lambda/(2g^2)$&$-25/4$\\\hline
$\mathcal{M}/g^2$&$24_{(\times 1)},\;0_{(\times 106)},\;-6_{(\times 21)}$\\\hline
$\mathcal{M}^{\rm vec}/g$&$4_{(\times 4)},\;3_{(\times 3)},\;3/2_{(\times 32)},\;1_{(\times 28)},$\\
&$0_{(\times 22)},\;-1/2_{(\times 32)},\;-3_{(\times 7)}$\\\hline
$A_1$&$11/4_{(\times 4)},\;-5/4_{(\times 4)},\;-7/4_{(\times 8)}$\\\hline
$A_3$&$33/4_{(\times 4)^*},\;17/4_{(\times 4)},\;11/4_{(\times 32)},$\\
&$9/4_{(\times 28)},\;3/4_{(\times 24)},\;-7/4_{(\times 28)},\;-21/4_{(\times 8)^*}$\\\hline
\end{tabular}
\end{eqnarray}

\chapter{A four-dimensional example}
\label{chapter4d}

For maximal ($N=8$) gauged supergravity in four dimensions with
compact gauge group $SO(8)$, the known nontrivial vacua are one with a
remaining symmetry of $SO(3)\times SO(3)$ and cosmological constant
$-14g^2$ \cite{Warner:du} as well as five further
ones for which it has been proven by exhaustive analysis of the
potential restricted to a submanifold of six scalars that they are the
only vacua with a remaining symmetry of at least $SU(3)$. It is
tempting to try to apply the tools presented here to go even further
and break $SO(8)$ down to a smaller subgroup. In particular, among the
different embeddings of $SO(3)$ into $SO(8)$, there is a very simple
one for which the $SO(3)$-invariant submanifold of the 70-dimensional
space $E_{7(7)}/SU(8)$ is ten-dimensional. This is given by the
following construction: under its maximal compact subgroup, $SU(8)$,
$E_{7(7)}$ decomposes into the adjoint representation as well as the
self-dual and anti-self-dual 4-forms ${\bf 133}\rightarrow {\bf
  63}+{\bf 35}_{\rm sd}+{\bf 35}_{\rm asd}$.  With respect to the
gauge group $SO(8)$, the ${\bf 35}_{\rm sd}$ and ${\bf 35}_{\rm asd}$
are just the symmetric traceless matrices over the spinors and
co-spinors, while ${\bf 63}$ decomposes into the adjoint ${\bf 28}$
and a further {\bf 35} given by the symmetric traceless matrices over
the vectors. Hence we split $E_{7(7)}$ adjoint indices via
${\tt A}\rightarrow(\underline{(\alpha\beta)},\underline{(\dot\alpha\dot\beta)}, \underline{(ab)}, \underline{[ij]})$.
If we furthermore split the fundamental representation
via ${\bf 56\rightarrow {\bf 28}+{\bf 28}}$, ${\tt P}\rightarrow(\underline{[ij]^a},\underline{[ij]^b})$
and introduce the auxiliary tensors
\bea
\Psi_{\underline{(\alpha\beta)}}^{ijkl}&:=&
\left\{\begin{array}{ll}
\gamma^{ijkl}_{\alpha\beta}\left(\delta^\alpha_n\delta^\beta_n-\delta^\alpha_{n+1}\delta^\beta_{n+1}\right)&\mbox{for $\underline{(\alpha\beta)}=n\in\{1\ldots7\}$}\\
\gamma^{ijkl}_{\alpha\beta}\left(\delta^\alpha_p\delta^\beta_q+\delta^\alpha_q\delta^\beta_p\right)&
\mbox{for $\left\{
\begin{array}{l}
\underline{(\alpha\beta)}=n\in\{8\ldots35\},\\
\underline{[pq]}=n-8,\\
p,q\;\mbox{such that}\;H_{\underline{[pq]}}^{pq}=1
\end{array}\right.$}
\end{array}
\right.\\
\dot\Psi_{\underline{(\dot\alpha\dot\beta)}}^{ijkl}&:=&\mbox{ditto, with $\gamma^{ijkl}_{\dot\alpha\dot\beta}$ in place of $\gamma^{ijkl}_{\alpha\beta}$}\\
\Omega_{{\underline{(ij)}}b}{}^c&=&
\left\{
\begin{array}{ll}
i\left(\delta^b_n\delta^c_n-\delta^b_{n+1}\delta^c_{n+1}\right)&\mbox{for $\underline{(ij)}=n\in\{1..7\}$}\\
i\left(\delta^b_p\delta^c_q+\delta^b_q\delta^c_p\right)&
\mbox{for 
$\left\{\begin{array}{l}\underline{(ij)}=n\in\{8..35\},\\
\underline{[pq]}=n-8,\\
p,q\;\mbox{such that}\;H_{\underline{[pq]}}^{pq}=1
\end{array}\right.$}
\end{array}\right.
\eea
then generators in the $E_{7(7)}$ fundamental representation can be obtained
from\footnote{What happens here is quite obvious, but considerably obfuscated
by the necessity of successive index mappings to get rid of all factor-two ambiguities:
in Matrix notation, the generators of $E_7$ in the fundamental representation
decompose into $28\times 28$ blocks; the right-top and left-bottom blocks
are formed from the self-dual and anti-self-dual $SU(8)$ 4-forms, which are
equivalent to the ${\bf 35}_s$ and ${\bf 35}_c$ from $SO(8)$,
while the blocks on the major diagonal are essentially given by $SU(8)$
generators and their conjugates; cf. $(3.10)$ in \cite{Warner:du}.
(Note that the generators in the right-bottom corner should be the complex conjugated ones there!)}
 (here, capital letters $I,J,\ldots$ also designate $SO(8)$ indices)
\bea
t_{\tt AP}{}^{\tt Q}&=&\phantom+\frac{1}{8}H_{\tt A}^{\underline{(\alpha\beta)}}H_{\underline{[ij]}}^{ij}H^{\underline{[ij]^a}}_{\tt Q}H^{\underline{[KL]}}_{KL}H_{\underline{[KL]^b}}^{\tt P}\Psi_{\underline{(\alpha\beta)}}^{i j K L}\nonumber\\
&&+\frac{1}{8}H_{\tt A}^{\underline{(\alpha\beta)}}H_{\underline{[kl]}}^{kl}H^{\underline{[kl]^b}}_{\tt Q}H^{\underline{[IJ]}}_{IJ}H_{\underline{[IJ]^a}}^{\tt P}\Psi_{\underline{(\alpha\beta)}}^{k l I J}\nonumber\\
&&+\frac{i}{8}H_{\tt A}^{\underline{(\dot\alpha\dot\beta)}}H_{\underline{[ij]}}^{ij}H^{\underline{[ij]^a}}_{\tt Q}H^{\underline{[KL]}}_{KL}H_{\underline{[KL]^b}}^{\tt P}\dot\Psi_{\underline{(\dot\alpha\dot\beta)}}^{i j K L}\nonumber\\
&&-\frac{i}{8}H_{\tt A}^{\underline{(\dot\alpha\dot\beta)}}H_{\underline{[kl]}}^{kl}H^{\underline{[kl]^b}}_{\tt Q}H^{\underline{[IJ]}}_{IJ}H_{\underline{[IJ]^a}}^{\tt P}\dot\Psi_{\underline{(\dot\alpha\dot\beta)}}^{k l I J}\\
&&+2H_{\tt A}^{\underline{(ab)}}H_{\underline{[ij]^a}}^{\tt Q}H^{\underline{[IJ]^a}}_{\tt P}\delta_{j'}^{J'}\delta^{ij}_{i'j'}\delta_{IJ}^{I'J'}\Omega_{{\underline{(ab)}}{i'}}{}^{I'}\nonumber\\
&&+2H_{\tt A}^{\underline{[ab]}}H_{\underline{[ij]^a}}^{\tt Q}H^{\underline{[IJ]^a}}_{\tt P}\delta_{j'}^{J'}\delta^{ij}_{i'j'}\delta_{IJ}^{I'J'}H_{\underline{[ab]}}^{ab} \delta^{i'}_a \delta^{I'}_b\nonumber\\
&&-2H_{\tt A}^{\underline{(ab)}}H_{\underline{[kl]^b}}^{\tt Q}H^{\underline{[KL]^b}}_{\tt P}\delta_{l'}^{L'}\delta^{kl}_{k'l'}\delta_{KL}^{K'L'}\Omega_{{\underline{(ab)}}{k'}}{}^{K'}\nonumber\\
&&+2H_{\tt A}^{\underline{[ab]}}H_{\underline{[kl]^b}}^{\tt Q}H^{\underline{[KL]^b}}_{\tt P}\delta_{l'}^{L'}\delta^{kl}_{k'l'}\delta_{KL}^{K'L'}H_{\underline{[ab]}}^{ab} \delta^{k'}_a \delta^{K'}_b\nonumber.
\eea

Structure constants of $E_{7(7)}$ then satisfy $t_{\tt AP}{}^{\tt Q} t_{\tt bq}{}^{\tt R}-t_{\tt BP}{}^{\tt Q} t_{\tt AQ}{}^{\tt R}=f_{\tt AB}{}^{\tt C} t_{\tt CP}{}^{\tt S}$.
With the normalization
\beq
\eta_{\tt AB}:=\frac{1}{288}\,f_{\tt AC}{}^{\tt D}\,f_{\tt BD}{}^{\tt C},
\eeq
metric tensor entries are $\pm 1$ on the diagonal and $\pm 1/2$
for neighbouring entries on diagonals on a ${\bf 35}$.

The potential for this model is given as follows: for an element of
$E_{7(7)}$ given as $56\times 56$ matrix $\cV^{\tt Q}{}_{\tt P}$, one
first forms the tensor $T_l{}^{kij}$ via
\bea
u_{ij}{}^{IJ}&:=&\cV^{\underline{[ij]}}{}_{\underline{[IJ]}}H^{ij}_{\underline{[ij]}^a}H_{IJ}^{\underline{[IJ]}^a}\nonumber\\
u^{kl}{}_{KL}&:=&\cV^{\underline{[kl]}}{}_{\underline{[KL]}}H^{kl}_{\underline{[kl]}^b}H_{KL}^{\underline{[KL]}^b}\nonumber\\
v_{ijKL}&:=&\cV^{\underline{[ij]}}{}_{\underline{[KL]}}H^{ij}_{\underline{[ij]}^a}H_{KL}^{\underline{[KL]}^b}\\
v^{klIJ}&:=&\cV^{\underline{[kl]}}{}_{\underline{[IJ]}}H^{kl}_{\underline{[kl]}^b}H_{IJ}^{\underline{[IJ]}^a}\nonumber\\
T_l{}^{kij}&:=&\frac{1}{8}\left(u^{ij}{}_{IJ}+v^{ijIJ}\right)\left(u_{lm}{}^{JK}u^{km}{}_{KI}-v_{lmJK}v^{kmKI}\right).\nonumber
\eea
Then, the $A_1$ and $A_2$ tensors are given by
\beq
A_1^{ij}=-\frac{4}{21}T_m{}^{ijm},\qquad A_{2l}{}^{ijk}=-\frac{4}{3}T_l{}^{i'j'k'}\delta_{i'j'k'}^{ijk}
\eeq
and the potential is
\beq\label{potentialE7}
V(\cV)=g^2\left(\frac{1}{24}A_{2i}^{jkl}A_{2i}^{*jkl}-\frac{3}{4}A_1^{ij}A_1^{*ij}\right).
\eeq

If we consider the $SO(3)$ subgroup leaving invariant the vector
coordinates $4\ldots 8$, then the $SO(3)$ invariant scalars are
\bea
S_n{}^{\tt Q}{}_{\tt P}&=&\frac{1}{16}\Psi^{123n}_{\underline{(\alpha\beta)}}H^{\underline{(\alpha\beta)}}_{\tt B}\eta^{\tt BA} t_{\tt AP}{}^{\tt Q}\nonumber\\
C_n{}^{\tt Q}{}_{\tt P}&=&\frac{1}{16}\dot\Psi^{123n}_{\underline{(\dot\alpha\dot\beta)}}H^{\underline{(\dot\alpha\dot\beta)}}_{\tt B}\eta^{\tt BA} t_{\tt AP}{}^{\tt Q}\\
&&\mbox{where $n\in\{4,5,6,7,8\}$}\nonumber
\eea
and the task is now to evaluate $(\ref{potentialE7})$ for an arbitrary linear combination
\[
V\left(\exp \left(\sum_{n=4}^8 (s_n S_n+ c_n C_n)\right)\right)
\]
with ten parameters $s_n, c_n$. Although it is easily possible to
analytically exponentiate every single one of the (noncommuting)
generators $S_n, C_n$, this is not the case for an arbitrary linear
combination, hence it is necessary to find a different parametrization
of this space more amenable to an analytic treatment. It is mostly the
complexity of these other parametrizations which allow analytic
treatment that make the resulting formulae so complicated, thus it is
natural to ask the question whether there might be an alternative
approach to this class of problems that avoids this step entirely.

Here, we will make use of $\cV(SX)=U(S)\cV(X)U(S^{-1})$ with $S\in
su(8)$ and start with $\exp\left(\rho S_4\right)$ to which we apply an alternating
sequence of $SU(8)$ rotations with real and imaginary generators that
turn part of the $4_s$-component into the $4_c$ component, then part
of the $4_s$ component into the $5_s$ component, then $5_s\rightarrow
5_c$, $4_s\rightarrow 6_s$, etc. In particular, if we define
\bea
w_n&:=&s_n+i c_n\nonumber\\
R^{SO(8)\,a}{}_b(j,k,\alpha)&:=&\delta_{ab}+\delta_{aj}\delta_{kb}\,\sin\alpha-\delta_{aj}\delta_{kb}\,\sin\alpha\nonumber\\
&&+\left(\delta_{aj}\delta_{bj}+\delta_{ak}\delta_{bk}\right)\left(\cos\alpha-1\right)\\
R^{SU(8)\,a}{}_b(j,k,\alpha)&:=&\delta_{ab}+\delta_{aj}\delta_{bj}\left(e^{i\alpha}-1\right)+\delta_{ak}\delta_{bk}\left(e^{-i\alpha}-1\right)\nonumber
\eea
and promote $SU(8)$ rotations to $E_7$ via
\bea
R^{\tt P}{}_{\tt Q}&=&\frac{1}{2}\left(R^k{}_iR^l{}_j H^{ij}_{\underline{[ij]}} H_{kl}^{\underline{[kl]}}H^{\tt P}_{\underline{[kl]}^a}H^{\tt Q}_{\underline{[ij]}^a}\right.\nonumber\\
&&+\left.\bar R^k{}_i\bar R^l{}_j H^{ij}_{\underline{[ij]}} H_{kl}^{\underline{[kl]}}H^{\tt P}_{\underline{[kl]}^b}H^{\tt Q}_{\underline{[ij]}^b}\right)
\eea
then we have
\beq\label{e7potential10}
V\left(\exp \left(\sum_{n=4}^8 (s_n S_n+ c_n C_n)\right)\right)=V\left(RA\exp\left(\rho S_4\right)A^{-1}R^{-1}\right)
\eeq
where
\beq
\begin{array}{l}
\begin{array}{lclclclclcl}
\tilde\phi_{45}&:=&\frac{1}{2}\Arg\frac{w_5}{w_4}&\;&\tilde\omega_{45}&:=&\atan\frac{w_5e^{-i\tilde\phi_{45}}}{w_4e^{i\tilde\phi_{45}}}&\;&w_{45}&:=&\frac{w_4e^{i\tilde\phi_{45}}}{\cos(\tilde\omega_{45})}\\
\tilde\phi_{46}&:=&\frac{1}{2}\Arg\frac{w_6}{w_{45}}&\;&\tilde\omega_{46}&:=&\atan\frac{w_6e^{-i\tilde\phi_{46}}}{w_{45}e^{i\tilde\phi_{46}}}&\;&w_{46}&:=&\frac{w_{45}e^{i\tilde\phi_{46}}}{\cos(\tilde\omega_{46})}\\
\tilde\phi_{47}&:=&\frac{1}{2}\Arg\frac{w_7}{w_{46}}&\;&\tilde\omega_{47}&:=&\atan\frac{w_7e^{-i\tilde\phi_{47}}}{w_{46}e^{i\tilde\phi_{47}}}&\;&w_{47}&:=&\frac{w_{46}e^{i\tilde\phi_{47}}}{\cos(\tilde\omega_{47})}\\
\tilde\phi_{48}&:=&\frac{1}{2}\Arg\frac{w_8}{w_{47}}&\;&\tilde\omega_{48}&:=&\atan\frac{w_8e^{-i\tilde\phi_{48}}}{w_{47}e^{i\tilde\phi_{48}}}&\;&w_{48}&:=&\frac{w_{47}e^{i\tilde\phi_{48}}}{\cos(\tilde\omega_{48})}\\
\alpha&:=&\Arg w_{48}&&\rho&:=&|w_{48}|&&\\
\phi_{\scriptscriptstyle \square}&:=&-\tilde\phi_{\scriptscriptstyle \square}&&\omega_{\scriptscriptstyle \square}&:=&-\tilde\omega_{\scriptscriptstyle \square}
\end{array}\\
\\
R:=R^{SU(8)}(4,5,\phi_45)R^{SO(8)}(4,5,\omega_{45})R^{SU(8)}(4,6,\phi_{46})\times\\
\phantom{R:=}\times R^{SO(8)}(4,6,\omega_{46})R^{SU(8)}(4,7,\phi_{47})R^{SO(8)}(4,7,\omega_{47})\times\\
\phantom{R:=}\times R^{SU(8)}(4,8,\phi_{48})R^{SO(8)}(4,8,\omega_{48})\\
\\
W:=2\,[S_4,C_4]\\
A^{\tt P}{}_{\tt Q}:=\delta^{\tt P}{}_{\tt Q}-W^{\tt P}{}_{\tt Q}\sin(\alpha/2)+W^{\tt P}{}_{\tt R}W^{\tt R}{}_{\tt Q}\left(\cos(\alpha/2)-1\right).
\end{array}
\eeq

With this parametrization, the calculation of the potential is
straightforward, but still a solid computational challenge, even if we
make use of the observation that it is independent of $\alpha$. By
trial and error, the author found out that forming the tensors $A_1$
and $A_2$ is unproblematic, but even after reducing the maximal number
of variables which the \LambdaTensor{} package can handle from 28 to
14 (in order to halve memory requirements to store terms), the
calculation of their abs-squares, in particular the component
$A_1^{44}$, exceeds memory limitations of 32-bit computer
architectures. This step was performed by splitting the corresponding
difficult terms into four pieces each, doing multiplication
component-wise and writing out intermediate quantities to
disk.\footnote{The advantages to have direct access to low-level
  details of the implementation of symbolic algebra should be
  obvious.} Even then, peak memory requirements well exceed 1 GB of
RAM. The result is quite long, but numerical checks indicate its
correctness. To conserve it in printed form, it is given in appendix
\ref{d=4potential}, where a special notation is also introduced to
present it. Furthermore, it has been made available in electronic form
\cite{Fischbacher:4dpotential}.

One may well consider this potential as defining the present upper
limit of what may be done with the \LambdaTensor{} package with
reasonable effort. Since memory requirements are much more a problem
than run time here, it would be possible to add code to the package
which makes use of the experience gained in this calculation to
automatically handle such situations. Corresponding functionality will
be included in a new version of the \LambdaTensor{} package as soon as
some related nontrivial design decisions have been resolved.
Remarkably, despite its complexity, it is possible to translate this
formula to a machine code version which can be evaluated in a
sufficiently short time\footnote{roughly 1.2 ms on a 1.8 GHz
  Pentium-IV} and with sufficient accuracy to make a numerical search
for vacua feasible. So far, there is only numerical evidence for one
further nontrivial stationary point of this potential with a
cosmological constant close to $-14g^2$ which also shows up when
setting $\phi_{47}=\omega_{47}=\phi_{48}=\omega_{48}=0$ and thereby
restricting this potential further to the submanifold of $SO(3)\times
SO(2)$-invariant scalars. Since there is a known vacuum with
$SO(3)\times SO(3)$ symmetry and cosmological constant $-14g^2$, no
attempt was made so analyze this candidate for a vacuum any further.

\chapter[Symbolic Algebra]{Specialized High-Performance Symbolic Algebra}
\label{LambdaTensorCh}

\section{The \LambdaTensor{} package}

Due to considerable complexity of the group-theoretic calculations
involved, as well as due to the large number of different individual
possible gaugings to be considered, there is a strong incentive to
employ computer aid in the study of supergravity scalar potentials.
Since none of the readily available packages for symbolic algebra
turned out to be powerful enough to perform computations on the level
of complexity required here, a new tool had to be developed for
interactive work with large Lie algebras and Lie groups on the
symbolic as well as numeric level. This approach eventually led to the
implementation of the \LambdaTensor{} package, designed for efficient
symbolic and numeric calculations on sparse and nonsparse higher-rank
tensors, which finally was released as a library under a free software
license (version 2.1 of the GNU Lesser General Public License), since
the expectation is that other areas of research may as well benefit
from an efficient implementation of this functionality.

This chapter is intended to explain not only why \LambdaTensor{} does
exist at all, but also give reasons underlying the design decisions
that give it its present form, hence providing important background
information for users of \LambdaTensor{}.

Concerning the original formulation of the problem, there are two
complementary ways how such a package could be designed: either as a
tool to perform symbolic manipulations on the level of tensor
equations themselves, or as a tool to operate on explicit coordinate
instantiations of tensor equations. For \LambdaTensor, the latter
approach was chosen for a variety of reasons: first, as desirable as a
toolkit for computer-controlled interactive tensor equation
manipulation may be, one important drawback when it comes to the
application of such a framework to the problem at hand is that it
would {\em a priori} not be possible to switch over to numerical
counter-checks of results obtained on the purely symbolic level, or to
do a fully numeric search for stationary points in the
high-dimensional potentials of gauged extended supergravity theories
which would not be accessible at the symbolic level. Second, since the
most promising method available at present to systematically obtain
information about vacua of these theories is to consider the potential
on a subspace of the full scalar manifold which is invariant under a
subgroup of the gauge group, and since one is in particular interested
in making this investigation as exhaustive as possible by making the
invariant subspace as large, and hence the subgroup of the gauge group
as small as is feasible with available resources, the number of terms
in such a purely analytic approach is bound to explode, removing some
of the aesthetic as well as computational advantages of this route.

Furthermore, the change in perspective induced by the choice to focus
on explicit coordinate representations of generators of exceptional
groups turned out to be very fruitful for the solution of the mundane
sub-problem of finding exact analytic expressions for Lie group
elements obtained by exponentiation of certain generators of special
relevance, practically rendering this important step trivial. For
investigations of the potential of $N=8$, $D=4$ supergravity, this step
is greatly simplified by the peculiar $SU(8)$ sub-structure of the
fundamental ${\bf 56}$ of $E_{7(7)}$ available there, while for
$E_{8(8)}$, no corresponding construction seems to exist. Since no
attempt was made to understand how the underlying structure of the
exceptional Lie algebras involved simplifies exponentiation of
generators given explicitly in coordinate representation, this is (for
this work) perhaps to be considered a lucky coincidence.

\section[\LambdaTensor{} Design]{Design and implementation}

Judging from the design of similarly specialized packages for
computer-aided calculations, such as GAP, LiE, Form, R, SPSS, and the
like, one very common approach is to implement a standalone executable
program which employs an interpreter for a simplistic programming
language (frequently also usable interactively), providing commands
that wrap internal functionality implementing the relevant
algorithms. For \LambdaTensor, an approach fundamentally different from
this conventional one was chosen, which therefore should be justified.

The task to implement an interactively usable symbolic algebra package
asks for a run-time system which takes care of dynamic memory
management, provides an interactive top-level, and allows to define
new values and functions at run-time. It turns out that these
requirements which one would like to impose on a run-time system are
common enough to be relevant for a much wider class of highly dynamic
programs, and therefore the question arises whether it is feasible to
do an efficient and aggressively optimized implementation of a bare
core of such a programmable run-time system providing a high-quality
implementation of widely used functionality, with the idea in mind
that having such a standardized system available should take away the
burden of writers of dynamic programs to build one themselves as the
foundation for their work. Instead, the idea is to evolve an existing
core system by supplementing it with additional definitions.  Although
it is therefore of vital importance that this core subsystem is
extensible and programmable at the level of its implementation itself
(in contrast to systems that provide an extension language built on
top of them), it should be seen an an advantage if it provided
metasyntactic capabilities but hardly any syntax by itself; any
existing syntax imposes restrictions on the language the application
writer is going to build on top of it, and it can not be anticipated
(and should not be tried to) what form the final application language
is going to have.

Such a programmable efficiently implemented minimal core system is
just what a COMMON LISP system is intended to provide; the fact that
LISP syntax looks so different from almost anything else is readily
explained in this context: LISP is not intended to have any syntax;
instead, a LISP program essentially is the syntax tree which in other
languages is generated from code during the first compilation
step.

A decision was made not to supplement \LambdaTensor{} with an own
package-specific language, as for example
MAXIMA -- which is also LISP-based -- does, since this
would only have meant restricting functionality available to the end
user to a subset of LISP, while adding the burden for the user to
learn yet another language\footnote{not to speak of the author to
design and implement and document one -- which certainly is quite far
away from the original research topic of investigating the extremal
structure of gauged extended supergravity!}  which he eventually is
bound to drop again, since advanced users will most likely sooner or
later want to work with \LambdaTensor{} at the level it was written
itself. Hence, since there is no reason to assume LISP to be worse
than any proprietary limited language specific to only a single
application, the application and extension language for
\LambdaTensor{} is also LISP.

The immediate price that has to be paid by implementing a system like
\LambdaTensor{} by extending a COMMON LISP system, in particular CMU
Common LISP (CMUCL), is that due to CMUCL containing an optimizing
machine code compiler with debugger as well as a base set of further
LISP- and Unix-related definitions of considerable size, a running
\LambdaTensor{} process will typically occupy more than 40 MB of RAM.
(This, however, has to be seen in relation to the size of typical
intermediate values that appear in calculations one uses this package
for, usually some 100 MB.) Furthermore, and perhaps even much more
important, CMUCL is (with a few exceptions) only well supported on
x86-based free unix platforms (Linux and BSD), hence in particular not
available for some workstation or supercomputer architectures one
would like to run \LambdaTensor{} on. This is of importance, since x86
is a 32-bit architecture, meaning that one has to deal with an address
space limit of 4 GB. As there is no realistic problem-induced limit to
which supergravity potential calculations could be taken using the
present approach, technical limitations how far one can go will be
either induced by a memory or computation time barrier. Tests have
shown that for these calculations, memory limitations are much more
severe than calculation time limitations.

There is a variant of CMU CL, called SBCL, which (with minor
adjustments) can be used to run \LambdaTensor{} on non-x86 Unix
workstations, like PPC- or Alpha-based architectures, which might at a
first glance seem especially attractive, since the 64-bit Alpha
processor offers a 43-bit virtual address space. Unfortunately, SBCL
only runs as 32-bit application on all 64-bit architectures it has
been ported to, so nothing is gained here. In fact, to the present
author's best knowledge, there does not exist a freely available
sufficiently evolved COMMON LISP implementation using 64-bit
addressing at the time of this writing.  For the LISP derivative
Scheme, such solutions do exist, e.g. Marc Feely's
Gambit\cite{GambitScheme}. Since the \LambdaTensor{} code employs only
a comparatively small subset of the COMMON LISP language, which was
deliberately chosen to simplify semi-automatic porting to other
LISP-like systems, migration to Scheme may become an interesting
option in the future.

\section{Central Algorithms}

The \LambdaTensor{} package is by orders of magnitude more efficient
than general-purpose symbolic algebra packages would be for the
specialized task it was designed for.\footnote{Therefore, it was
  nominated as one of the best three contributions to the {\em Heinz
    Billing Award for the Advancement of Scientific Computation} in
  2002\cite{HeinzBilling}.}  Due to its novel ways to handle symbolic
algebra, we will give an overview over the central algorithms that
make it fast. This section should also help to clarify questions on
which optimizations are performed automatically by \LambdaTensor{} and
which are not. For more details, the reader should also consult
\cite{Fischbacher:2002hm}, the \LambdaTensor{} documentation, and the
LISP source.

As explained in \cite{Fischbacher:2002hm}, one fundamentally important
idea is to utilize the observation that tensors showing up in group
theory calculations frequently are very sparsely occupied -- for
example, in the conventions used here, structure constants of the
largest (248-dimensional) exceptional Lie group $E_8$
$f_{\mathcal{A}\mathcal{B}}{}^\mathcal{C}$ contain only $49\,440$ out
of $248^3=15\,252\,992$ nonzero entries -- and hence, we can make good
use of efficient implementations of abstract algorithms that can
handle sparsely occupied higher-rank tensors. Efficient code working
on sparse matrices is widely used and readily available; the
appropriate algorithms for handling higher-rank tensors are also quite
well-known, albeit in a very different context: relational databases.

In particular, at the level of explicit tensor entries, forming a
quantity like
\begin{equation}
        M_{abc}=N_{gha}P^{gh}{}_{bc}
\end{equation}
translates as follows into the language of relational databases (SQL syntax used here):
\begin{verbatim}
SELECT t1.index3 as index1,
       t2.index3 as index2,
       t2.index4 as index3,
       SUM(t1.val*t2.val)
   FROM tensor1 t1, tensor2 t2
   WHERE t1.index1=t2.index1 AND t1.index2=t2.index2
   GROUP BY t1.index3, t2.index3, t2.index4;
\end{verbatim}

Unfortunately, it is not feasible to just connect to an existing SQL
data\-base system (like PostgreSQL), create relations for tensors, and
use existing implementations of these algorithms by doing all the
calculations in the database, for various reasons. Besides
considerations concerning the efficiency of communication, and
considerable additional computational overhead due to data\-bases
having different aims, one major problem is that extending the
database system to abstract from the implementation of sum and product
here, as is necessary as soon as we want to work with data types not
natively supported by the database (which are frequently limited to
integers and floatingpoint numbers) would bring along too many
technical problems. Hence, what is required is a re-implementation of
the underlying database algorithms with numerical as well as symbolic tensor
computations as applications in mind. Furthermore, this implementation
has to be abstract enough to allow all relevant arithmetic operations
to be provided as parameters, so that one may switch between
approximate numerics, exact (i.e. rational number) numerics, and
symbolic calculations.\footnote{The ability to implement and use
arbitrary arithmetics on tensor entries has proven to be of great
value during the debugging phase of the symbolic algebras provided
within this package. For example, it is easy to (even automatically)
lift an existing implementation of arithmetic operations on symbolic
terms to an implementation working on pairs of terms and numerical
values of these terms for a given occupation of variables that signals
an error whenever a discrepancy between these values shows up.}

The underlying data structure chosen for the implementation of sparse
tensors is that of a multidimensional hash. Just as an array can be
regarded as a dictionary mapping every index from a given range of
natural numbers to a corresponding value with fast (i.e. independent
of the size of the array) lookups, a hash is a generalization of this
concept where the set of keys of the dictionary is not restricted to a
range of natural numbers, but may be an arbitrary set of values which
can effectively be compared for equality.
(In our case, a vector of indices.) In brief, the idea behind
implementations of hashing algorithms\footnote{see e.g. \cite{Knuth}}
in its simplest (one-dimensional) form is to make a naive
implementation of a dictionary which does lookups by traversing a list
of key-value pairs (and hence has $O(N)$ time complexity for lookups
in a dictionary of $N$ elements) fast by splitting this key-value list
into an array of such key-value lists of limited average length, where
the first lookup step consists of finding the right array entry
(called a {\em hash bucket}) holding the key-value list in which a
given key, if present in the hash, is bound to lie. This is done by
the help of a hash function which maps keys to bucket numbers. Such
hash functions should be easy to calculate and distribute keys evenly
between buckets. Whenever a new entry is made into a hash, its key is
also first mapped to a bucket number, and then, the corresponding
key-value-pair is either appended to the key-value-list in that
bucket, or merged with a previously existing entry for this
key. Should the average occupation of a hash grow beyond a given limit
(like, two key-value pairs per bucket), then a rehash is initiated,
which means that the underlying array is deleted and all the data it
contained is transferred to another array larger than the previous one
by a given factor. This process is invisible to the user of the hash data
structure. One-dimensional hashing is easily generalized
to the multi-dimensional case (using a higher-rank array of buckets),
which is not explained in detail here. (Readers wanting to know more
may find it instructive to use the LISP inspector\footnote{{\tt (inspect $\langle$TENSOR$\rangle$)}}
to get an impression how this is done in the underlying implementation.
At least, it should be noted here that although LISP has built-in
support for higher-rank arrays, it would do considerable harm to
efficiency to actually use these instead of implementing higher-rank arrays
on flattened rank-1 arrays, despite recommendations in \cite{cltl2}.
The reason is that LISP's method of array access for higher-rank arrays
whose rank is not known at compile time (as in our case)
via {\tt (apply \#'aref \dots)}, resp.
{\tt (setf (apply \#'aref \dots) \dots)} would cause an intolerable
amount of unnecessary consing.)


While for usual applications, the effectivity of hashing will increase
with increasing number of buckets (and hence, decreasing average
occupation), one frequent operation in database-related applications
(like tensor multiplication in the example above) is to iterate over
all entries of a hash. This, of course, works best if the number of
unoccupied hash buckets is small, so creating overly sparse hashes
also hinders performance. Note that version 1.0 of \LambdaTensor{} does
not shrink hashes automatically for which occupation density falls
below a given threshold, so e.g. calculations with an empty (i.e. all
entries zero) tensor which was created from a sparsely occupied one by
setting entries to zero will at the average not be much faster than
calculations with the original sparsely occupied tensor. (In most
cases, this is not an issue.) If more than a certain share of all
possible entries of a sparse tensor are set to nonzero values
(currently, about one-quarter), \LambdaTensor{} will internally rehash
the tensor to a conventional nonsparse representation. 

The tensor index hash function must map a dimension-$N$ index from the
range $0\ldots (N-1)$ to the range of hash buckets.%
\footnote{In LISP, it is conventional to start counting of indices
at zero, just as in C. One important advantage of such a convention
is that linear indices corresponding to flattened versions of
higher-rank arrays are more naturally expressed in the original indices
than with the FORTRAN convention to start array index counting at 1.
}

For \LambdaTensor{}, the hash function currently employed
is simple integer modulus. At a first glance,
this may seem to be a very dangerous design decision,
since a good hash function should try to distribute entries
evenly among buckets, and therefore apply some perturbing
operations on the hash key before taking the modulus.
Here, the idea is that in most cases, group theory itself
will take care of the task of even distribution of elements
among buckets (since there usually are no preferred coordinates,
or obvious special relations causing hash bucket clashes).%
\footnote{Note that simple integer modulus is also the hash function
for integer numbers in CMU CL and GCL, though not in CLISP.}
In fact, looking in detail at a typical higher-rank tensor in the
current implementation, like the $49\,440$ nonzero entries of the
structure constants of $E_{8(8)}$ {\tt e8-fabc},%
\footnote{Using package-private internal functions, this information can be extracted via\hfill
{\tt (array-dimensions (lambdatensor::sp-array-data e8-fabc))} and\hfill
{\tt (sort (seq-statistics (map 'simple-array \#'length (lambdatensor::sp-array-linear-data e8-fabc))) \#'< :key \#'car)}
}
one discovers that these are stored in a sparse
$41\times41\times27$ array with hash bucket occupation statistics
$0_{\times15948}\,1_{\times 15740}\,2_{\times9143}\,3_{\times3230}\,4_{\times1019}\,5_{\times222}\,6_{\times 63}\,7_{\times 16}\,8_{\times 6}$,
which is reasonably close to a typical random distribution on available hash buckets.%
\footnote{Things look worse for tensors like {\tt so8-sigma-ijkl-ab} or {\tt epsilon8},
but this may in part be traced back to the small index range in every coordinate,
which is supposed to cause major distortions for every hash function mapping
coordinates to a smaller range. Still, the quality of this naive
and easy to calculate hash function turns out to be sufficient
even in these cases, although there is clearly room for improvement.}

Since the sparse array handling part of \LambdaTensor{} has to deal
with sparse and nonsparse arrays without exposing this to the user, in
particular provide tensor arithmetic operations for the cases
nonsparse-nonsparse, nonsparse-sparse and sparse-sparse, and
furthermore supports specialized efficient implementations for certain
types of sparse and nonsparse tensors which employ machine arithmetics
where available (i.e. for {\tt double-float} and {\tt (complex
double-float)} tensors), this module contains a considerable amount of
code to cover all the possible cases resulting from the product of
these distinctions.

One further notable feature of \LambdaTensor{} is that in products of
multiple tensors, the order in which multiplications are executed is
scheduled in such a way as to minimize the total number of
multiplications to be performed. Let us briefly illustrate this in a
very simplistic matrix example: if $A$ is a nonsparse $10\times2$
matrix, $B$ a nonsparse $2\times4$ matrix, and $C$ a nonsparse
$4\times8$ matrix, then the $10\times8$ matrix $ABC$ can be calculated
by performing the first or second product first. Calculating $(AB)C$
will require $10\cdot2\cdot4$ multiplications for the calculation of
the $10\times4$ matrix $AB$ and $10\cdot4\cdot8$ multiplications to
form the other matrix product, for a total of 400
multiplications. Likewise, calculation of $A(BC)$ only requires
$2\cdot4\cdot8+10\cdot2\cdot8=224$ multiplications. \LambdaTensor{} is
aware of such optimizations even for products of multiple tensors with
arbitrary contractions between tensors and will automatically try to
choose that particular multiplication order which results in the
smallest number of arithmetic operations. For sparse tensors, this is
done heuristically by assuming independent equidistribution of tensor
entries. (Should this heuristic approach fail due to non-independence
of the distribution of entries and produce overly expensive
calculations, one may always resort to determining multiplication
order by splitting such products manually.)

Even with efficient tensor arithmetics available, one further problem
is the symbolic complexity of analytic expressions involved in
supergravity potential calculations. The widespread approach to first
introduce coordinates on certain special submanifolds of the symmetric
space $E_{8(8)}/SO(16)$ by a procedure reminiscent of Euler angle
parametrizations of $SO(3)$ inevitably generates very complicated
analytic expressions for the $\mathcal{V}$-matrix and all further
intermediate quantities derived from it whose complexity will explode
with the number of (compact or noncompact) angular
coordinates. Therefore, it is also of vital importance to have an
efficient machine representation of such terms in order to make such
calculations feasible on manifolds with enough dimensions to produce
interesting results (i.e. new vacua).

\begin{figure}
\includegraphics{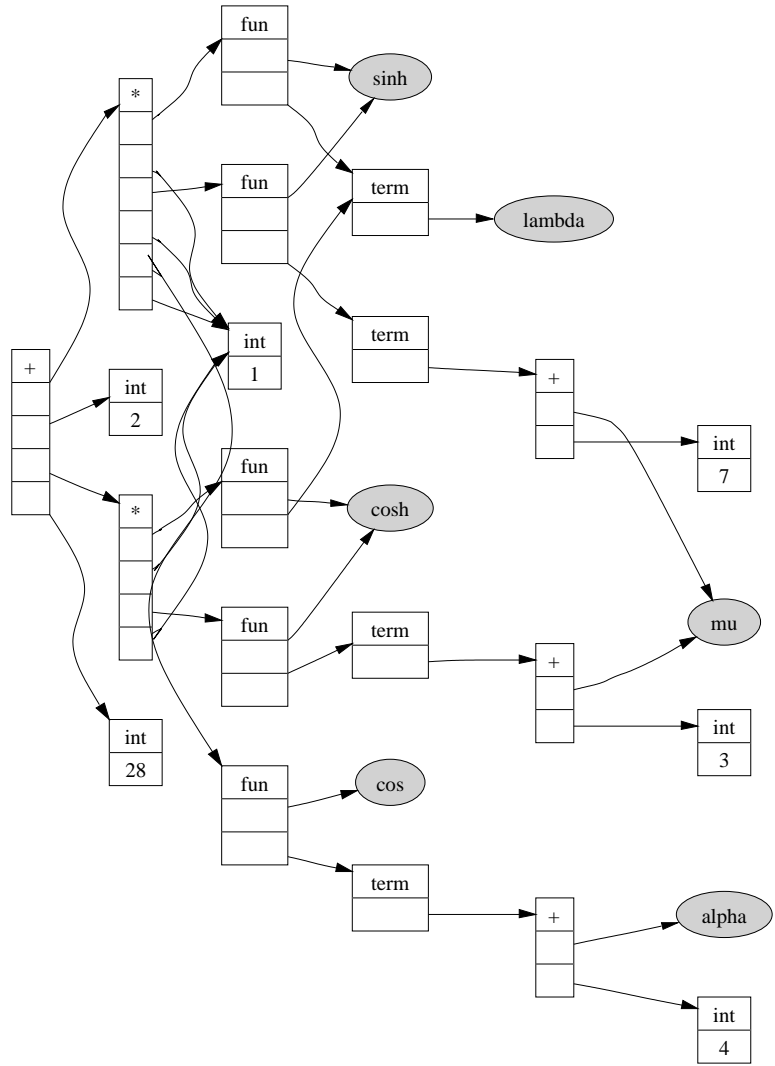}
\caption{Maple's internal term representation (example)}
\label{mapleterm}
\end{figure}

The conventional way to represent an analytic expression, in this
particular example Maple's internal representation of the term
\begin{equation}\label{two_example_terms}
2\,\sinh\lambda\,\sinh7\mu\,\cos4\alpha+28\,\cosh\lambda\cosh3\mu,
\end{equation}
is shown in figure\footnote{Diagrams have been generated with {\tt graphviz} (Trademark by AT\&T).}
$(\ref{mapleterm})$.\footnote{Cf. \cite{AndreHeckMaple} for details
concerning Maple's term representations.}  The general
underlying idea here is to use a representation general enough to
handle arbitrary terms, but to try to save space by not duplicating
subterms.\footnote{One way to implement this is to use a hash of weak
pointers on all currently known terms to map a newly generated
expression to a pre-existing memory representation of that
expression.} The major drawback of such general representations of
terms is that for specialized applications, in particular for the
problem at hand, they tend to conceal some possible simplifications or
reductions which take a more natural form in a term representation
that fits the application. In this particular case, all that a
general-purpose symbolic algebra package like Maple can do is to
factor out common subterms or perform partial reductions by applying
simple trigonometric identities.\footnote{Indeed, only a very small
subset of all possible reductions is found and applied, as can be seen
by transforming a potential like $(\ref{g2singlets})$ to expanded
exponential form via {\tt simplify(normal(expand(convert(Phi,exp))))}
and then trying to reduce this to an expression of comparable length
using Maple's builtin simplification functions.}

\begin{figure}
\includegraphics{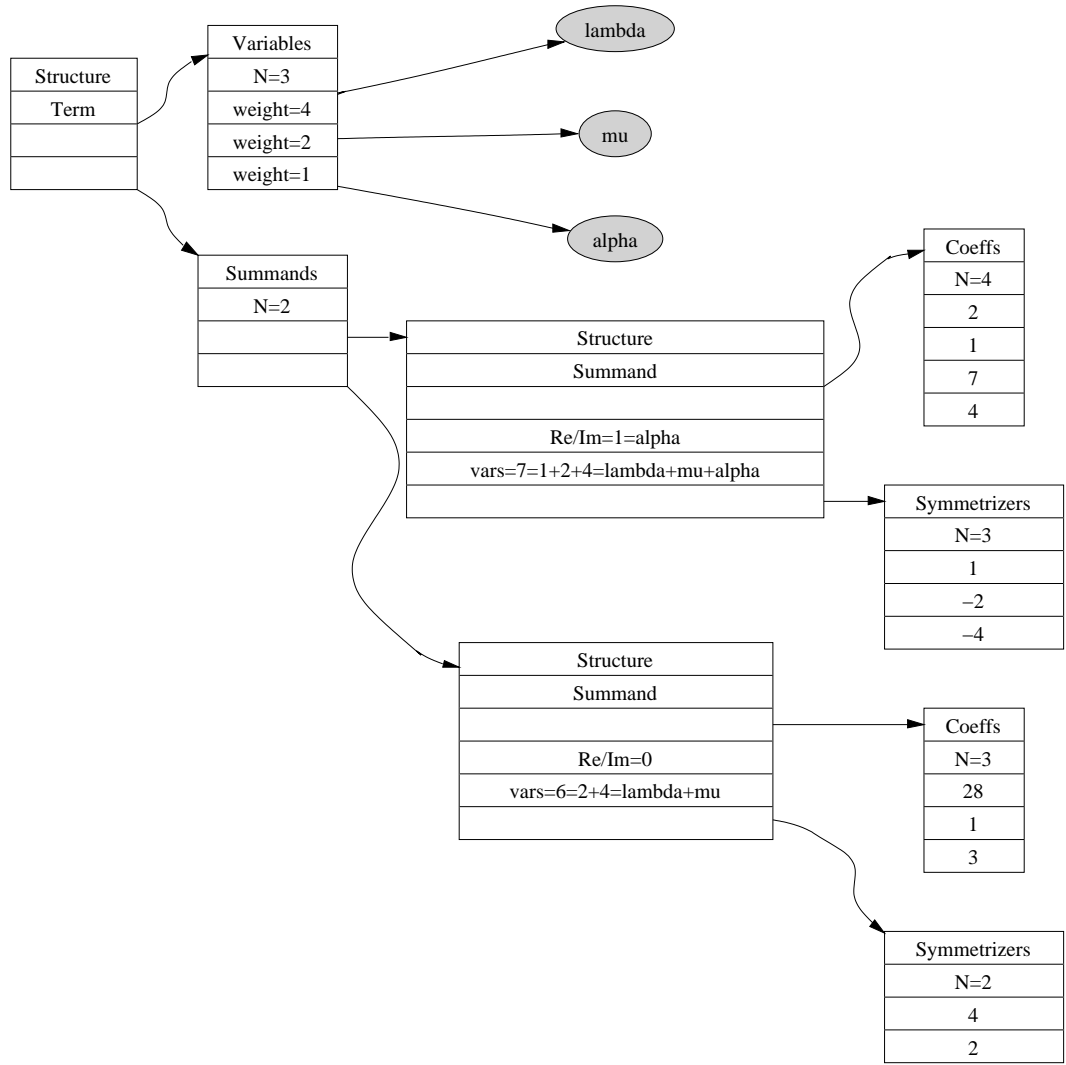}
\caption{Specialized trigonometric term representation (example)}
\label{poexpterm}
\end{figure}

Concerning trigonometric manipulations, the most useful term
representation for doing calculations would be
\begin{equation}
\langle{\rm Term}\rangle=\sum_j k_j\exp\left(\sum_{j_k} c_{j_k} v_{j_k}\right)
\end{equation}
where the $k_j$ and $c_{j_k}$ are (complex) rational numbers and
$v_{j_k}$ variables. Although calculations involving only terms of
this form are conceptually simple, the big drawback of this
representation is the large number of individual summands; even an
expression as simple as $(\ref{two_example_terms})$ would consist of
$12$ different summands if written in this form, which come in two
groups of almost identical summands that differ only in signs of
coefficients. Hence, the idea suggests itself to use this explicit
exponential form but store terms in a packed format where every such
group of summands that are identical up to sign flips is represented
by the first term in a suitable lexicographical ordering and
information about additional symmetrizations and anti-symmetrizations.
(As a further optimization, one observes that coefficients are either
real and rational, or purely imaginary and rational, so it suffices to
store only a real number and one bit of information whether the
corresponding coefficient is real or imaginary.) Such a term
representation turns out to be of remarkably compact form, yet contain
enough information to do efficient calculations that also yield terms
of such compact form -- which conventional symbolic algebra cannot do
with similar efficiency. The internal representation of the term
$(\ref{two_example_terms})$ in this format is shown in figure
$(\ref{poexpterm})$. In both memory representation diagrams (on 32-bit
architectures), one rectangular box corresponds to one 32-bit
cell.\footnote{This will also be changed in subsequent versions,
which will only allow a maximum of $14$ different angular variables,
but considerably reduce memory requirements.}

Addition of such terms is straightforward to implement. Multiplication
is much more involved, since product forming of summands with
overlapping symmetrizers involves partial unpacking of one factor. The
most difficult step for multiplication is the combination of resulting
contributions into packed terms. Indeed, there are some situations
involving symmetrizers that cover multiple variables at once in which
possible symmetrizers overlap in such an unfortunate way that the
optimal form can not be found. Experiments have shown that this rarely
happens for the calculation of potentials of four variables, but may
become more common the larger the number of variables. Nevertheless,
degradation of this method due to this effect is a minor issue in most
applications.

Further attempts to use memoization or subterm identification as
performed by Maple with this new scheme to implement symbolic algebra
have not resulted in noticeable improvements in performance or memory
requirements so far.

\vfill

\section[\LambdaTensor{} user interface]{Provided functionality and the user interface}

The main user interface to multilinear algebra, given by the function
{\tt sp-x}, evolved into its present form through a process of
experimentation with a considerable number of different approaches.
The basic idea is to subsume the elementary operations of index
re-ordering, tensor contraction and tensor product forming in one
function that resembles conventional formalism as closely as possible.
Since the underlying tensor operations themselves typically require a
large number of machine operations to be executed, it is perhaps
affordable here to add an overhead of a few extra operations to
provide a decent interface. This approach turned out to be much more
convenient to use than the one implemented by Maple, which only
provides bare operations like tensor contraction and product forming
of two (not multiple) tensors and in which indices have to be labeled
by position, not by symbolic names. In contrast, \LambdaTensor{} index
names can be arbitrary LISP values which are compared for equality by
EQUALP.\footnote{Index names of the form {\tt (cons :fix
    $\langle$number$\rangle$)} play a special role and denote indices
  fixed to a number.}

Although it is not difficult to implement the index matching code in
such a way that upper and lower indices are discerned and summation
can only take place between an upper and a lower index, a deliberate
choice was made not to do so, in part due to the prominent role of the
special orthogonal groups in supergravity, for which there is no need
to discern between upper and lower indices (and hence, doing so would
only introduce superfluous $\delta^{ij}$-metric tensors both in
formulae and in code), and in part since it is more advantageous in
this case to build code whose main purpose is to catch user errors on
top of a library providing bare functionality than into it. (Making it
mandatory would only limit the potential given to users that know how
and when to deliberately break the rules.)

It would clearly be very desirable to have a close one-to-one
correspondence between formulae in a machine representation suited for
calculations and conventional notation in physics (especially since
manual conversion of a large number of formulae can be a considerable
source of errors), and one is tempted to try to use LISP's
forementioned metasyntactic capabilities to implement such a
`language' that resembles conventional formalism as close as possible,
but despite considerable effort, any previous attempts in that
direction have not produced satisfactory results so far. The
underlying problem seems to be that casting the somewhat casual
practice of omitting simple embedding tensors by the introduction of
(sequences of) index splitting conventions into the framework of a set
of strict rules suited for a computer so far only produced very
elaborate rulesets with many special cases that turned out to be
difficult to handle and by far not as convenient as usual physical
notation.

\section{Future developments}

Since the first public release of this package
\cite{Fischbacher:2002hm}, this package has considerably evolved.
Besides a few bug fixes to the documentation and functions of minor
importance that did not endanger the validity of any of the results
obtained with this package, as well as necessary simple improvements
in a few places like the serializer, a considerable amount of new
functionality has been implemented that will be part of the next
release. One important improvement is the extension of LISP number
arithmetics to finite-dimensional field extensions (or even
nondivision algebras) over the rational numbers. It is a happy
coincidence that a lot of data can be extracted from the $D=3$ maximal
gauged supergravity potentials using almost exclusively rational
arithmetics; there nevertheless are also quite some cases where
rational arithmetics is not sufficient, and having access to
a direct implementation of finite-dimensional algebras over
$\mathbb{Q}$ that are specified via a multiplication table
can for some tasks circumvent common problems of conventional symbolic
algebra.\footnote{For example, MapleVR5.1 does not automatically
simplify $\frac{97+56\,\sqrt{3}}{\left(2+\sqrt{3}\right)^4}$ to $1$.
Even making the denominator a rational number only produces
$-\left(97+56\,\sqrt{3}\right)\left(-97+56\,\sqrt{3}\right)$.
While one may get rid of such artefacts by applying
{\tt x$\rightarrow$expand(rationalize(x))} 
or {\tt radnormal} from the library with the same name
to end results, this default behaviour certainly does not help to
reduce expression swell in intermediate quantities.}

In a certain sense complementary to the large sparse tensor
functionality provided by the initial release is the group-theoretic
approach based on roots and weight vectors that is implemented by LiE
\cite{LeCoLi92}; Since it is occasionally very useful to have this
functionality directly available when working with explicit tensors,
and not only via a detour through another program, the next version
will also implement more abstract group theory.\footnote{A strong
incentive to implement some important group-theoretic algorithms was
given by research that culminated in a paper with H. Nicolai
on the structure of $E_{10}$ and $E_{11}$ \cite{Nicolai:2003fw}.}

As was already mentioned, the next release will change the internal
representation of trigonometric term arithmetics to a much denser
format, limiting the number of different names for angular variables
in a calculation to $14$; this limit is perhaps not too unreasonable.
(It can be changed back if necessary.)

On the more experimental side, the next release will also contain
first rudimentary support for distributed computing where asynchronous
I/O is used to broadcast and receive calculation requests and results
via a TCP stream socket.

Finally, the installation procedure for two major commercial Linux
distributions which do not use the Debian package format will be
simplified.

\chapter{Conclusion and Outlook}
\label{chConclusion}

As the analysis of a few example cases with interesting structure has
shown, the tools presented in this thesis greatly simplify the task of
the determination of nontrivial stationary points of gauged extended
supergravity theories. The underlying mathematical ideas of the
approach taken here (i.e. restriction of the potential to maximal
submanifolds of the whole nonlinear space of scalars which are
invariant under some given subgroup of the gauge group) have been in
use for many years and still are considered the most promising ansatz
to at least obtain information about vacua with a certain amount of
unbroken gauge symmetry. By employing and inventing new methods in
symbolic algebra, it was possible to take this approach to previously
unreached heights, concerning both the study of a large variety of
different related models in a reasonable amount of time and the level
of detail to which these investigations can be carried.

On the computational side of this work, three major advances conspire
to make this technological jump possible: first, a reasonably
efficient implementation of multilinear algebra on sparsely occupied
tensors of higher rank and large dimension, necessary to effectively
handle explicit realisations of exceptional Lie groups (not algebras).
This part is quite straightforward, since the underlying algorithms by
now have been known for decades in the context of relational
databases, only that an implementation of these algorithms with group
theory as an application in mind has to the author's best knowledge
not been available so far. What should be considered as a new
approach, however, is to use $\lambda$ abstraction to fully
parametrize the database part by the implementation of arithmetics on
tensor entries, allowing maximal flexibility in the application of
this framework. Second, a highly memory efficient problem specific
encoding of symbolic expressions that appear in these calculations as
tensor entries. This is noteworthy since here it is possible for terms
of sufficient complexity to outperform conventional methods for the
memory representation of symbolic expressions by more than one order
of magnitude. Third, related to the novel problem-specific tight
encoding of information carried by symbolic expressions, term algebra
can be implemented in such a way that far more simplifying
trigonometric identities are discovered than with a conventional
approach. This greatly helps in reducing intermediate expression
swell.\footnote{To give an explicit example, the calculations presented in
\cite{Fischbacher:2002hg} originally were performed by employing
a term representation comparable to that used by conventional
symbolic algebra packages, but with additional provisions to recognize
some types of non-local reductions of summands typically not performed by
usual symbolic algebra systems. The original form of the five-parameter
potential as it dropped out of the tensor calculation filled 475 pages,
but could then in a separate step be reduced down to just one single
page\cite{Fischbacher:billing2002}. Due to exponential swell in
expression complexity, calculations with eight- or even nine-parameter
potentials, as presented here, would have been entirely impossible
with such an `usual' implementation of term algebra.
On the other hand, it is hard to imagine how far out of reach
of conventional symbolic algebra a potential like $(\ref{potentialE7})$
which we obtained by these novel methods and which would fill
well over 200 pages in a less compact notation is.}

The algorithmic tools developed for this work are probably at present
the most efficient ones available for working with explicit coordinate
representations of large Lie groups; since they might be useful in a
much broader context, they have been made publicly available in source
form under a free software license \cite{Fischbacher:2002hm}.

As is exemplified by a dozen new possible vacua that have been found
in this work, the extremal structure of gauged three-dimensional
supergravities is far richer than that of any higher-dimensional
supergravity model. It should be emphasized that these models are
likely to still have many more vacua than the ones given here. In
particular, we mostly have been concerned with vacua on submanifolds
that are invariant under subgroups which are embedded into $\Eee$ in a
very simple fashion. In particular, we did not consider a single case
here where the gauge group $SO(8)\times SO(8)$ is broken down to a
diagonal subgroup whose definition makes use of an extra triality
rotation. Also for the four- and five-dimensional case, a much broader
investigation would be possible. Concerning especially the
four-dimensional case, there is an embedding of $SO(3)$ into $SO(8)$
which gives rise to a $20$-dimensional invariant submanifold of the
scalar manifold - certainly too large for an analytic treatment.
Nevertheless, a numerical search for vacua which employ direct
numerical exponentiation of $56\times 56$ matrices should be feasible
in a highly distributed environment (maybe using computer power from
volunteers on the internet). Another question which has not been
answered so far that might be solved by a computer-aided exhaustive
search is whether there are any semisimple nonmaximal subgroups of
$\Eee$ that can be promoted to gauge groups.

One of the big pending problems is to clarify the relation between the
$N=16$ $D=3$ gauged supergravity models and higher-dimensional
supergravity; at present one can only speculate whether this might
lead to a theory beyond eleven-dimensional supergravity. The $N=16$
models may play a role in the supergravity description of matrix
string theories \cite{Itzhaki:1998dd,Morales:2002ys}; this has to be
clarified in the future. In particular, not much is known about the
(super)conformal theories corresponding to the AdS solutions of gauged
$N=16$ supergravity.

\appendix
\chapter[Potentials]{Potentials too lengthy to be given in the main text}
\label{chPotentials}

\section{The $SO(6)_{\rm diag}$ invariant submanifold}

For the gauge group $SO(6,2)\times SO(6,2)$,
the potential corresponding to the parametrization
$(\ref{so6-sl3xsl2})$ reads:
{\small
\setlongtables
\begin{longtable}{lll}\label{potential_so6diag_so62}
$-8g^{-2}V$&$=$&\hfill
\\
&&$27+3\,\cosh(4\,z)+3\,\cosh(4\,z)\,\cos(2\,r_2)-3\,\cosh(4\,z)\,\cos(2\,r_1)$\\
&&$-3\,\cosh(4\,z)\,\cos(2\,r_1)\,\cos(2\,r_2)+\frac{1}{4}\,\cosh(4\,s)$\\
&&$+\frac{1}{4}\,\cosh(4\,s)\,\cos(2\,r_2)-\frac{1}{4}\,\cosh(4\,s)\,\cos(2\,r_1)$\\
&&$-\frac{1}{4}\,\cosh(4\,s)\,\cos(2\,r_1)\,\cos(2\,r_2)+9\,\cosh(2\,s)\,\cosh(2\,z)$\\
&&$+\frac{3}{4}\,\cosh(2\,s)\,\cosh(6\,z)-3\,\cos(2\,r_3)\,\sinh(2\,z)\,\sinh(2\,s)$\\
&&$+\frac{1}{4}\,\cos(2\,r_3)\,\sinh(6\,z)\,\sinh(2\,s)$\\
&&$-3\,\cosh(2\,s)\,\cosh(2\,z)\,\cos(2\,r_2)-\frac{1}{4}\,\cosh(2\,s)\,\cosh(6\,z)\,\cos(2\,r_2)$\\
\end{longtable}
}

If we use the same parametrization for the $SO(7,1)\times SO(7,1)$ gauged theory, we get:
{\small
\setlongtables
\begin{longtable}{lll}\label{potential_so6diag_so71}
$-8g^{-2}V$&$=$&\hfill
\\
&&$\frac{1663}{64}\,+\frac{43}{64}\,\cos(2\,r_5)+\frac{9}{64}\,\cos(4\,r_3)$\\
&&$-\frac{3}{64}\,\cos(4\,r_3)\,\cos(2\,r_5)-\frac{3}{16}\,\cos(2\,r_2)$\\
&&$+\frac{21}{64}\,\cos(4\,r_2)+\frac{1}{16}\,\cos(2\,r_2)\,\cos(2\,r_5)$\\
&&$-\frac{7}{64}\,\cos(4\,r_2)\,\cos(2\,r_5)+\frac{3}{16}\,\cos(2\,r_2)\,\cos(4\,r_3)$\\
&&$+\frac{3}{64}\,\cos(4\,r_2)\,\cos(4\,r_3)-\frac{1}{16}\,\cos(2\,r_2)\,\cos(4\,r_3)\,\cos(2\,r_5)$\\
&&$-\frac{1}{64}\,\cos(4\,r_2)\,\cos(4\,r_3)\,\cos(2\,r_5)-\frac{9}{64}\,\cos(2\,r_1)$\\
&&$+\frac{3}{64}\,\cos(2\,r_1)\,\cos(2\,r_5)-\frac{15}{64}\,\cos(2\,r_1)\,\cos(4\,r_3)$\\
\end{longtable}
}

\section{The $\left(SO(5)\times SO(2)\right)_{\rm diag}$ invariant submanifold}

For the gauge group $SO(5,3)\times SO(5,3)$,
the potential corresponding to the parametrization
$(\ref{so5xso2param})$ reads:
{\small
\setlongtables
\begin{longtable}{l}\label{potential_so5diag_so53}
$-8g^{-2}V$=\hfill
\\
$23-5\,\cosh(\frac{12}{5}\,v)-\frac{15}{2}\,\cosh(\frac{2}{5}\,v+4\,z)+\frac{15}{8}\,\cosh(\frac{4}{5}\,v-4\,z)$\\
$-\frac{5}{4}\,\cosh(\frac{8}{5}\,v+4\,z)+\frac{3}{2}\,\cosh(2\,v-4\,z)-\frac{1}{8}\,\cosh(4\,v+4\,z)$\\
$+\frac{5}{2}\,\cosh(\frac{2}{5}\,v+4\,z)\,\cos(2\,r_2)+\frac{15}{8}\,\cosh(\frac{4}{5}\,v-4\,z)\,\cos(2\,r_2)$\\
$-\frac{5}{4}\,\cosh(\frac{8}{5}\,v+4\,z)\,\cos(2\,r_2)-\frac{1}{2}\,\cosh(2\,v-4\,z)\,\cos(2\,r_2)$\\
$-\frac{1}{8}\,\cosh(4\,v+4\,z)\,\cos(2\,r_2)+\frac{5}{2}\,\cosh(\frac{2}{5}\,v+4\,z)\,\cos(2\,r_1)$\\
$+\frac{15}{8}\,\cosh(\frac{4}{5}\,v-4\,z)\,\cos(2\,r_1)-\frac{5}{4}\,\cosh(\frac{8}{5}\,v+4\,z)\,\cos(2\,r_1)$\\
$-\frac{1}{2}\,\cosh(2\,v-4\,z)\,\cos(2\,r_1)-\frac{1}{8}\,\cosh(4\,v+4\,z)\,\cos(2\,r_1)$\\
\end{longtable}
}

We skip the case of $SO(6,2)\times SO(6,2)$ gauging, since with this
parametrization, $SO(6)$ would be implemented on the spinor indices
$1,2,3,4,5,8$, and this relabelling is a bit inconvenient. If we use this
parametrization with the $SO(7,1)\times SO(7,1)$ gauging, we get:
{\small
\setlongtables
\begin{longtable}{l}\label{potential_so5diag_so71}
$-8g^{-2}V$=\hfill
\\
$23-5\,\cosh(\frac{12}{5}\,v)+\frac{15}{2}\,\cosh(\frac{2}{5}\,v+4\,z)$\\
$+\frac{15}{8}\,\cosh(\frac{4}{5}\,v-4\,z)-\frac{5}{4}\,\cosh(\frac{8}{5}\,v+4\,z)$\\
$-\frac{3}{2}\,\cosh(2\,v-4\,z)-\frac{1}{8}\,\cosh(4\,v+4\,z)$\\
$-\frac{5}{2}\,\cosh(\frac{2}{5}\,v+4\,z)\,\cos(2\,r_2)$\\
$+\frac{15}{8}\,\cosh(\frac{4}{5}\,v-4\,z)\,\cos(2\,r_2)$\\
$-\frac{5}{4}\,\cosh(\frac{8}{5}\,v+4\,z)\,\cos(2\,r_2)+\frac{1}{2}\,\cosh(2\,v-4\,z)\,\cos(2\,r_2)$\\
$-\frac{1}{8}\,\cosh(4\,v+4\,z)\,\cos(2\,r_2)-\frac{5}{2}\,\cosh(\frac{2}{5}\,v+4\,z)\,\cos(2\,r_1)$\\
\end{longtable}
}
Finally, for the compact gauge group $SO(8)\times SO(8)$, the potential reads:
{\small
\setlongtables
\begin{longtable}{l}\label{potential_so5diag}
$-8g^{-2}V$=\hfill
\\
$23+5\,\cosh(\frac{12}{5}\,v)+\frac{15}{2}\,\cosh(\frac{2}{5}\,v+4\,z)$\\
$+\frac{15}{8}\,\cosh(\frac{4}{5}\,v-4\,z)+\frac{5}{4}\,\cosh(\frac{8}{5}\,v+4\,z)+\frac{3}{2}\,\cosh(2\,v-4\,z)-\frac{1}{8}\,\cosh(4\,v+4\,z)$\\
$-\frac{5}{2}\,\cosh(\frac{2}{5}\,v+4\,z)\,\cos(2\,r_2)+\frac{15}{8}\,\cosh(\frac{4}{5}\,v-4\,z)\,\cos(2\,r_2)$\\
$+\frac{5}{4}\,\cosh(\frac{8}{5}\,v+4\,z)\,\cos(2\,r_2)-\frac{1}{2}\,\cosh(2\,v-4\,z)\,\cos(2\,r_2)$\\
$-\frac{1}{8}\,\cosh(4\,v+4\,z)\,\cos(2\,r_2)-\frac{5}{2}\,\cosh(\frac{2}{5}\,v+4\,z)\,\cos(2\,r_1)$\\
$+\frac{15}{8}\,\cosh(\frac{4}{5}\,v-4\,z)\,\cos(2\,r_1)+\frac{5}{4}\,\cosh(\frac{8}{5}\,v+4\,z)\,\cos(2\,r_1)$\\
$-\frac{1}{2}\,\cosh(2\,v-4\,z)\,\cos(2\,r_1)-\frac{1}{8}\,\cosh(4\,v+4\,z)\,\cos(2\,r_1)$\\
\end{longtable}
}

\section{The $SU(3)\times SU(3)$ invariant submanifold}

The parametrization $(\ref{su3xsu3param})$ yields for the 
$G_{2(-14)}\times F_{4(-20)}$ gauged model the following
potential:
{\small
\setlongtables
\begin{longtable}{l}\label{potential_su3su3_g2xf4}
$-8g^{-2}V$=\hfill
\\
$\frac{24125}{2048}\,+\frac{9}{2048}\,\cos(8\,r_5)+\frac{9}{2048}\,\cos(8\,r_2)+\frac{115}{128}\,\sin(4\,r_5)\,\sin(4\,r_2)$\\
$-\frac{27}{2048}\,\cos(8\,r_2)\,\cos(8\,r_5)-\frac{1}{64}\,\cos(4\,r_2)\,\cos(4\,r_3-4\,r_6)\,\cos(4\,r_5)$\\
$-\frac{9}{2048}\,\cos(8\,r_1-8\,r_4)-\frac{9}{2048}\,\cos(8\,r_1-8\,r_4)\,\cos(8\,r_5)$\\
$-\frac{1}{64}\,\cos(4\,r_1-4\,r_4)\,\cos(4\,r_3-4\,r_6)+\frac{1}{64}\,\sin(4\,r_5)\,\sin(4\,r_3-4\,r_6)\,\sin(4\,r_1-4\,r_4)$\\
$-\frac{9}{2048}\,\cos(8\,r_1-8\,r_4)\,\cos(8\,r_2)-\frac{9}{512}\,\cos(4\,r_1-4\,r_4)\,\sin(8\,r_5)\,\sin(8\,r_2)$\\
$+\frac{115}{128}\,\cos(4\,r_1-4\,r_4)\,\cos(4\,r_2)\,\cos(4\,r_5)-\frac{9}{2048}\,\cos(8\,r_1-8\,r_4)\,\cos(8\,r_2)\,\cos(8\,r_5)$\\
$+\frac{1}{64}\,\sin(4\,r_3-4\,r_6)\,\sin(4\,r_2)\,\sin(4\,r_1-4\,r_4)$\\
\end{longtable}
}

Using this parametrization with the embedding tensor of $SO(8)\times SO(8)$ gives:
{\small
\setlongtables
\begin{longtable}{l}\label{potential_su3su3}
$-8g^{-2}V$=\hfill
\\
$\frac{8391}{256}+\frac{39}{256}\,\cos(8\,r_5)+\frac{39}{256}\,\cos(8\,r_4)+\frac{39}{256}\,\cos(8\,r_4)\,\cos(8\,r_5)$\\
$+\frac{39}{256}\,\cos(8\,r_2)-\frac{9}{256}\,\cos(8\,r_2)\,\cos(8\,r_5)-\frac{9}{256}\,\cos(8\,r_2)\,\cos(8\,r_4)$\\
$-\frac{9}{256}\,\cos(8\,r_2)\,\cos(8\,r_4)\,\cos(8\,r_5)+\frac{39}{256}\,\cos(8\,r_1)$\\
$-\frac{9}{256}\,\cos(8\,r_1)\,\cos(8\,r_5)-\frac{9}{256}\,\cos(8\,r_1)\,\cos(8\,r_4)$\\
$-\frac{9}{256}\,\cos(8\,r_1)\,\cos(8\,r_4)\,\cos(8\,r_5)+\frac{39}{256}\,\cos(8\,r_1)\,\cos(8\,r_2)$\\
$-\frac{9}{256}\,\cos(8\,r_1)\,\cos(8\,r_2)\,\cos(8\,r_5)-\frac{9}{256}\,\cos(8\,r_1)\,\cos(8\,r_2)\,\cos(8\,r_4)$\\
$+\frac{45}{8}\,\cos(4\,r_1)\,\cos(4\,r_2)\,\cos(4\,r_4)\,\cos(4\,r_5)$\\
\end{longtable}
}

\section{The $D=4$ potential on a manifold of ten $SO(3)$ singlets}
\label{d=4potential}
For the potential whose construction was described in ch.\ref{chapter4d},
we introduce the following abbreviations:

\begin{equation}
\begin{array}{lclclcl}
A&:=&\cos(2\,\omega_{48})&\quad&B&:=&\cos(4\,\omega_{48})\\
C&:=&\cos(6\,\omega_{48})&\quad&D&:=&\cos(8\,\omega_{48})\\
E&:=&\cos(2\,\omega_{47})&\quad&F&:=&\cos(4\,\omega_{47})\\
G&:=&\cos(6\,\omega_{47})&\quad&H&:=&\cos(8\,\omega_{47})\\
I&:=&\cos(2\,\phi_{47}-4\,\phi_{48})&\quad&J&:=&\cos(2\,\omega_{46})\\
K&:=&\cos(4\,\omega_{46})&\quad&L&:=&\cos(6\,\omega_{46})\\
M&:=&\cos(8\,\omega_{46})&\quad&N&:=&\cos(2\,\phi_{46}-4\,\phi_{47})\\
O&:=&\cos(2\,\phi_{46}-2\,\phi_{47}-4\,\phi_{48})&\quad&P&:=&\cos(4\,\omega_{45})\\
Q&:=&\cos(8\,\omega_{45})&\quad&R&:=&\cos(4\,\phi_{45})\\
S&:=&\cos(2\,\phi_{45})&\quad&T&:=&\cos(4\,\phi_{46})\\
U&:=&\cos(2\,\phi_{46}+4\,\phi_{47})&\quad&V&:=&\cos(2\,\phi_{46}+2\,\phi_{47}+4\,\phi_{48})\\
W&:=&\cos(2\,\omega_{45})&\quad&X&:=&\sin(4\,\phi_{46})\\
Y&:=&\sin(2\,\phi_{45})&\quad&Z&:=&\cos(6\,\omega_{45})\\
a&:=&\sin(2\,\phi_{46}+4\,\phi_{47})&\quad&b&:=&\sin(2\,\phi_{46}+2\,\phi_{47}+4\,\phi_{48})\\
c&:=&\cosh(1/2\,\rho)&\quad&d&:=&\cosh(\rho)\\
e&:=&\cosh(3/2\,\rho)&\quad&f&:=&\cosh(2\,\rho)
\end{array}
\end{equation}

Furthermore, since every denominator in this potential is a power of
two, we define $a\%b:=a/2^b$. Then, the $12\,240$ terms of the
potential are given on the following eight pages.

\newpage
\includegraphics{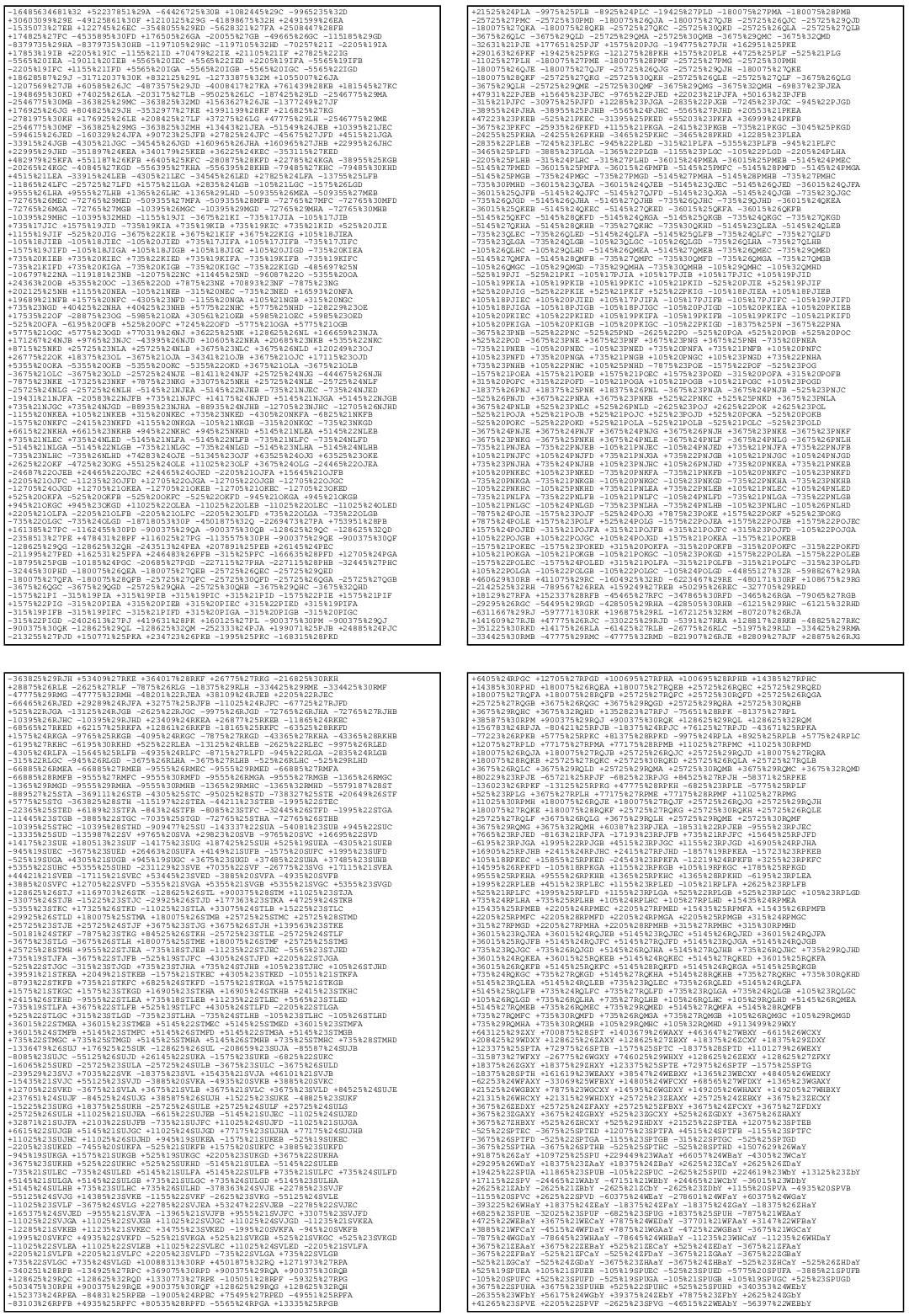}
\newpage
\includegraphics{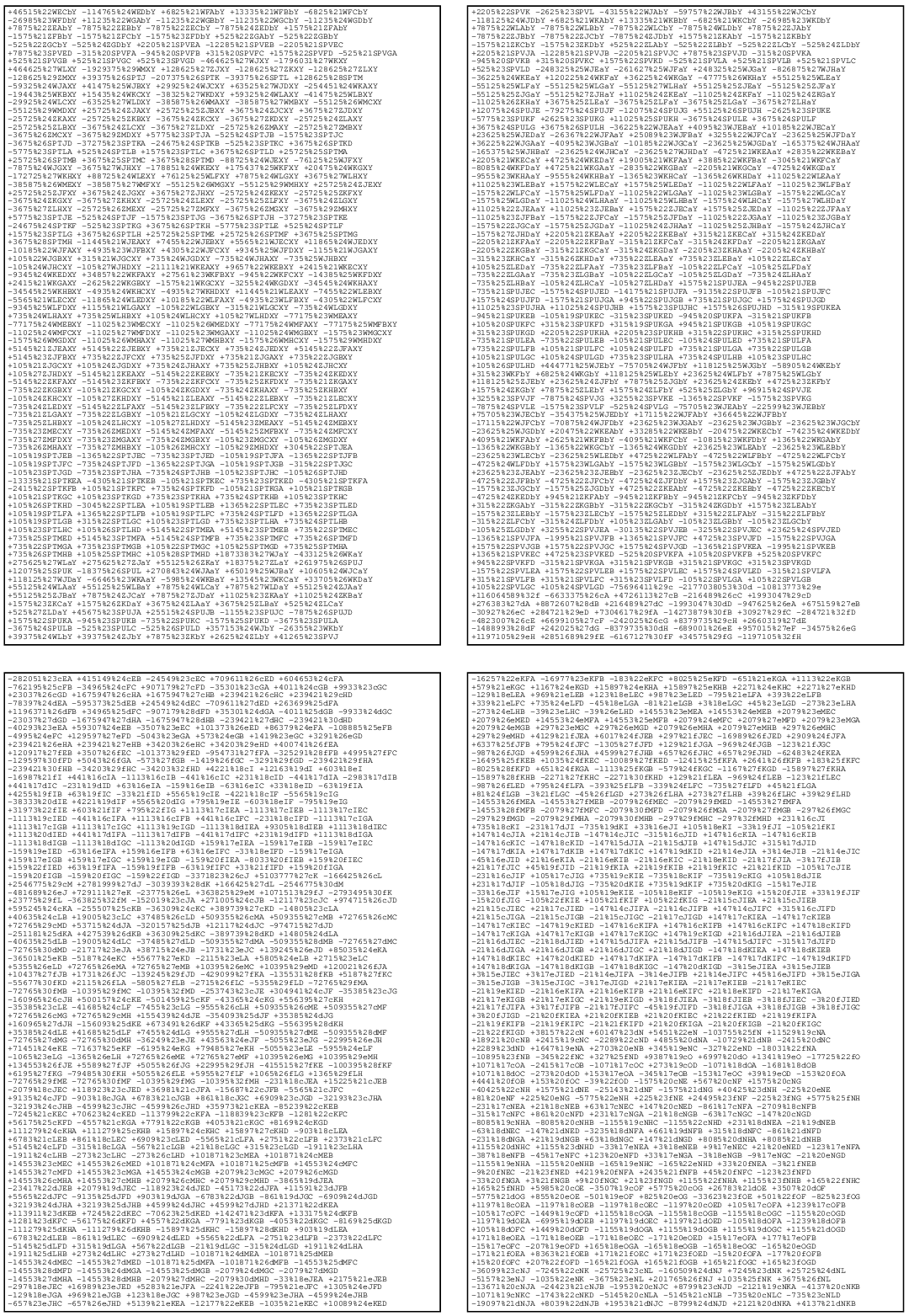}
\newpage
\includegraphics{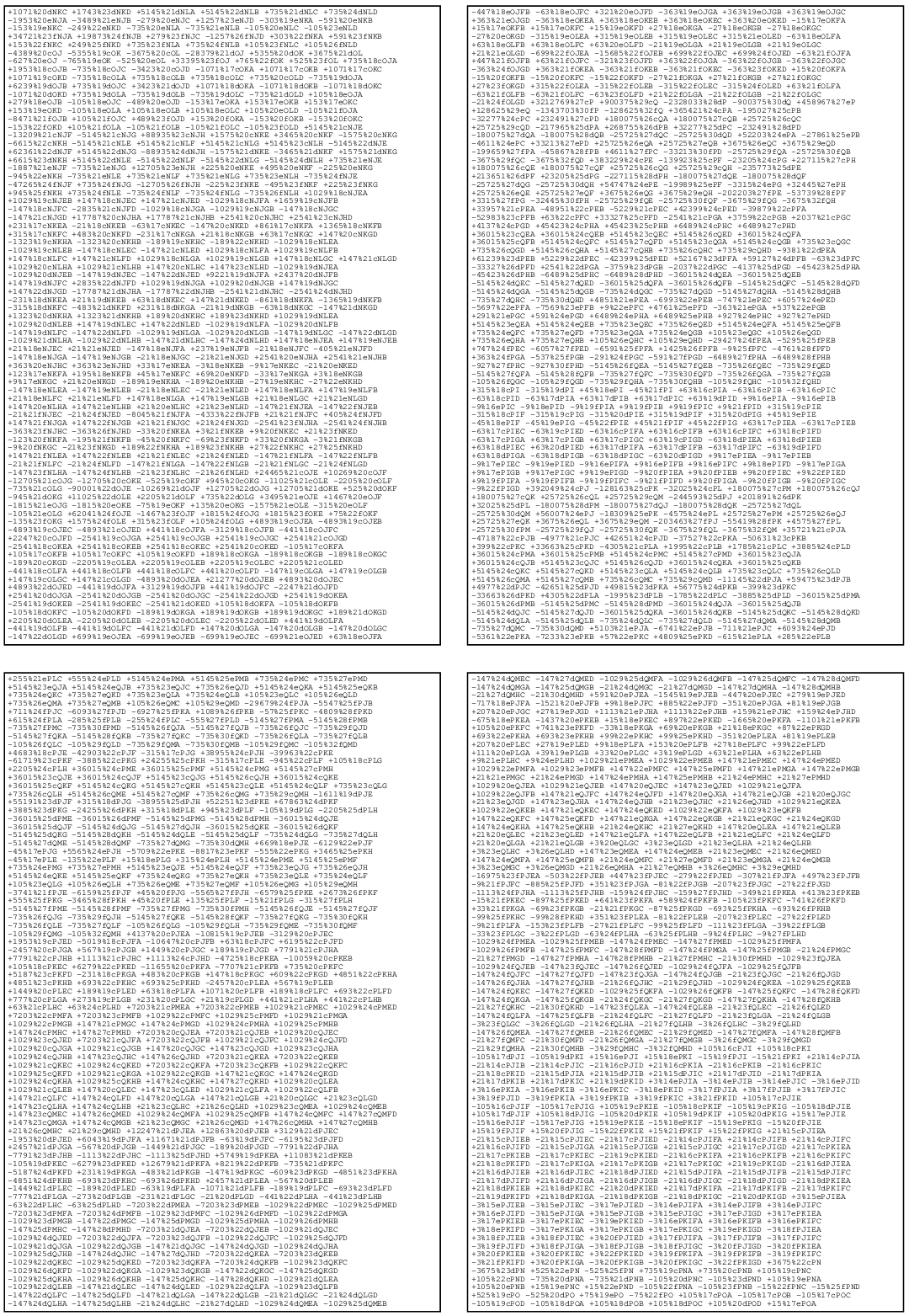}
\newpage
\includegraphics{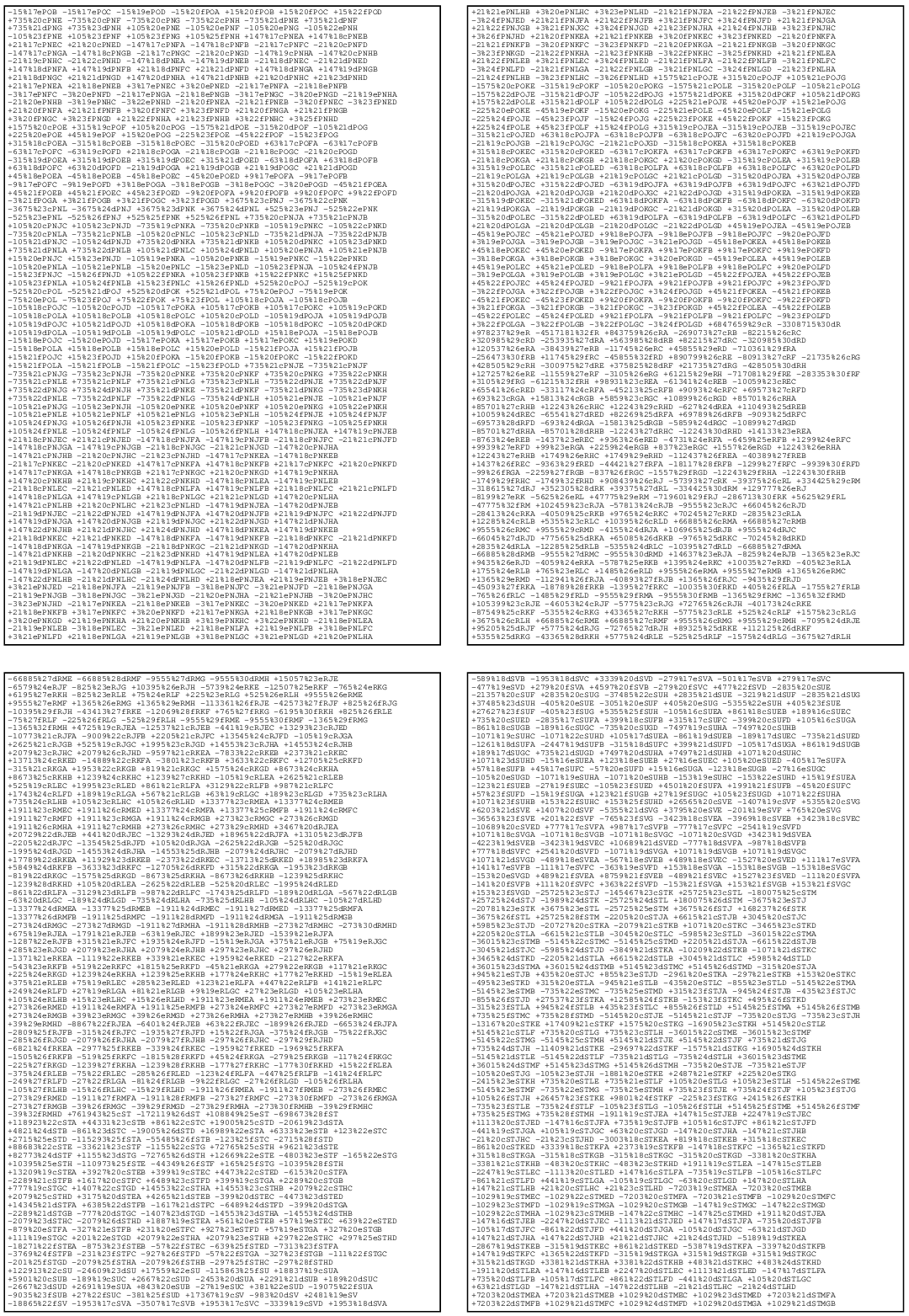}
\newpage
\includegraphics{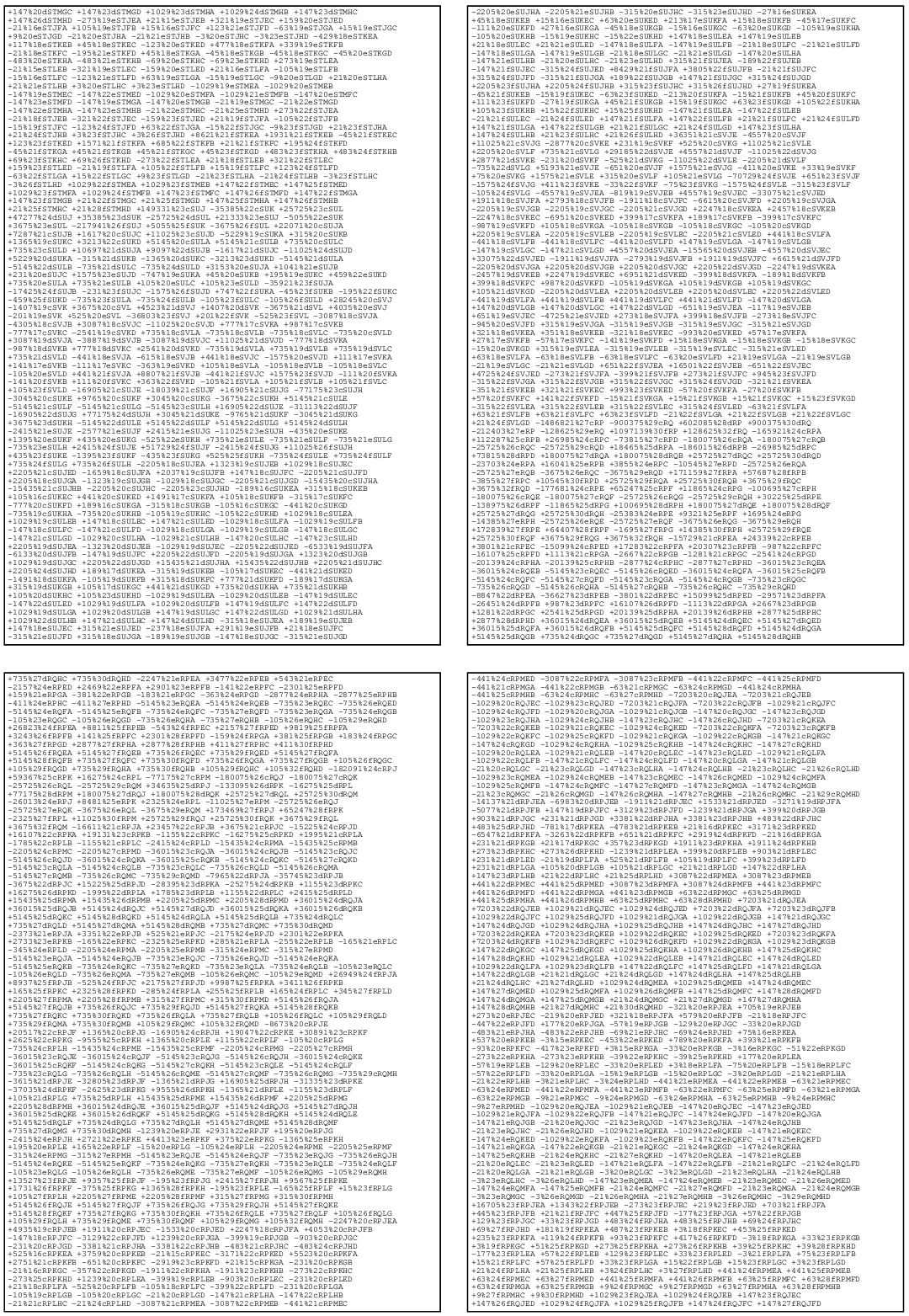}
\newpage
\includegraphics{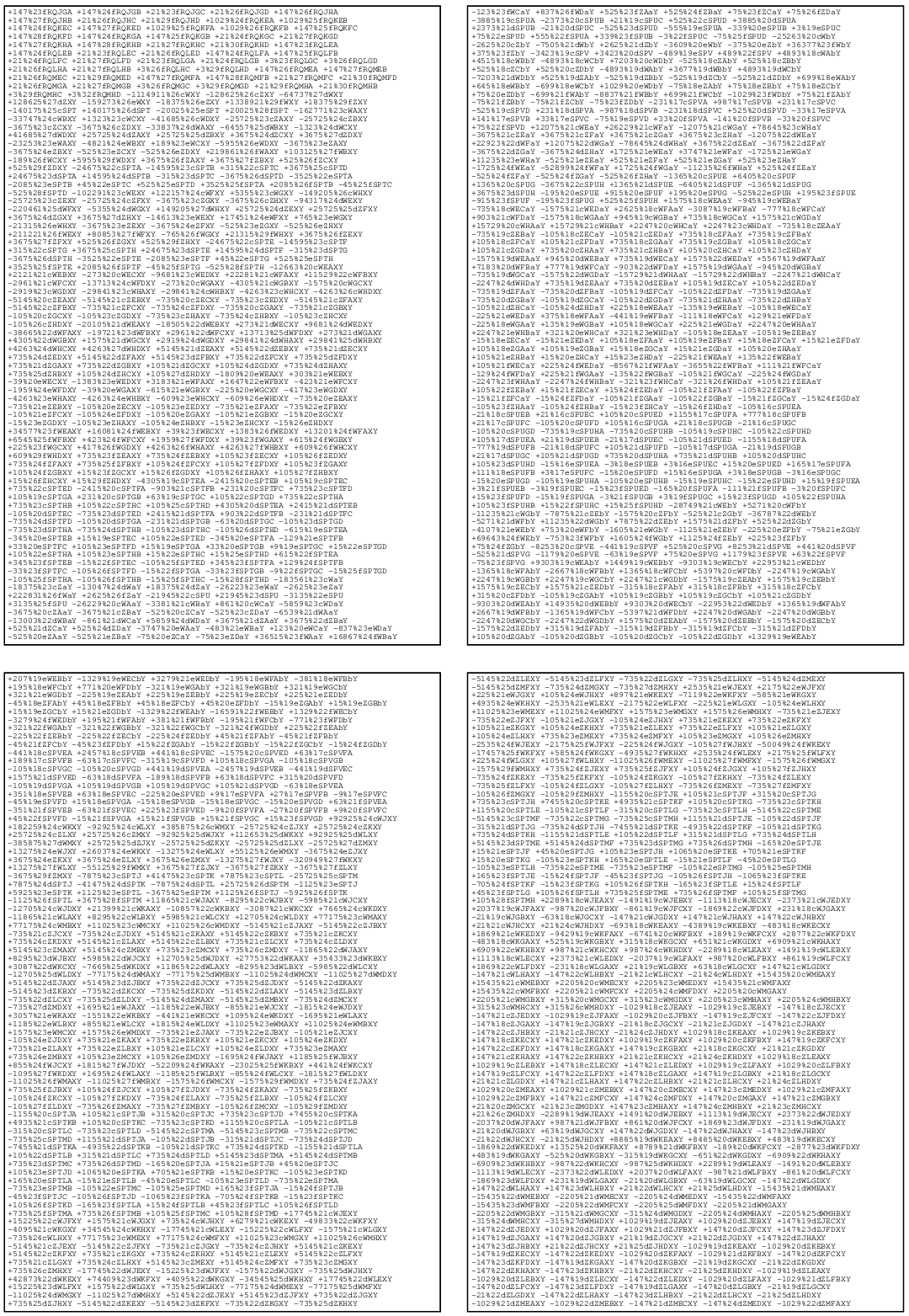}
\newpage
\includegraphics{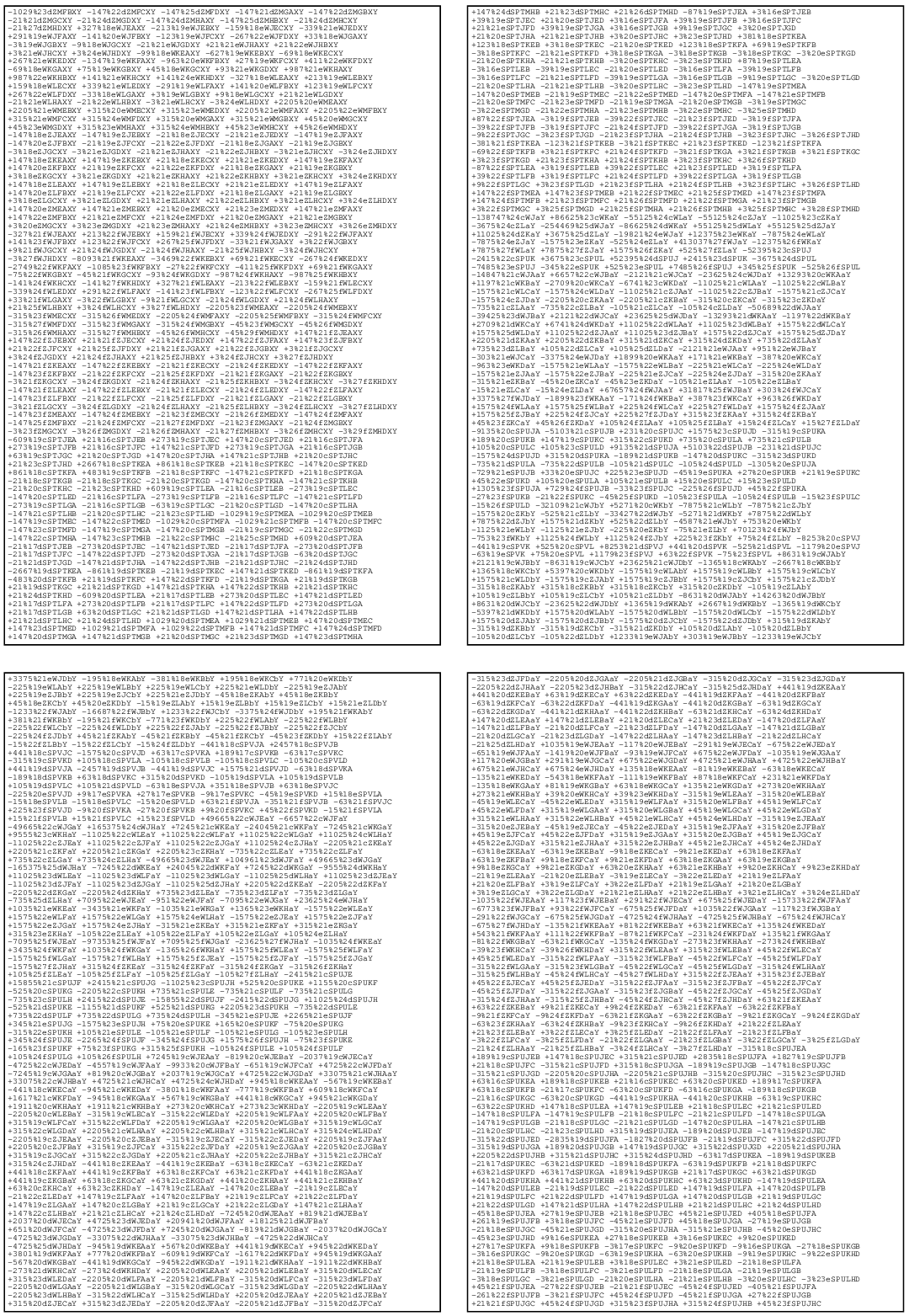}
\newpage
\includegraphics{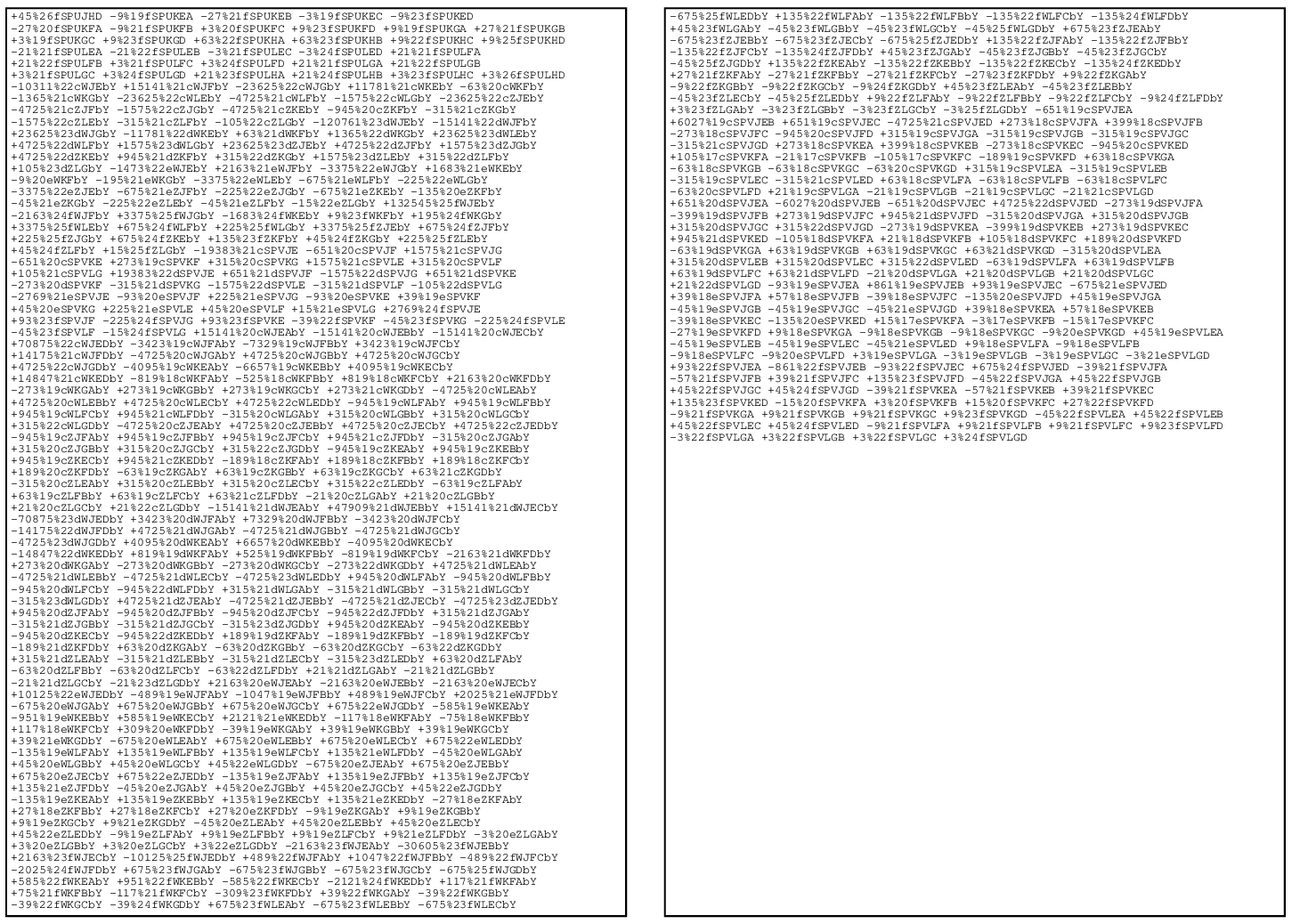}
\newpage

\chapter[LISP Definitions]{Explicit definitions reproducing all major results given in the text}
\label{chLISPcode}

For the sake of reproducibility, \LambdaTensor{} definitions that
reproduce the main results of this work are given here. This may also
serve as a starting point for further investigations of the potentials
of gauged maximal supergravities and furthermore gives many examples
how to work with the \LambdaTensor{} package. In order not to make
this appendix overly long, most consistency cross-checks originally
present in the code have been omitted here. Furthermore, due to
typographical reasons (LISP code easily tends to become quite wide),
and since this appendix presupposes quite deep knowledge of LISP, the
decision was made to typeset it in a rather small font. Although the
\LambdaTensor{} package (which itself is much too large to be
presented here, consisting of $20\,000+$ lines of code and
documentation) should be considered an integral part of this work, no
knowledge of LISP is necessary to use the data on vacua obtained here.
However, it should perhaps be noted that, despite considerable
diligence on the author's side, there still is some chance that in the
process of transliteration from LISP to physical notation, which still
is not fully automatized and hence prone to human errors, occasional
mistakes occurred, or some ambiguity remained. Thus, for any formula in
question, the LISP definition should be considered as known good and
machine checked.

\section{The partial $SU(3)_{\rm diag}$ potential}

{\tiny
\begin{verbatim}

(eval-when (compile load eval)
  (progn
    (require :e8-supergravity)
    (use-package :lambdatensor)
    (use-package :tf-spellbook)))

;; First of all, we have to set up the set of variables
;; we are going to use

(setf *poexp-vars*
      (map 'simple-array #'identity
           '(w z lambda1 lambda2 a phi theta psi)))

(setf (sp-arith-converter *sp-arith-poexp*)
      #'(lambda (x) (poexp-converter x *poexp-vars*)))


;; The embedding tensor

(defvar theta-so8xso8 (e8-theta-so-p 8))

;; E7 is parametrized as in [Phys.\ Lett.\ B {\bf 128} (1983) 169.]
;; These singlets correspond to G1+ and G2+ given there.

(defvar ac-g1+
  (sp-x `(e8-a)
        1/4
        `(,ixmap-e8-so16-alphabeta e8-a alpha beta)
        `(,(sp+
            (sp-x `(alpha beta)
                  `(,so8-sigma-ijkl-ab (:fix 0) (:fix 1) (:fix 2) (:fix 3)
                                       alpha beta))
            (sp-x `(alpha beta)
                  `(,so8-sigma-ijkl-ab (:fix 0) (:fix 1) (:fix 4) (:fix 5)
                                       alpha beta))
            (sp-x `(alpha beta)
                  `(,so8-sigma-ijkl-ab (:fix 0) (:fix 1) (:fix 6) (:fix 7)
                                       alpha beta)))
          alpha beta)))


(defvar ac-g2+
  (sp-x `(e8-a)
        1/4
        `(,ixmap-e8-so16-alphabeta e8-a alpha beta)
        `(,(sp+
            `(-1 ,(sp-x `(alpha beta)
                        `(,so8-sigma-ijkl-ab (:fix 0) (:fix 2) (:fix 4) (:fix 6)
                                             alpha beta)))
            (sp-x `(alpha beta)
                  `(,so8-sigma-ijkl-ab (:fix 0) (:fix 2) (:fix 5) (:fix 7)
                                       alpha beta))
            (sp-x `(alpha beta)
                  `(,so8-sigma-ijkl-ab (:fix 0) (:fix 3) (:fix 4) (:fix 7)
                                       alpha beta))
            (sp-x `(alpha beta)
                  `(,so8-sigma-ijkl-ab (:fix 0) (:fix 3) (:fix 5) (:fix 6)
                                       alpha beta)))
          alpha beta)))

;; Next, we have to lift anti-hermitian SU(8) generators
;; built out of LISP numbers to the corresponding SO(16) generators

(defun translate-su8-so16 (su8)
  (declare (optimize (speed 3)))
  (let ((so16 (make-sp-array '(16 16))))
    (sp-do
     #'(lambda (pos val)
         (declare (type (simple-array fixnum (2)) pos))
         (let* ((p1 (aref pos 0))
                (p2 (aref pos 1))
                (p1+8 (i+ p1 8))
                (p2+8 (i+ p2 8))
                ;; NOTE: if we were using some more generic algebra,
                ;; we would have to employ (sp-arith-conj arith) here.
                (vr (realpart val))
                (vi (imagpart val)))
           (declare (fixnum p1 p2 p1+8 p2+8))
           (sp-set! so16 vr p1 p2)
           (sp-set! so16 vr p1+8 p2+8)
           (sp-set! so16 vi  p1+8 p2) ; real p2 turned into imag p1
           (sp-set! so16 (- vi)  p1 p2+8) ; imag turned into -real
           ))
     su8)
    (sp-scale so16 1/2)))

(defun translate-su8-gen-e8 (su8)
  (let ((so16 (translate-su8-so16 su8)))
    (sp-x `(e8-M e8-N)
          `(,so16 I J)
          `(,ixmap-e8-so16-i-j e8-IJ I J)
          `(,e8-fabc e8-IJ e8-N e8-M))))

;; With these, we define the E8 analoga of the a,phi,theta,psi
;; rotation generators.

(defvar gen-a
  (let ((su8 (make-sp-array '(8 8)
                :with-entries
                '((-3 6 6) (-3 7 7)))))
    (i-dotimes (j 6)
       (setf (sp-ref su8 j j) 1))
  (translate-su8-gen-e8 (sp-scale su8 (complex 0 -1/4)))))


(defvar gen-phi
  (let ((su8 (make-sp-array '(8 8)
                :with-entries
                '((1 6 6) (-1 7 7)))))
    (translate-su8-gen-e8 (sp-scale su8 (complex 0 1)))))

(defvar gen-theta
  (let ((su8 (make-sp-array '(8 8)
                :with-entries
                '((-1 6 7) (1 7 6)))))
    (translate-su8-gen-e8 su8)))

(defvar gen-psi gen-theta)


;; Parametrization of SL(2) 

(defvar ac-W
  (sp-x `(e8-a)
        `(,ixmap-e8-so16-i-j e8-a j16 k16)
        `(,ixmap-so16-i16-i j16 j)
        `(,ixmap-so16-i16-i* k16 k)
        `(,(sp-scale (sp-id 8) 1/4) j k)))

(defvar ac-z
  (sp-x `(e8-a)
        `(,ixmap-e8-so16-alphabeta e8-a alpha beta)
        `(,(sp-scale (sp-id 8) 1/4) alpha beta)))

;; Note that there are four further singlets from the (56,2)
;; in 248 -> (133,1) + (56,2) + (1,3) which we simply
;; ignore here to reduce complexity. Hence,
;; We will probably not obtain all stationary points
;; that break symmetry down to SU(3)_diag,
;; and furthermore, we will probably get other false positives
;; than coordinate system artefacts, which have to be eliminated.

;; This corresponds to the potential (3.9) from hep-th/0207206,
;; but here with the SL(2) part fully parametrized.

(defvar phi-so8xso8-e7xsl2-and-tensors
  (labels
      ((e8-gen (x)
          (sp-ac-to-generator e8-fabc x))
       (poexp-simplify (x)
          (sp-map #'(lambda (z) (poexp* z 1)) x)))
    (let* ((rot-w (poexp-make-rot (e8-gen ac-w) 'w))
           (rot-z (poexp-make-rot (e8-gen ac-z) 'z))
           (rot-a (poexp-make-rot gen-a 'a))
           (rot-phi (poexp-make-rot gen-phi 'phi))
           (rot-theta (poexp-make-rot gen-theta 'theta))
           (rot-psi (poexp-make-rot gen-psi 'psi))
           (rot-g1+ (poexp-make-rot (e8-gen ac-g1+) 'lambda1))
           (rot-g2+ (poexp-make-rot (e8-gen ac-g2+) 'lambda2))
           (rot-g12+ (poexp-simplify (sp* rot-g1+ rot-g2+)))
           (rot-compact-e7
            (poexp-simplify (sp* (poexp-simplify (sp* rot-a rot-phi))
                                 (poexp-simplify (sp* rot-theta rot-psi)))))
           (rot-sl2 (poexp-simplify (sp* rot-w rot-z)))
           (rot-e7  (poexp-simplify (sp* rot-compact-e7 rot-g12+)))
           (v-matrix (poexp-simplify (sp* rot-sl2 rot-e7)))
           ;; Due to local SO(16) invariance of the potential, we can ignore
           ;; the compact rhs factors.
           *e8-last-computed-a1*
           *e8-last-computed-a2*
           *e8-last-computed-t-tensor*)
      (declare (special *e8-last-computed-a1*
                        *e8-last-computed-a2*
                        *e8-last-computed-t-tensor*))
      (let ((potential (e8-potential-from-v-theta v-matrix theta-so8xso8)))
        `((:potential . ,potential)
          (:a1 . ,*e8-last-computed-a1*)
          (:a2 . ,*e8-last-computed-a2*)
          (:t-tensor . ,*e8-last-computed-t-tensor*))
        ))))

\end{verbatim}
}

\section{The $SO(6)_{\rm diag}$ potentials}

{\tiny
\begin{verbatim}

(eval-when (compile load eval)
  (progn
    (require :e8-supergravity)
    (use-package :lambdatensor)
    (use-package :tf-spellbook)))

;; XXX still relics in the code for forming an embedding tensor!

(defun e8-so-p-8-p-diag-gen-acs (p &key (only-first nil))
  (let ((gen-acs-1 nil)
        (gen-acs-2 nil))
    (dotimes (k p)
      (dotimes (j k)
        (ppush gen-acs-1
               (cons +1
                     (sp-x `(e8)
                           `(,ixmap-e8-so16-ij e8 ij16)
                           `(,ixmap-so16-ij-i-j ij16 i16 j16)
                           `(,ixmap-so16-I16-i i16 i8)
                           `(,ixmap-so16-I16-i j16 j8)
                           `(,so8-sigma-ij-ab i8 j8 (:fix ,j) (:fix ,k))
                           )))
        (ppush gen-acs-2
               (cons -1
                     (sp-x `(e8)
                           `(,ixmap-e8-so16-ij e8 ij16)
                           `(,ixmap-so16-ij-i-j ij16 i16 j16)
                           `(,ixmap-so16-I16-i* i16 i8*)
                           `(,ixmap-so16-I16-i* j16 j8*)
                           `(,so8-sigma-ij-ab i8* j8* (:fix ,j) (:fix ,k))
                           )))))
    (if (not only-first)
        (do ((j p (i1+ j)))
            ((i= j 8))
          (do ((k (i1+ j) (i1+ k)))
              ((i= k 8))
            (ppush gen-acs-2
                   (cons -1
                         (sp-x `(e8)
                               `(,ixmap-e8-so16-ij e8 ij16)
                               `(,ixmap-so16-ij-i-j ij16 i16 j16)
                               `(,ixmap-so16-I16-i i16 i8)
                               `(,ixmap-so16-I16-i j16 j8)
                               `(,so8-sigma-ij-ab i8 j8 (:fix ,j) (:fix ,k))
                               )))
            (ppush gen-acs-1
                   (cons +1
                         (sp-x `(e8)
                               `(,ixmap-e8-so16-ij e8 ij16)
                               `(,ixmap-so16-ij-i-j ij16 i16 j16)
                               `(,ixmap-so16-I16-i* i16 i8*)
                               `(,ixmap-so16-I16-i* j16 j8*)
                               `(,so8-sigma-ij-ab i8* j8* (:fix ,j) (:fix ,k))
                               ))))))
    (map 'simple-array #'(lambda (x y) (sp+ (cdr x) (cdr y))) gen-acs-1 gen-acs-2)))

(defvar diag6 (e8-so-p-8-p-diag-gen-acs 6))

(defvar diag6-singlets (sp-ac-singlets e8-fabc diag6 e8-scalars))
(defvar diag6-singlets+so16 (sp-ac-singlets e8-fabc diag6 e8-scalars+so16))
(defvar diag6a (e8-so-p-8-p-diag-gen-acs 6 :only-first t))
(defvar diag6a-singlets (sp-ac-singlets e8-fabc diag6a e8-scalars)) ;; seven
(defvar diag6a-singlets+so16 (sp-ac-singlets e8-fabc diag6a e8-scalars+so16)) ; eleven (SL(3) x SL(2) = 8+3)

(defvar u1-generator (sp+ (aref diag6-singlets+so16 5) (aref diag6-singlets+so16 8)))

(defvar sl2
  (make-array 3 :initial-contents
              (list (sp+ `(1/4 ,u1-generator))
                    (sp-x `(e8)
                          `(,ixmap-e8-so16-alphabeta e8 alpha beta)
                          `(,(make-sp-array '(8 8) :with-entries '((+1/2 6 6) (-1/2 7 7))) alpha beta))
                    (sp-x `(e8)
                          `(,ixmap-e8-so16-alphabeta e8 alpha beta)
                          `(,(make-sp-array '(8 8) :with-entries '((+1/2 6 7) (+1/2 7 6))) alpha beta)))))

(defvar sl3
  (let* ((sl3-v0 (sp-lin-indep-ac-commutators e8-fabc diag6-singlets+so16 diag6-singlets+so16))
         (s (copy-seq sl3-v0)))
    (setf (aref s 6) (sp+ `(-1 ,(aref sl3-v0 6)) `(1/4 ,(aref sl3-v0 7))))
    (setf (aref s 7) (sp+ `(+1 ,(aref sl3-v0 6)) `(1/4 ,(aref sl3-v0 7))))
    (map 'simple-array
         #'(lambda (x scale) (sp-scale x scale)) s '(-1/6 -1/6 -1/2 1/6 1/6 1/2 1/2 1/2))
    ;; The minus signs for the first three generators make the structure constants equal +epsilon3.
    ))


(defvar sl3-g_ab
  (let ((mx (make-sp-array '(8 8))))
    (dotimes (j 8)
      (dotimes (k 8)
        (let ((sprod (* 1/60
                        (sp-ref (sp-x `() `(,e8-g_ab a b) `(,(aref sl3 j) a) `(,(aref sl3 k) b))))))
          (sp-set! mx sprod j k))))
    mx))

(defvar sl3-g^ab (sp-invert sl3-g_ab))

(defun sl3-decompose (e8-ac)
  (let ((v (make-sp-array '(8))))
    (dotimes (c 8)
      (let ((coeff (* (sp-ref sl3-g^ab c c)
                      (sp-ref (sp-x `() `(,e8-eta p q) `(,(aref sl3 c) p) `(,e8-ac q))))))
        (sp-set! v coeff c)))
    v))
  
(defvar sl3-fabc
      (let ((fabc (make-sp-array '(8 8 8))))
        (dotimes (a 8)
          (dotimes (b 8)
            (let* ((ab (e8-ac-[] (aref sl3 a) (aref sl3 b)))
                   (vc (sl3-decompose ab)))
              (sp-do #'(lambda (p v) (sp-set! fabc v a b (aref p 0))) vc))))
        fabc))


;; (sp-check-jacobi sl3-fabc) ==> T


(defvar data-so62
  (let* ((delta8-ij (sp-generate-indexsplit-tensor 16 '(8)))
         (delta8-ij-8 (sp-generate-indexsplit-tensor 16 '(8) :offset 8))
         (p1 (sp-x `(e8-c e8-b)
                   1/2 1/2
                   `(,(make-sp-array '(8 8) :with-entries '((1 0 6) (1 1 7) (-1 2 5) (1 3 4))) i j)
                   `(,(sp+ `(+1 ,(sp-x `(Ix Jx i j) `(,delta8-ij Ix i) `(,delta8-ij-8 Jx j)))
                           `(-1 ,(sp-x `(Ix Jx i j) `(,delta8-ij-8 Jx i) `(,delta8-ij Ix j)))
                           `(+1 ,(sp-x `(Ix Jx i j) `(,delta8-ij-8 Ix i) `(,delta8-ij Jx j)))
                           `(-1 ,(sp-x `(Ix Jx i j) `(,delta8-ij Jx i) `(,delta8-ij-8 Ix j)))
                           )
                     Ix Jx i j)
                   `(,ixmap-so16-ij-i-j IJ Ix Jx)
                   `(,ixmap-e8-so16-ij e8-a IJ)
                   `(,e8-fabc e8-a e8-b e8-c)))
         (p2 (sp-x `(e8-c e8-b)
                   1/2 1/2
                   `(,(sp-id 8) i j)
                   `(,(sp+ `(+1 ,(sp-x `(Ix Jx i j) `(,delta8-ij-8 Ix i) `(,delta8-ij Jx j)))
                           `(-1 ,(sp-x `(Ix Jx i j) `(,delta8-ij Ix i) `(,delta8-ij-8 Jx j)))) Ix Jx i j)
                   `(,ixmap-e8-so16-ij e8-a ij)
                   `(,ixmap-so16-ij-i-j ij Ix Jx)
                   `(,e8-fabc e8-a e8-b e8-c)))

         (p3 (sp-x `(e8-c e8-b)
                   -1/2 1/2
                   `(,(make-sp-array '(8 8) :with-entries '((1 0 6) (1 1 7) (-1 2 5) (1 3 4))) i j)
                   `(,(sp+ `(+1 ,(sp-x `(Ix Jx i j) `(,delta8-ij Ix i) `(,delta8-ij Jx j)))
                           `(-1 ,(sp-x `(Ix Jx i j) `(,delta8-ij Jx i) `(,delta8-ij Ix j)))
                           `(-1 ,(sp-x `(Ix Jx i j) `(,delta8-ij-8 Ix i) `(,delta8-ij-8 Jx j)))
                           `(+1 ,(sp-x `(Ix Jx i j) `(,delta8-ij-8 Jx i) `(,delta8-ij-8 Ix j)))
                           )
                     Ix Jx i j)
                   `(,ixmap-so16-ij-i-j IJ Ix Jx)
                   `(,ixmap-e8-so16-ij e8-a IJ)
                   `(,e8-fabc e8-a e8-b e8-c)))
         (p4 (sp-x `(e8-c e8-b)
                   1/2
                    `(,(make-sp-array '(8 8) :with-entries '((1 0 1) (-1 2 3) (1 4 5) (1 6 7))) c* d*)
                    `(,(sp+ `(+1 ,(sp-x `(a* b* c* d*) `(,(sp-id 8) a* c*) `(,(sp-id 8) b* d*)))
                            `(-1 ,(sp-x `(a* b* c* d*) `(,(sp-id 8) b* c*) `(,(sp-id 8) a* d*))))
                      a* b* c* d*)
                    `(,ixmap-e8-so16-alpha*beta* e8-a a* b*)
                    `(,e8-fabc e8-a e8-b e8-c)))
         (p5 (sp-x `(e8-c e8-b)
                   1/2
                   `(,(sp-id 8) a* b*)
                   `(,ixmap-e8-so16-alpha*beta* e8-a a* b*)
                   `(,e8-fabc e8-a e8-b e8-c)))
         (p6 (sp-x `(e8-c e8-b)
                   `(,(make-sp-array '(8 8) :with-entries '((1 6 7) (-1 7 6))) a b)
                   `(,ixmap-e8-so16-alphabeta e8-a a b)
                   `(,e8-fabc e8-a e8-b e8-c)))
         (p7 (sp-x `(e8-c e8-b)
                   1/2
                   `(,(sp-id 8) a b)
                   `(,ixmap-e8-so16-alphabeta e8-a a b)
                   `(,e8-fabc e8-a e8-b e8-c)))
         (p8
          (sp+ p7
               `(-2
                 ,(sp-x `(e8-c e8-b)
                        `(,(make-sp-array '(8 8) :with-entries '((1 6 6) (+1 7 7))) a b)
                        `(,ixmap-e8-so16-alphabeta e8-a a b)
                        `(,e8-fabc e8-a e8-b e8-c)))))
         ;; ==========================================================
         (q1 (sp-x `(e8-c e8-b)
                   1/4 1/2
                   `(,(make-sp-array '(8 8) :with-entries '((1 0 6) (1 1 7) (-1 2 5) (1 3 4))) i j)
                   `(,(sp+ `(+1 ,(sp-x `(Ix Jx i j) `(,delta8-ij Ix i) `(,delta8-ij Jx j)))
                           `(-1 ,(sp-x `(Ix Jx i j) `(,delta8-ij Jx i) `(,delta8-ij Ix j)))
                           `(+1 ,(sp-x `(Ix Jx i j) `(,delta8-ij-8 Ix i) `(,delta8-ij-8 Jx j)))
                           `(-1 ,(sp-x `(Ix Jx i j) `(,delta8-ij-8 Jx i) `(,delta8-ij-8 Ix j)))
                           )
                     Ix Jx i j)
                   `(,ixmap-so16-ij-i-j IJ Ix Jx) ; note to include a factor 1/2 since we are mapping I J -> IJ!
                   `(,ixmap-e8-so16-ij e8-a IJ)
                   `(,e8-fabc e8-a e8-b e8-c)))
         (q2 (sp-x `(e8-c e8-b)
                   1/2
                   `(,(make-sp-array '(8 8) :with-entries '((1 6 6) (-1 7 7))) a b)
                   `(,ixmap-e8-so16-alphabeta e8-a a b)
                   `(,e8-fabc e8-a e8-b e8-c)))
         (q3 (sp-x `(e8-c e8-b)
                   1/2
                   `(,(make-sp-array '(8 8) :with-entries '((1 6 7) (1 7 6))) a b)
                   `(,ixmap-e8-so16-alphabeta e8-a a b)
                   `(,e8-fabc e8-a e8-b e8-c))))
    `((:sl3 . ,(vector p1 p2 p3 p4 p5 p6 p7 p8))
      (:sl2 . ,(vector q1 q2 q3)))))

#|
(dotimes (j 8)
  (format t "~A: ~A  " j
          (sp-multiple-p (sp-ac-to-generator e8-fabc (aref sl3 0))
                         (aref (cav data-so62 :sl3) 0))))
=> 0: 1  1: 1  2: 1  3: 1  4: 1  5: 1  6: 1  7: 1
|#



;; dr is the defining representation.

(defvar sl3-dr
  (map 'simple-array #'(lambda (e) (make-sp-array '(3 3) :with-entries e))
       '(
         ((1 1 2) (-1 2 1))
         ((1 0 1) (-1 1 0))
         ((1 2 0) (-1 0 2))
         ;; ---
         ((1 1 2) (1 2 1))
         ((1 0 1) (1 1 0))
         ((-1 2 0) (-1 0 2))
         ((-1 0 0) (1 1 1))
         ((1 0 0) (1 1 1) (-2 2 2)))))

(defvar sl2-dr
  (map 'simple-array #'(lambda (e) (make-sp-array '(2 2) :with-entries e))
       `(
         ((1/2 0 1) (-1/2 1 0))
         ((1/2 0 0) (-1/2 1 1))
         ((1/2 0 1) ( 1/2 1 0))
         )))

(defvar sl3-g_ab-dr
  (let ((mx (make-sp-array '(8 8))))
    (dotimes (j 8)
      (dotimes (k 8)
        (let ((sprod (sp-trace (sp* (aref sl3-dr j) (aref sl3-dr k)))))
          (sp-set! mx sprod j k))))
    mx))

(defvar sl3-g^ab-dr (sp-invert sl3-g_ab-dr))

(defvar sl3-fabc-dr
  (labels
      ((sl3-decompose-dr (gen)
         (let ((v (make-sp-array '(8))))
           (dotimes (c 8)
             (let ((coeff (* (sp-ref sl3-g^ab-dr c c) (sp-trace (sp* gen (aref sl3-dr c))))))
               (sp-set! v coeff c)))
           v)))
    (let ((fabc (make-sp-array '(8 8 8))))
      (dotimes (a 8)
        (dotimes (b 8)
          (let* ((ab (sp-[] (aref sl3-dr a) (aref sl3-dr b)))
                 (vc (sl3-decompose-dr ab)))
              (sp-do #'(lambda (p v) (sp-set! fabc v a b (aref p 0))) vc))))
        fabc)))
   
;; (sp+ sl3-fabc `(-1 ,sl3-fabc-dr)) => 0

;; The generic SL(3) coset element (mod right SO(16) rotation),
;; extended by a SL(2) rotation

(defvar v-rot-so62
  (let* ((rot-s (poexp-make-rot (sp-ac-to-generator e8-fabc (sp-scale (aref sl3 6) -1)) 's))
         ;; Note: this corresponds to diag(1,-1,0)
         (rot-z (poexp-make-rot (sp-ac-to-generator e8-fabc (aref sl3 7)) 'z))
         (rot-sz (sp* rot-s rot-z))
         (rot-x1 (poexp-make-rot (sp-ac-to-generator e8-fabc (sp-scale (aref sl3 0) +1)) 'r1))
         (rot-x2 (poexp-make-rot (sp-ac-to-generator e8-fabc (sp-scale (aref sl3 2) +1)) 'r2))
         (rot-x3 (poexp-make-rot (sp-ac-to-generator e8-fabc (sp-scale (aref sl3 1) +1)) 'r3))
         (rot-x123-sz (sp* rot-x1 (sp* rot-x2 (sp* rot-x3 rot-sz))))
         (rot-v (poexp-make-rot (sp-ac-to-generator e8-fabc (aref sl2 1)) 'v))
         (rot-x5 (poexp-make-rot (sp-ac-to-generator e8-fabc (aref sl2 0)) 'r5))
         (rot-sl2 (sp* rot-x5 rot-v))
         (rot-sl3xsl2 (sp* rot-x123-sz rot-sl2))
         )
    (sp-map #'(lambda (x) (poexp* 1 x)) rot-sl3xsl2)))

(defvar phi-so6-so62 (poexp* 1 (e8-potential-from-v-theta v-rot-so62 (e8-theta-so-p 6))))
(defvar phi-so6-so71 (poexp* 1 (e8-potential-from-v-theta v-rot-so62 (e8-theta-so-p 7))))
(defvar phi-so6-so8 (poexp* 1 (e8-potential-from-v-theta v-rot-so62 (e8-theta-so-p 8))))


\end{verbatim}
}

\section{The $SO(5)_{\rm diag}$ potentials}

The code given here is a bit more elaborate than for the other cases
to demonstrate the process used to identify the $SL(3)$ generators.

{\tiny
\begin{verbatim}

(eval-when (compile load eval)
  (progn
    (require :e8-supergravity)
    (use-package :lambdatensor)
    (use-package :tf-spellbook)))

(setf *bytes-consed-between-gcs* 40000000 *gc-verbose* nil)

(defun e8-so-p-diag-gen-acs (p &key (start 0))
  (let ((gen-acs-1 nil)
        (gen-acs-2 nil))
    (dotimes (k p)
      (dotimes (j k)
        (ppush gen-acs-1
               (sp-x `(e8)
                     `(,ixmap-e8-so16-ij e8 ij16)
                     `(,ixmap-so16-ij-i-j ij16 i16 j16)
                     `(,ixmap-so16-I16-i i16 i8)
                     `(,ixmap-so16-I16-i j16 j8)
                     `(,so8-sigma-ij-ab i8 j8 (:fix ,(+ j start)) (:fix ,(+ k start)))
                     ))
        (ppush gen-acs-2
               (sp-x `(e8)
                     `(,ixmap-e8-so16-ij e8 ij16)
                     `(,ixmap-so16-ij-i-j ij16 i16 j16)
                     `(,ixmap-so16-I16-i* i16 i8*)
                     `(,ixmap-so16-I16-i* j16 j8*)
                     `(,so8-sigma-ij-ab i8* j8* (:fix ,(+ j start)) (:fix ,(+ k start)))
                     ))))
    (map 'simple-array #'(lambda (x y) (sp+ x y)) gen-acs-1 gen-acs-2)))

(defvar so5_diag-v-ac (e8-so-p-diag-gen-acs 5))
(defvar so3_diag-v-ac (e8-so-p-diag-gen-acs 3 :start 5))
(defvar so2_diag-v-ac (e8-so-p-diag-gen-acs 2 :start 5))

(defvar so5xso3-singlets
  (sp-ac-singlets e8-fabc (concatenate 'list so5_diag-v-ac so3_diag-v-ac) e8-scalars))

(defvar v-rot-so5xso3
  (let* ((*HEURISTIC-FACTORIZE-POLYNOME-ALSO-TRY-THESE-ZEROES* '(1/4))
         (ac-W
          (sp-x `(e8-a)
                `(,ixmap-e8-so16-i-j e8-a j16 k16)
                `(,ixmap-so16-i16-i j16 j)
                `(,ixmap-so16-i16-i* k16 k)
                `(,(sp-scale (sp-id 8) 1/4) j k)))
         (ac-Z
          (sp-x `(e8-a)
                `(,ixmap-e8-so16-alphabeta e8-a alpha beta)
                `(,(sp-scale (sp-id 8) 1/4) alpha beta)))
         (ac-M
          (sp-x `(e8-a)
                `(,ixmap-e8-so16-alphabeta e8-a alpha beta)
                `(,(sp-scale
                    (make-sp-array '(8 8) :with-entries
                       '((3 0 0) (3 1 1) (3 2 2) (3 3 3) (3 4 4)
                         (-5 5 5) (-5 6 6) (-5 7 7)))
                    1/4) alpha beta))))
    (declare (special *HEURISTIC-FACTORIZE-POLYNOME-ALSO-TRY-THESE-ZEROES*))
    (sp* (poexp-make-rot (sp-ac-to-generator e8-fabc ac-M) 's)
         (poexp-make-rot (sp-ac-to-generator e8-fabc ac-W) 'w)
         (poexp-make-rot (sp-ac-to-generator e8-fabc ac-Z) 'z))))

(defvar phi-so5xso3-so5
  (poexp* 1 (e8-potential-from-v-theta v-rot-so5xso3
                                       (e8-theta-so-p 5))))


(defvar phi-so5xso3-so8
  (poexp* 1 (e8-potential-from-v-theta v-rot-so5xso3
                                       (e8-theta-so-p 8))))

(defvar so5xso2-singlets
  (sp-ac-singlets e8-fabc
                  (concatenate 'list so5_diag-v-ac so2_diag-v-ac)
                  e8-scalars))

(defvar so5xso2-singlets+so16
  (sp-ac-singlets e8-fabc
                  (concatenate 'list so5_diag-v-ac so2_diag-v-ac)
                  e8-scalars+so16))

;; These are ten.

;; (length (sp-lin-indep-ac-commutators e8-fabc so5xso2-singlets+so16 so5xso2-singlets+so16))
;; => 8; 3c+5nc. This must be a SL(3).

;; (length (sp-ac-commutator-closure e8-fabc so5xso2-singlets))
;; => 9

;; (sp-total-dimension (concatenate 'list (sp-lin-indep-ac-commutators e8-fabc so5xso2-singlets+so16 so5xso2-singlets+so16) so5xso2-singlets))
;; => there must be a singlet not contained in SL(3). Hence, 5 Singlets from SL(5) and a single extra one. This is the picture.

(defvar so3-in-sl3
  (sp-lin-indep-ac-commutators e8-fabc so5xso2-singlets so5xso2-singlets))

(defvar non-sl3-singlet
  (aref (sp-ac-singlets e8-fabc so3-in-sl3 so5xso2-singlets) 0)) ; only one

(defvar sl3-noncompact
  (map 'simple-array #'sp-from-array
       (lintrans-core-and-complement
        #'(lambda (v)
            (vector (sp-ref (sp-x `() `(,e8-eta a b)
                                  `(,non-sl3-singlet a)
                                  `(,(sp-from-array v) b)))))
        (map 'simple-array #'sp-to-array so5xso2-singlets))))

(defvar full-sl3 (sp-ac-commutator-closure e8-fabc sl3-noncompact))

;; (length (sp-ac-commutator-closure e8-fabc sl3-noncompact)) => 8
;; Hence, we now really have SL(3).

#|

This shows that our generators are orthonormal.
And all except the first one can easily be normalized to +-2;
for the first one, 6 is natural,
so we may want to identify this with diag(1,1,-2).

(dotimes (j 8)
  (dotimes (k 8)
    (format t "~A ~A: ~A~%" j k
            (sp-ref (sp-x `() `(,e8-eta a b)
                          `(,(aref full-sl3 j) a)
                          `(,(aref full-sl3 k) b))))))

|#

(defvar properly-normalized-sl3
  (map 'simple-array
       #'(lambda (x)
           (let ((sqrt-abs-norm^2/2
                  (sqrt
                   (* 1/2 (abs
                           (sp-ref (sp-x `()
                                         `(,e8-eta a b)
                                         `(,x a) `(,x b))))))))
             (format t "sqrt-abs-norm^2/2=~A~%" sqrt-abs-norm^2/2)
             (if (= (floor sqrt-abs-norm^2/2) sqrt-abs-norm^2/2)
                 (sp-scale x (/ (floor sqrt-abs-norm^2/2)))
               x)))
       full-sl3))

(defvar sl3-pn properly-normalized-sl3)

(defvar sl3-pn-g_ab
  (let ((mx (make-sp-array '(8 8))))
    (dotimes (j 8)
      (dotimes (k 8)
        (let ((sprod (sp-ref (sp-x `() `(,e8-eta a b) `(,(aref sl3-pn j) a) `(,(aref sl3-pn k) b)))))
          (sp-set! mx sprod j k))))
    mx))

(setf sl3-pn-g^ab (sp-invert sl3-pn-g_ab))

(defun sl3-pn-decompose (e8-ac)
  (let ((v (make-sp-array '(8))))
    (dotimes (c 8)
      (let ((coeff (* (sp-ref sl3-pn-g^ab c c)
                      (sp-ref (sp-x `() `(,e8-eta p q)
                                    `(,(aref sl3-pn c) p) `(,e8-ac q))))))
        (sp-set! v coeff c)))
    v))


(defvar sl3-pn-fabc
      (let ((fabc (make-sp-array '(8 8 8))))
        (dotimes (a 8)
          (dotimes (b 8)
            (let* ((ab (e8-ac-[] (aref sl3-pn a) (aref sl3-pn b)))
                   (vc (sl3-pn-decompose ab)))
              (sp-do #'(lambda (p v) (sp-set! fabc v a b (aref p 0))) vc))))
        fabc))

;; By educated guess and trial and error, the following permutation was found:

(setf sl3-reordered-fabc
  (let ((re-ordering
         (make-sp-array '(8 8)
                        :with-entries
                        '((1 0 7) (1 7 1) (1 1 4) (1 2 6)
                          (1 3 5) (1 4 3) (1 5 2) (1 6 0)))))
    (sp-x `(ar br cr)
          `(,re-ordering a ar) `(,re-ordering b br) `(,re-ordering c cr)
          `(,sl3-pn-fabc a b c))))

;; ...and with this re-ordering, we just have the same sl3 fabc commutation relations as in the SO(6)_diag case.

(defvar sl3
  (map 'simple-array #'(lambda (n) (aref sl3-pn n))
       #(6 7 5 4 1 3 2 0)))

(defvar sl3-g_ab
  (let ((mx (make-sp-array '(8 8))))
    (dotimes (j 8)
      (dotimes (k 8)
        (let ((sprod (sp-ref (sp-x `() `(,e8-eta a b) `(,(aref sl3 j) a) `(,(aref sl3 k) b)))))
          (sp-set! mx sprod j k))))
    mx))

(defvar sl3-g^ab (sp-invert sl3-g_ab))

(defun sl3-decompose (e8-ac)
  (let ((v (make-sp-array '(8))))
    (dotimes (c 8)
      (let ((coeff (* (sp-ref sl3-g^ab c c)
                      (sp-ref (sp-x `()
                                    `(,e8-eta p q)
                                    `(,(aref sl3 c) p)
                                    `(,e8-ac q))))))
        (sp-set! v coeff c)))
    v))


(defvar sl3-fabc
      (let ((fabc (make-sp-array '(8 8 8))))
        (dotimes (a 8)
          (dotimes (b 8)
            (let* ((ab (e8-ac-[] (aref sl3 a) (aref sl3 b)))
                   (vc (sl3-decompose ab)))
              (sp-do #'(lambda (p v) (sp-set! fabc v a b (aref p 0))) vc))))
        fabc))

;; (sp+ sl3-fabc `(-1 ,sl3-reordered-fabc)) => #<empty>
;; Inspection shows that fabc is also the same as for the SO(6) case.

(defvar v-rot-so52
  (let* ((rot-s (poexp-make-rot (sp-ac-to-generator e8-fabc (sp-scale (aref sl3 6) -1)) 's))
         (rot-z (poexp-make-rot (sp-ac-to-generator e8-fabc (aref sl3 7)) 'z))
         (rot-sz (sp* rot-s rot-z))
         (rot-x1 (poexp-make-rot (sp-ac-to-generator e8-fabc (sp-scale (aref sl3 0) +1)) 'r1))
         (rot-x2 (poexp-make-rot (sp-ac-to-generator e8-fabc (sp-scale (aref sl3 2) +1)) 'r2))
         (rot-x3 (poexp-make-rot (sp-ac-to-generator e8-fabc (sp-scale (aref sl3 1) +1)) 'r3))
         (rot-x123-sz (sp* rot-x1 (sp* rot-x2 (sp* rot-x3 rot-sz))))
         (rot-v (poexp-make-rot (sp-ac-to-generator e8-fabc non-sl3-singlet) 'v))
         (rot-sl3xR (sp* rot-x123-sz rot-v))
         )
    (sp-map #'(lambda (x) (poexp* 1 x)) rot-sl3xR)))


(defvar phi-so5-so8
  (poexp* 1 (e8-potential-from-v-theta v-rot-so52 (e8-theta-so-p 8))))

(defvar phi-so5-so5
  (poexp* 1 (e8-potential-from-v-theta v-rot-so52 (e8-theta-so-p 5))))

\end{verbatim}
}

\section{The $SO(4)_{\rm diag}$ potentials}

{\tiny
\begin{verbatim}

(eval-when (compile load eval)
  (progn
    (require :e8-supergravity)
    (use-package :lambdatensor)
    (use-package :tf-spellbook)))

(eval-when (compile load eval)
  (progn
    (setf lambdatensor::*poexp-vars*
          (coerce '(V S W Z X R0 R1 R2 R3 R4 R5) '(simple-array * (*))))
    (setf (lambdatensor::sp-arith-converter lambdatensor::*sp-arith-poexp*)
          #'(lambda (x) (lambdatensor::poexp-converter x lambdatensor::*poexp-vars*)))
    ))

(setf *bytes-consed-between-gcs* 40000000 *gc-verbose* nil)

(defun e8-so-p-diag-gen-acs (p &key (start 0))
  [as in the SO(5) case])

(defvar so4a_diag-v-ac (e8-so-p-diag-gen-acs 4))
(defvar so3a_diag-v-ac (e8-so-p-diag-gen-acs 3))
(defvar so4b_diag-v-ac (e8-so-p-diag-gen-acs 4 :start 4))
(defvar so3b_diag-v-ac (e8-so-p-diag-gen-acs 3 :start 4))
(defvar so2b_diag-v-ac (e8-so-p-diag-gen-acs 2 :start 4))

(defvar so4xso4-singlets
  (sp-ac-singlets e8-fabc
                  (concatenate 'list so4a_diag-v-ac so4b_diag-v-ac)
                  e8-scalars))

(defvar so4xso3-singlets
  (sp-ac-singlets e8-fabc
                  (concatenate 'list so4a_diag-v-ac so3b_diag-v-ac)
                  e8-scalars))

(defvar so3xso3-singlets
  (sp-ac-singlets e8-fabc
                  (concatenate 'list so3a_diag-v-ac so3b_diag-v-ac)
                  e8-scalars))
;; ten

(defvar so4xso2-singlets
  (sp-ac-singlets e8-fabc
                  (concatenate 'list so4a_diag-v-ac so2b_diag-v-ac)
                  e8-scalars))
;; eleven

(defvar so4xso3-singlets+so16
  (sp-ac-singlets e8-fabc
                  (concatenate 'list so4a_diag-v-ac so3b_diag-v-ac)
                  e8-scalars+so16))
;; seven

(defvar so3-singlet
  (aref (sp-ac-singlets e8-fabc so4xso4-singlets so4xso3-singlets) 0))


(defun v-rot-so43 (v w s z x &key (type :symbolic))
  (let* ((ac-W
          (sp-x `(e8-a)
                `(,ixmap-e8-so16-i-j e8-a j16 k16)
                `(,ixmap-so16-i16-i j16 j)
                `(,ixmap-so16-i16-i* k16 k)
                `(,(sp-scale (sp-id 8) 1/4) j k)))
         (ac-Z
          (sp-x `(e8-a)
                `(,ixmap-e8-so16-alphabeta e8-a alpha beta)
                `(,(sp-scale (sp-id 8) 1/4) alpha beta)))
         (e4 (make-sp-array '(8 8) :with-entries
                            '((1 0 0) (1 1 1) (1 2 2) (1 3 3))))
         (f4 (make-sp-array '(8 8) :with-entries
                            '((1 4 4) (1 5 5) (1 6 6) (1 7 7))))
         (ac-S1
          (sp-x `(e8-a)
                `(,(sp+ `(1/4 ,e4) `(-1/4 ,f4)) a b)
                `(,ixmap-e8-so16-alphabeta e8-a a b)))
         (ac-S2
          (sp-x `(e8-a)
                `(,(sp+ `(1/4 ,e4) `(-1/4 ,f4)) a* b*)
                `(,ixmap-e8-so16-alpha*beta* e8-a a* b*)))
         (ac-V (e8-ac-[] ac-s1 ac-s2))
         (ac-X (sp-x `(e8)
                     `(,ixmap-e8-so16-alphabeta e8 alpha beta)
                     `(,(make-sp-array '(8 8)
                                       :with-entries '((1 4 4) (1 5 5) (1 6 6) (-3 7 7)))
                       alpha beta))))
    ;; this is just so3-singlet
    (if (eq type :symbolic)
        (let* ((rot-wz
                (sp* (poexp-make-rot (sp-ac-to-generator e8-fabc ac-W) w)
                     (poexp-make-rot (sp-ac-to-generator e8-fabc ac-Z) z)))
               (rot-vs
                (sp* (poexp-make-rot (sp-ac-to-generator e8-fabc ac-V) v)
                     (poexp-make-rot (sp-ac-to-generator e8-fabc ac-S1) s)))
               (rot-x (poexp-make-rot (sp-ac-to-generator e8-fabc ac-X) x)))
          (sp* (sp* rot-vs rot-wz) rot-x))
      (labels
          ((gen (g x)
             (sp-exp
              (sp-scale
               (sp-change-arith (sp-ac-to-generator e8-fabc g) *sp-arith-double-float*)
               x))))
        (let* ((rot-wzw (sp* (gen ac-W w) (gen ac-Z z) (gen ac-W (- w))))
               (rot-vsv (sp* (gen ac-V v) (gen ac-S1 s) (gen ac-V (- v))))
               (rot-x (gen ac-X x)))
          (sp* rot-wzw rot-vsv rot-x))))))

(defvar phi-so43
  (poexp* 1 (e8-potential-from-v-theta
             (v-rot-so43 'v 'w 's 'z 'x :type :symbolic)
             (e8-theta-so-p 8))))

(defvar phi-so43-so4
  (poexp* 1 (e8-potential-from-v-theta
             (v-rot-so43 'v 'w 's 'z 'x :type :symbolic)
             (e8-theta-so-p 4))))

(defvar phi-so43-so7
  (poexp* 1 (e8-potential-from-v-theta
             (v-rot-so43 'v 'w 's 'z 'x :type :symbolic)
             (e8-theta-so-p 7))))

\end{verbatim}
}

\section{The $G_{2(-14)}\times F_{4(-20)}$ potential}

{\tiny
\begin{verbatim}

(eval-when (compile load eval)
  (progn
    (setf lambdatensor::*poexp-vars*
          (coerce '(V S W Z X R0 R1 R2 R3 R4 R5 R6) '(simple-array * (*))))
    (setf (lambdatensor::sp-arith-converter lambdatensor::*sp-arith-poexp*)
          #'(lambda (x) (lambdatensor::poexp-converter x lambdatensor::*poexp-vars*)))
    ))


(defvar ac-g2
  (map 'simple-array
       #'(lambda (x)
           (sp-scale 
            (sp-x `(e8)
                  `(,ixmap-e8-so16-i-j e8 i16 j16)
                  `(,ixmap-so16-i16-i i16 i)
                  `(,ixmap-so16-i16-i j16 j)
                  `(,x i j))
            1/2))
       g2-v8-vs))

(defvar ac-f4 (sp-ac-singlets e8-fabc ac-g2 e8-scalars+so16))


(defvar theta-f4
  (reduce #'(lambda (sf x) (sp+ sf (sp-x `(a b) `(,x a) `(,e8-eta c b) `(,x c)))) ac-f4
          :initial-value (make-sp-array '(248 248))))

;; Now, for g2, we have to trick a bit...

(defvar ac-g2o
  (map 'simple-array #'(lambda (x) (sp-from-vector x '(248)))
       (orthogonal-basis (map 'simple-array #'sp-to-vector ac-g2))))

;; Note that theta-g2 has an extra factor -1 sneaking in from the division by (negative) generator "length^2".

(defvar theta-g2
  (reduce #'(lambda (sf x)
              (let ((iprod (sp-ref (sp-x `() `(,x a) `(,e8-eta a b) `(,x b)))))
                (sp+ sf (sp-scale (sp-x `(a b) `(,x a) `(,e8-eta c b) `(,x c)) (/ 1 iprod)))))
              ac-g2o :initial-value (make-sp-array '(248 248))))

(defvar theta-g2xf4 (sp+ theta-f4 `(+3/2 ,theta-g2)))

;; (e8-p27000 theta-g2xf4) => 0

(setf explicit-theta-g2xf4
  (let*
      ((delta7 (sp-antisymmetrizer 7 2))
       (proj7 (sp+ (sp-id 8) (make-sp-array `(8 8) :with-entries '((-2 7 7)))))
       (delta78 (sp-x `(i j k l)
                      `(,delta7 i7 j7 k7 l7)
                      `(,proj7 i7 i) `(,proj7 j7 j)
                      `(,proj7 k7 k) `(,proj7 l7 l)))
       (traceless
        (sp+ (sp-id 8) (make-sp-array `(8 8) :with-entries `((-7 7 7))))))
    (sp+
     (sp-x `(e8-a e8-b)
           `(,ixmap-e8-so16-alphabeta e8-a alpha beta)
           `(,ixmap-e8-so16-alphabeta e8-b gamma delta)
           `(,(sp-id 8) alpha gamma)
           `(,(sp-id 8) beta (:fix 7)) `(,(sp-id 8) delta (:fix 7)))
     (sp-x `(e8-a e8-b)
           `(,ixmap-e8-so16-alpha*beta* e8-a alpha* beta*)
           `(,ixmap-e8-so16-alpha*beta* e8-b gamma* delta*)
           `(,(sp-id 8) alpha* gamma*)
           `(,(sp-id 8) beta* (:fix 7)) `(,(sp-id 8) delta* (:fix 7)))
     (sp-x `(e8-a e8-b)
           -1
           `(,ixmap-e8-so16-ij e8-a ij)
           `(,ixmap-e8-so16-ij e8-b kl)
           `(,ixmap-so16-ij-i-j ij i16 j16)
           `(,ixmap-so16-ij-i-j kl k16 l16)
           `(,ixmap-so16-i16-i i16 i)
           `(,ixmap-so16-i16-i* j16 j)
           `(,ixmap-so16-i16-i k16 k)
           `(,ixmap-so16-i16-i* l16 l)
           `(,(sp-id 8) j l)
           `(,(sp-id 8) i (:fix 7))
           `(,(sp-id 8) k (:fix 7)))
     (sp-x `(e8-a e8-b)
           -1/2
           `(,ixmap-e8-so16-ij e8-a ij)
           `(,ixmap-e8-so16-ij e8-b kl)
           `(,ixmap-so16-ij-i-j ij i16 j16)
           `(,ixmap-so16-ij-i-j kl k16 l16)
           `(,ixmap-so16-i16-i* i16 i)
           `(,ixmap-so16-i16-i* j16 j)
           `(,ixmap-so16-i16-i* k16 k)
           `(,ixmap-so16-i16-i* l16 l)
           `(,(sp-antisymmetrizer 8 2) i j k l))
     (sp-x `(e8-a e8-b)
           1/16
           `(,ixmap-e8-so16-ij e8-a ij)
           `(,ixmap-e8-so16-ij e8-b kl)
           `(,ixmap-so16-ij-i-j ij i16 j16)
           `(,ixmap-so16-ij-i-j kl k16 l16)
           `(,ixmap-so16-i16-i i16 i)
           `(,ixmap-so16-i16-i j16 j)
           `(,ixmap-so16-i16-i k16 k)
           `(,ixmap-so16-i16-i l16 l)
           `(,(sp+ `(8 ,delta78)
                   `(-1/7 ,(sp-x `(i j k l)
                                 `(,so8-sigma-ijkl-ab i j k l a b)
                                 `(,traceless a b)))
                   `(-1/7 ,(sp-x `(i j k l)
                                 `(,so8-sigma-ijkl-a*b* i j k l a b)
                                 `(,traceless a b))))
             i j k l)))))


;; (sp+ `(-1 ,explicit-theta-g2xf4) theta-g2xf4)
;; => #<sparse (248 248) array [0/61504 entries, hash space=96, hash density=0.000]>




;; Checks show that these indeed do form 8-dimensional semisimple groups.

(defvar ac-su3-left
      (map 'simple-array 
       #'(lambda (x)
           (sp-scale 
            (sp-x `(e8)
                  `(,ixmap-e8-so16-i-j e8 i16 j16)
                  `(,ixmap-so16-i16-i i16 i)
                  `(,ixmap-so16-i16-i j16 j)
                  `(,x i j))
            1/2))
       (sp-heuristic-singlets (list (sp-from-vector #(0 0 0 0 0 0 1 0) '(8)))
                              g2-v8-vs
                              :action #'(lambda (vec gen) (sp-x `(a) `(,gen a b) `(,vec b))))))


(defvar ac-su3-right
      (map 'simple-array 
       #'(lambda (x)
           (sp-scale 
            (sp-x `(e8)
                  `(,ixmap-e8-so16-i-j e8 i16 j16)
                  `(,ixmap-so16-i16-i* i16 i)
                  `(,ixmap-so16-i16-i* j16 j)
                  `(,x i j))
            1/2))
       (sp-heuristic-singlets (list (sp-from-vector #(0 0 0 0 0 0 1 0) '(8)))
                              g2-v8-vs
                              :action #'(lambda (vec gen) (sp-x `(a) `(,gen a b) `(,vec b))))))

;; Note that these are part of G2 and F4:
;;
;; (sp-total-dimension ac-su3-left ac-g2) => 14
;; (sp-total-dimension ac-su3-right ac-f4) => 52

(defvar su3-singlets (sp-ac-singlets e8-fabc (map 'simple-array #'sp+ ac-su3-left ac-su3-right)
                                     e8-scalars)) ; 12

(defvar su3-singlets+so16 (sp-ac-singlets e8-fabc (map 'simple-array #'sp+ ac-su3-left ac-su3-right)
                                          e8-scalars+so16)) ; 22

;; Since this is a bit hard to do: alternative suggestion: study breaking to SU(3)xSU(3).

(defvar su33-singlets (sp-ac-singlets e8-fabc (concatenate 'list ac-su3-left ac-su3-right)
                                      e8-scalars)) ; 8

(defvar su33-singlets+so16 (sp-ac-singlets e8-fabc (concatenate 'list ac-su3-left ac-su3-right)
                                           e8-scalars+so16)) ; 16

(defvar su33-so16 (sp-ac-singlets e8-fabc (concatenate 'list ac-su3-left ac-su3-right) e8-so16)) ; 8

#|

;; Note: the derivative of the SU(3)^2 singlets 16 is also a 16, hence no U(1) factors.

(length (SP-LIN-INDEP-AC-COMMUTATORS e8-fabc su33-so16 su33-so16)) ; => 6

And these stay 6 under further derivative forming. Hence, SO(3)xSO(3).

|#

(setf a0 (SP-LIN-INDEP-AC-COMMUTATORS e8-fabc su33-so16 su33-so16))
(setf so3-a (list (sp+ (aref a0 0) `(-1 ,(aref a0 5)))
                  (sp+ (aref a0 1) `(-1 ,(aref a0 4)))
                  (sp+ (aref a0 2) (aref a0 3))))
(setf so3-b (list (sp+ (aref a0 0) (aref a0 5))
                  (sp+ (aref a0 1) (aref a0 4))
                  (sp+ (aref a0 2) `(-1 ,(aref a0 3)))))

(setf su21-a (sp-ac-commutator-closure e8-fabc
                                       (concatenate 'list so3-a
                                                    (sp-lin-indep-ac-commutators e8-fabc so3-a su33-singlets))))
(setf su21-b (sp-ac-commutator-closure e8-fabc
                                       (concatenate 'list so3-b
                                                    (sp-lin-indep-ac-commutators e8-fabc so3-b su33-singlets))))


(defvar v-rot-su33
  (let* ((rot-s (poexp-make-rot (sp-ac-to-generator e8-fabc (aref su21-a 4)) 's))
         (rot-z (poexp-make-rot (sp-ac-to-generator e8-fabc (aref su21-b 4)) 'z))
         (rot-sz (sp* rot-s rot-z))
         (rot-x1 (poexp-make-rot (sp-ac-to-generator e8-fabc (aref su21-a 0)) 'r1))
         (rot-x2 (poexp-make-rot (sp-ac-to-generator e8-fabc (aref su21-a 1)) 'r2))
         (rot-x3 (poexp-make-rot (sp-ac-to-generator e8-fabc (aref su21-a 2)) 'r3))
         (rot-x4 (poexp-make-rot (sp-ac-to-generator e8-fabc (aref su21-b 0)) 'r4))
         (rot-x5 (poexp-make-rot (sp-ac-to-generator e8-fabc (aref su21-b 1)) 'r5))
         (rot-x6 (poexp-make-rot (sp-ac-to-generator e8-fabc (aref su21-b 2)) 'r6))
         (rot-x123 (sp* rot-x1 (sp* rot-x2 rot-x3)))
         (rot-x456 (sp* rot-x4 (sp* rot-x5 rot-x6)))
         (rot-xsz (sp* rot-x123 (sp* rot-x456 rot-sz))))
    (sp-map #'(lambda (x) (poexp* 1 x)) rot-xsz)))

(map 'simple-array
     #'sp-multiple-p
     (vector (aref su21-a 4) (aref su21-b 4)
             (aref su21-a 0) (aref su21-a 1) (aref su21-a 2)
             (aref su21-b 0) (aref su21-b 1) (aref su21-b 2)
             )
     (vector
      (sp-x `(e8)
            -1/2
            `(,ixmap-e8-so16-alpha*beta* e8 alpha* beta*)
            `(,(sp+ `(+1 ,(sp-x `(alpha* beta*) `(,(sp-id 8) alpha* (:fix 1)) `(,(sp-id 8) beta* (:fix 7))))
                    `(-1 ,(sp-x `(alpha* beta*) `(,(sp-id 8) alpha* (:fix 7)) `(,(sp-id 2) beta* (:fix 1)))))
                   alpha* beta*))
      (sp-x `(e8)
            -1/2
            `(,ixmap-e8-so16-alpha*beta* e8 alpha* beta*)
            `(,(sp+ `(+1 ,(sp-x `(alpha* beta*) `(,(sp-id 8) alpha* (:fix 1)) `(,(sp-id 8) beta* (:fix 7))))
                    `(+1 ,(sp-x `(alpha* beta*) `(,(sp-id 8) alpha* (:fix 7)) `(,(sp-id 2) beta* (:fix 1)))))
                   alpha* beta*))
      (sp-x `(e8)
            2
            `(,ixmap-e8-so16-ij e8 ij)
            `(,ixmap-so16-ij-i-j ij i16 j16)
            `(,ixmap-so16-i16-i i16 i)
            `(,ixmap-so16-i16-i* j16 j)
            `(,(sp+ `(+1 ,(sp-x `(i j) `(,(sp-id 8) i (:fix 6)) `(,(sp-id 8) j (:fix 7))))
                    `(-1 ,(sp-x `(i j) `(,(sp-id 8) i (:fix 7)) `(,(sp-id 8) j (:fix 6)))))
              i j))
      (sp-x `(e8)
            2
            `(,ixmap-e8-so16-ij e8 ij)
            `(,ixmap-so16-ij-i-j ij i16 j16)
            `(,ixmap-so16-i16-i i16 i)
            `(,ixmap-so16-i16-i* j16 j)
            `(,(sp+ `(+1 ,(sp-x `(i j) `(,(sp-id 8) i (:fix 6)) `(,(sp-id 8) j (:fix 6))))
                    `(+1 ,(sp-x `(i j) `(,(sp-id 8) i (:fix 7)) `(,(sp-id 8) j (:fix 7)))))
              i j))
      (sp-x `(e8)
            2
            `(,ixmap-e8-so16-ij e8 ij)
            `(,ixmap-so16-ij-i-j ij i16 j16)
            `(,(sp+ `(+1 ,(sp-x `(i16 j16)
                                `(,ixmap-so16-i16-i i16 i)
                                `(,ixmap-so16-i16-i j16 j)
                                `(,(sp-id 8) i (:fix 6)) `(,(sp-id 8) j (:fix 7))))
                    `(-1 ,(sp-x `(i16 j16)
                                `(,ixmap-so16-i16-i* i16 i)
                                `(,ixmap-so16-i16-i* j16 j)
                                `(,(sp-id 8) i (:fix 6)) `(,(sp-id 8) j (:fix 7)))))
              i16 j16))
      (sp-x `(e8)
            -2
            `(,ixmap-e8-so16-ij e8 ij)
            `(,ixmap-so16-ij-i-j ij i16 j16)
            `(,ixmap-so16-i16-i i16 i)
            `(,ixmap-so16-i16-i* j16 j)
            `(,(sp+ `(+1 ,(sp-x `(i j) `(,(sp-id 8) i (:fix 6)) `(,(sp-id 8) j (:fix 7))))
                    `(+1 ,(sp-x `(i j) `(,(sp-id 8) i (:fix 7)) `(,(sp-id 8) j (:fix 6)))))
              i j))
      (sp-x `(e8)
            -2
            `(,ixmap-e8-so16-ij e8 ij)
            `(,ixmap-so16-ij-i-j ij i16 j16)
            `(,ixmap-so16-i16-i i16 i)
            `(,ixmap-so16-i16-i* j16 j)
            `(,(sp+ `(+1 ,(sp-x `(i j) `(,(sp-id 8) i (:fix 6)) `(,(sp-id 8) j (:fix 6))))
                    `(-1 ,(sp-x `(i j) `(,(sp-id 8) i (:fix 7)) `(,(sp-id 8) j (:fix 7)))))
              i j))
      (sp-x `(e8)
            -2
            `(,ixmap-e8-so16-ij e8 ij)
            `(,ixmap-so16-ij-i-j ij i16 j16)
            `(,(sp+ `(+1 ,(sp-x `(i16 j16)
                                `(,ixmap-so16-i16-i i16 i)
                                `(,ixmap-so16-i16-i j16 j)
                                `(,(sp-id 8) i (:fix 6)) `(,(sp-id 8) j (:fix 7))))
                    `(+1 ,(sp-x `(i16 j16)
                                `(,ixmap-so16-i16-i* i16 i)
                                `(,ixmap-so16-i16-i* j16 j)
                                `(,(sp-id 8) i (:fix 6)) `(,(sp-id 8) j (:fix 7)))))
              i16 j16))
      ))
;; => #(1 1 1 1 1 1 1 1)


(defvar phi-g2xf4 (poexp* 1 (e8-potential-from-v-theta v-rot-su33 theta-g2xf4)))
(defvar phi-g2xf4-so8 (poexp* 1 (e8-potential-from-v-theta v-rot-su33 (e8-theta-so-p 8))))
;; (defvar phi-g2xf4-so7 (poexp* 1 (e8-potential-from-v-theta v-rot-su33 (e8-theta-so-p 7))))

\end{verbatim}
}

\section{The $E_{7(-5)}\times SU(2)$ potential}

{\tiny
\begin{verbatim}

(eval-when (compile load eval)
  (require :e8-supergravity)
  (require :cartan-dynkin)
  (use-package :tf-spellbook)
  (use-package :lambdatensor))
  
(setf *bytes-consed-between-gcs* 40000000)

(defvar r+ (sp-find-root-operator e8-fabc e8-v-cartan-ac #(-1/2 -1/2 -1/2 -1/2  -1/2 -1/2 -1/2 -1/2)))
(defvar r- (sp-map #'conjugate r+))

(setf r1 (sp+ `(1/2 ,r+) `(1/2 ,r-)))
(setf r2 (sp+ `(#c(0 1/2) ,r+) `(#c(0 -1/2) ,r-)))
(setf r0 (e8-ac-[] r1 r2))

;; These indeed do close and form a SL(2).

;; Let us add  #(-1/2 -1/2 -1/2 -1/2 1/2 1/2 1/2 1/2), which are also not in e7xsu2:

(defvar r2+ (sp-find-root-operator e8-fabc e8-v-cartan-ac #(-1/2 -1/2 -1/2 -1/2  1/2 1/2 1/2 1/2)))
(defvar r2- (sp-map #'conjugate r2+))

(defvar r21 (sp+ `(1/2 ,r2+) `(1/2 ,r2-)))
(defvar r22 (sp+ `(#c(0 1/2) ,r2+) `(#c(0 -1/2) ,r2-)))
(defvar r20 (e8-ac-[] r21 r22))

(defvar v-rot-sl2^2
  (let* ((r0-scaled (sp-scale r0 1/2)) ; then, [r0,r1]=-2r2, [r1,r2]=2r0, [r2,r0]=-2r1
         (r20-scaled (sp-scale r20 1/2))
         (rot-vs
          (sp* (poexp-make-rot (sp-ac-to-generator e8-fabc r0-scaled) 'v)
               (poexp-make-rot (sp-ac-to-generator e8-fabc r1) 's)))
         (rot-wz
          (sp* (poexp-make-rot (sp-ac-to-generator e8-fabc r20-scaled) 'w)
               (poexp-make-rot (sp-ac-to-generator e8-fabc r21) 'z))))
    (sp* rot-vs rot-wz)))

(setf theta-e7xsu2
      (let* ((8-4a (sp-generate-indexsplit-tensor 8 '(4)))
             (8-4b (sp-generate-indexsplit-tensor 8 '(4) :offset 4))
             (part-abcd
              (sp-x `(p a q b)
                    1/2
                    `(,(sp-id 8) p q)
                    `(,(sp+ (sp-id 8)
                            (sp-x `(a b) `(,so8-sigma-ijkl-ab (:fix 0) (:fix 1) (:fix 2) (:fix 3) a b))) a b)))
             (part-a*b*c*d*
              (sp-x `(p a q b)
                    1/2
                    `(,(sp-id 8) p q)
                    `(,(sp+ (sp-id 8)
                            (sp-x `(a b) `(,so8-sigma-ijkl-a*b* (:fix 0) (:fix 1) (:fix 2) (:fix 3) a b))) a b)))
             (part-ijkl
              (sp+ `(+1/2 ,(sp-x `(a8 b8 c8 d8)
                               `(,(sp-antisymmetrizer 4 2) a4 b4 c4 d4)
                               `(,8-4a a8 a4) `(,8-4a b8 b4) `(,8-4a c8 c4) `(,8-4a d8 d4)))
                   `(-1/2 ,(sp-x `(a8 b8 c8 d8)
                               `(,(sp-antisymmetrizer 4 2) a4 b4 c4 d4)
                               `(,8-4b a8 a4) `(,8-4b b8 b4) `(,8-4b c8 c4) `(,8-4b d8 d4)))
                   `(+12 ,(sp-x `(i j k l) `(,(sp-antisymmetrizer 8 4) i j k l (:fix 0) (:fix 1) (:fix 2) (:fix 3))))))
             (part-i*j*k*l* (sp-scale (sp-antisymmetrizer 8 2) -1/2))
             (part-ij*kl*
              (sp-x `(i j k l)
                    -1
                    `(,(sp-id 8) j l)
                    `(,(sp-id 4) i4 k4)
                    `(,8-4b i i4) `(,8-4b k k4))))
        (sp+
         (sp-x `(e8-a e8-b)
               `(,ixmap-e8-so16-alphabeta e8-a alpha beta)
               `(,ixmap-e8-so16-alphabeta e8-b gamma delta)
               `(,part-abcd alpha beta gamma delta))
         (sp-x `(e8-a e8-b)
               `(,ixmap-e8-so16-alpha*beta* e8-a alpha* beta*)
               `(,ixmap-e8-so16-alpha*beta* e8-b gamma* delta*)
               `(,part-a*b*c*d* alpha* beta* gamma* delta*))
         (sp-x `(e8-a e8-b)
               `(,ixmap-e8-so16-ij e8-a ij) `(,ixmap-e8-so16-ij e8-b kl)
               `(,ixmap-so16-ij-i-j ij i16 j16) `(,ixmap-so16-ij-i-j kl k16 l16)
               `(,ixmap-so16-i16-i i16 i) `(,ixmap-so16-i16-i j16 j)
               `(,ixmap-so16-i16-i k16 k) `(,ixmap-so16-i16-i l16 l)
               `(,part-ijkl i j k l))
         (sp-x `(e8-a e8-b)
               `(,ixmap-e8-so16-ij e8-a ij) `(,ixmap-e8-so16-ij e8-b kl)
               `(,ixmap-so16-ij-i-j ij i16 j16) `(,ixmap-so16-ij-i-j kl k16 l16)
               `(,ixmap-so16-i16-i* i16 i) `(,ixmap-so16-i16-i* j16 j)
               `(,ixmap-so16-i16-i* k16 k) `(,ixmap-so16-i16-i* l16 l)
               `(,part-i*j*k*l* i j k l))
         (sp-x `(e8-a e8-b)
               `(,ixmap-e8-so16-ij e8-a ij) `(,ixmap-e8-so16-ij e8-b kl)
               `(,ixmap-so16-ij-i-j ij i16 j16) `(,ixmap-so16-ij-i-j kl k16 l16)
               `(,ixmap-so16-i16-i  i16 i) `(,ixmap-so16-i16-i* j16 j)
               `(,ixmap-so16-i16-i  k16 k) `(,ixmap-so16-i16-i* l16 l)
               `(,part-ij*kl* i j k l)))))

(defvar phi-e7xsu2 (e8-potential-from-v-theta v-rot-sl2^2 theta-e7xsu2))

#|

* PHI-E7XSU2
#<poexp-term 39/4+1/4 cosh(4 z)+1/4 cosh(4 s)+6 cosh(2 s) cosh(2 z)-1/4 cosh(4 s) cosh(4 z)>

|#

(defvar v-stationary-sl2^2
  (let* ((k (* 1/2 (log (+ 2 (sqrt 3.0d0)))))
         (rot (sp-exp
               (sp-change-arith
                (sp-ac-to-generator e8-fabc (sp+ `(,k ,r1) `(,k ,r21)))
                *sp-arith-double-float*))))
    rot))

;; (e8-numeric-potential-all-derivs v-stationary-sl2^2 theta-e7xsu2) => truly stationary.
;; A123 condition also shows this.

\end{verbatim}
}

\section{The $D=4$ $N=8$ potential}

Since this problem is so complex that it requires splitting of
sub-problems, the code given here is especially instructive when
machine limitations have to be overcome.

{\tiny

}

\newpage

{\begingroup\raggedright
\endgroup
}
\newpage

\thispagestyle{empty}
{\Large Acknowledgments}

\bigbreak

First of all, I want to thank my advisor, Hermann Nicolai, for his
advice and many insightful discussions during my work on this thesis,
and for being a constant source of challenging and beautiful problems.
Furthermore, I want to thank Henning Samtleben for fruitful
collaborations and inspiring discussions. Financial support of the
Max-Planck-Gesellschaft is gratefully acknowledged. Also, I want to
thank my colleagues at the Albert-Einstein-Institute, especially
Kilian Koepsell, Sebastian Silva, Markus P\"ossel, Hanno Sahlmann,
Stefan Fredenhagen, Jan Plefka, Matthias Staudacher, Thomas Thiemann,
Bianca Dittrich, Johannes Brunnemann, Niklas Beisert, and Thomas
Klose, to name only a few, for nice and stimulating discussions and
for contributing to the friendly atmosphere at the institute. It is a
pleasure to thank my mathematical friends Klaus Aehlig and Marc
Sch\"afer for countless nice (and lengthy) discussions about
functional programming. Finally, I want to thank my family
for support, and especially Helena for her love.

\bigbreak

\noindent This work is dedicated to my closest friend with whom I shared so
much, and whom I always will be and always have been very proud of.
Shortly before the defense of her MD thesis, Yvonne Schie\ss{}ler
unexpectedly fell sick of with lethal meningitis and left this world
at the untimely age of 26. I miss you dearly, Yvonne!


\begin{thebibliography}{10}

\bibitem{Berends:wu}
F.~A.~Berends, J.~W.~van Holten, P.~van Nieuwenhuizen and B.~de Wit,
``On Spin 5/2 Gauge Fields,''
Phys.\ Lett.\ B {\bf 83}, 188 (1979)
[Erratum-ibid.\  {\bf 84B}, 529 (1979)].

\bibitem{Berg:2001ty}
M.~Berg and H.~Samtleben,
``An exact holographic RG flow between 2d conformal fixed points,''
JHEP {\bf 0205} (2002) 006
[arXiv:hep-th/0112154].


\bibitem{Bianchi:2001de}
M.~Bianchi, D.~Z.~Freedman and K.~Skenderis,
``How to go with an RG flow,''
JHEP {\bf 0108} (2001) 041
[arXiv:hep-th/0105276].

\bibitem{HeinzBilling} The main web page for the {\em Heinz Billing Award for the Advancement of Scientific Computation} is {\tt http://www.mpg.de/billing/eng/billinguk.html}

\bibitem{Breitenlohner:jf}
P.~Breitenlohner and D.~Z.~Freedman,
``Stability In Gauged Extended Supergravity,''
Annals Phys.\  {\bf 144} (1982) 249.

\bibitem{Coleman:ad}
S.~R.~Coleman and J.~Mandula,
``All Possible Symmetries Of The S Matrix,''
Phys.\ Rev.\  {\bf 159} (1967) 1251.

\bibitem{Cremmer:1978km}
E.~Cremmer, B.~Julia and J.~Scherk,
``Supergravity Theory In 11 Dimensions,''
Phys.\ Lett.\ B {\bf 76} (1978) 409.


\bibitem{Cremmer:1980zb}
E.~Cremmer,
``N=8 Supergravity,''
LPTENS 80/9
{\it Presented at the Europhysics Study Conf. on Unification of the Fundamental Interactions, Erice, Italy, Mar 17-24, 1980}

\bibitem{Cremmer:1997ct}
E.~Cremmer, B.~Julia, H.~Lu and C.~N.~Pope,
``Dualisation of dualities. I,''
Nucl.\ Phys.\ B {\bf 523} (1998) 73
[arXiv:hep-th/9710119].


\bibitem{Deser:1976eh}
S.~Deser and B.~Zumino,
``Consistent Supergravity,''
Phys.\ Lett.\ B {\bf 62} (1976) 335.

\bibitem{Deser:uq}
S.~Deser and B.~Zumino,
Phys.\ Rev.\ Lett.\  {\bf 38} (1977) 1433.

\bibitem{deWit:1981eq}
B.~de Wit and H.~Nicolai,
``N=8 Supergravity With Local SO(8) X SU(8) Invariance,''
Phys.\ Lett.\ B {\bf 108} (1982) 285.

\bibitem{deWit:1982ig}
B.~de Wit and H.~Nicolai,
``N=8 Supergravity,''
Nucl.\ Phys.\ B {\bf 208} (1982) 323.

\bibitem{deWit:1983xe}
B.~de Wit, P.~G.~Lauwers, R.~Philippe and A.~Van Proeyen,
``Noncompact N=2 Supergravity,''
Phys.\ Lett.\ B {\bf 135} (1984) 295.

\bibitem{deWit:2002vt}
B.~de Wit, H.~Samtleben and M.~Trigiante,
``On Lagrangians and gaugings of maximal supergravities,''
Nucl.\ Phys.\ B {\bf 655} (2003) 93
[arXiv:hep-th/0212239].

\bibitem{Ferrara:1976fu}
S.~Ferrara and P.~van Nieuwenhuizen,
``Consistent Supergravity With Complex Spin 3/2 Gauge Fields,''
Phys.\ Rev.\ Lett.\  {\bf 37} (1976) 1669.

\bibitem{Freedman:1976aw}
D.~Z.~Freedman and A.~Das,
``Gauge Internal Symmetry In Extended Supergravity,''
Nucl.\ Phys.\ B {\bf 120} (1977) 221.

\bibitem{Fischbacher:2002hm}
T.~Fischbacher,
``Introducing LambdaTensor1.0 - A package for explicit symbolic and numeric Lie algebra and Lie group calculations,''
arXiv:hep-th/0208218.
(See also {\tt http://www.cip.physik.uni-muenchen.de/\textasciitilde{}tf/lambdatensor/})

\bibitem{Fischbacher:2002fx}
T.~Fischbacher, H.~Nicolai and H.~Samtleben,
``Vacua of maximal gauged D = 3 supergravities,''
arXiv:hep-th/0207206.

\bibitem{FNS}
T.~Fischbacher, H.~Nicolai and H.~Samtleben, work in progress.


\bibitem{Fischbacher:2002hg}
T.~Fischbacher,
``Some stationary points of gauged N = 16 D = 3 supergravity,''
Nucl.\ Phys.\ B {\bf 638} (2002) 207
[arXiv:hep-th/0201030].

\bibitem{Fischbacher:billing2002}
T.~Fischbacher, ``LambdaTensor -- Numerische und symbolische multilineare Algebra auf d\"unn besetzten Tensoren'',
talk given at the 10th Heinz Billing award event, available at
{\tt http://www.cip.physik.uni-muenchen.de/\textasciitilde{}tf/billing2002/} (in german).

\bibitem{Fischbacher:4dpotential}
{\tt http://www.cip.physik.uni-muenchen.de/\textasciitilde{}tf/supergravity/E7/}

\bibitem{Freedman:1976xh}
D.~Z.~Freedman, P.~van Nieuwenhuizen and S.~Ferrara,
``Progress Toward A Theory Of Supergravity,''
Phys.\ Rev.\ D {\bf 13} (1976) 3214.

\bibitem{Freedman:1999gp}
D.~Z.~Freedman, S.~S.~Gubser, K.~Pilch and N.~P.~Warner,
``Renormalization group flows from holography supersymmetry and a  c-theorem,''
Adv.\ Theor.\ Math.\ Phys.\  {\bf 3} (1999) 363
[arXiv:hep-th/9904017].

\bibitem{GambitScheme}
The Gambit Scheme Webpage is {\small {\tt http://www.iro.umontreal.ca/\textasciitilde{}gambit/}}

\bibitem{Girardello:1999bd}
L.~Girardello, M.~Petrini, M.~Porrati and A.~Zaffaroni,
``The supergravity dual of N = 1 super Yang-Mills theory,''
Nucl.\ Phys.\ B {\bf 569} (2000) 451
[arXiv:hep-th/9909047].

\bibitem{Glashow:ep}
S.~L.~Glashow and M.~Gell-Mann,
``Gauge Theories Of Vector Particles,''
Annals Phys.\  {\bf 15} (1961) 437.

\bibitem{Golfand}
Yu. A. Gol'fand and E.P. Likhtman, JETP Lett. 13 (1971) 452 (English p. 323).


\bibitem{Gunaydin:1984qu}
M.~Gunaydin, L.~J.~Romans and N.~P.~Warner,
``Gauged N=8 Supergravity In Five-Dimensions,''
Phys.\ Lett.\ B {\bf 154} (1985) 268.


\bibitem{Haag:1974qh}
R.~Haag, J.~T.~Lopuszanski and M.~Sohnius,
``All Possible Generators Of Supersymmetries Of The S Matrix,''
Nucl.\ Phys.\ B {\bf 88} (1975) 257.

\bibitem{AndreHeckMaple}
A. Heck
``Introduction to Maple, 2nd ed.''
Springer-Verlag New York, Inc. 1990
ISBN 0-387-94535-0

\bibitem{Hull:vg}
C.~M.~Hull,
``Noncompact Gaugings Of N=8 Supergravity,''
Phys.\ Lett.\ B {\bf 142} (1984) 39.

\bibitem{Hull:qz}
C.~M.~Hull,
``More Gaugings Of N=8 Supergravity,''
Phys.\ Lett.\ B {\bf 148} (1984) 297.

\bibitem{Itzhaki:1998dd}
N.~Itzhaki, J.~M.~Maldacena, J.~Sonnenschein and S.~Yankielowicz,
``Supergravity and the large N limit of theories with sixteen  supercharges,''
Phys.\ Rev.\ D {\bf 58} (1998) 046004
[arXiv:hep-th/9802042].

\bibitem{Knuth}
D.~E.~Knuth,
``The Art of Computer Programming, vol. 3: Searching and Sorting'', pp. 506

\bibitem{Maldacena:1997re}
J.~M.~Maldacena,
``The large N limit of superconformal field theories and supergravity,''
Adv.\ Theor.\ Math.\ Phys.\  {\bf 2} (1998) 231
[Int.\ J.\ Theor.\ Phys.\  {\bf 38} (1999) 1113]
[arXiv:hep-th/9711200].

\bibitem{Marcus:1983hb}
N.~Marcus and J.~H.~Schwarz,
``Three-Dimensional Supergravity Theories,''
Nucl.\ Phys.\ B {\bf 228} (1983) 145.

\bibitem{Mezincescu:ev}
L.~Mezincescu and P.~K.~Townsend,
``Stability At A Local Maximum In Higher Dimensional Anti-De Sitter Space And Applications To Supergravity,''
Annals Phys.\  {\bf 160} (1985) 406.

\bibitem{Morales:2002ys}
J.~F.~Morales and H.~Samtleben,
``Supergravity duals of matrix string theory,''
JHEP {\bf 0208} (2002) 042
[arXiv:hep-th/0206247].

\bibitem{GrScWi87}
M.~Green, J.~Schwarz, and E.~Witten, { Superstring theory}.
\newblock Cambridge University Press, Cambridge, 1987.

\bibitem{Nicolai:2000sc}
H.~Nicolai and H.~Samtleben,
``Maximal gauged supergravity in three dimensions,''
Phys.\ Rev.\ Lett.\  {\bf 86} (2001) 1686
[arXiv:hep-th/0010076].

\bibitem{Nicolai:2001sv}
H.~Nicolai and H.~Samtleben,
``Compact and noncompact gauged maximal supergravities in three  dimensions,''
JHEP {\bf 0104} (2001) 022
[arXiv:hep-th/0103032].

\bibitem{Nicolai:2003fw}
H.~Nicolai and T.~Fischbacher,
``Low level representations for E(10) and E(11),''
arXiv:hep-th/0301017.

\bibitem{Pernici:ju}
M.~Pernici, K.~Pilch and P.~van Nieuwenhuizen,
Nucl.\ Phys.\ B {\bf 259} (1985) 460.

\bibitem{cltl2}
Guy L. Steele, Thinking Machines, Inc.,
``Common Lisp the Language, 2nd edition''
Digital Press, 1990
ISBN 1-55558-041-6,
Web version at 
{\tiny{\tt http://www-2.cs.cmu.edu/afs/cs.cmu.edu/project/ai-repository/ai/html/cltl/cltl2.html}}

\bibitem{LeCoLi92}
M.~van Leeuwen, A.~Cohen, and B.~Lisser, { {LiE}, a computer algebra package
  for {L}ie group computations},  { Computer Algebra Nederland, Amsterdam}
  (1992).

\bibitem{VanNieuwenhuizen:ae}
P.~Van Nieuwenhuizen,
``Supergravity,''
Phys.\ Rept.\  {\bf 68} (1981) 189.

\bibitem{Volkov:jx}
D.~V.~Volkov and V.~P.~Akulov,
``Possible Universal Neutrino Interaction,''
JETP Lett.\  {\bf 16} (1972) 438.

\bibitem{Warner:du}
N.~P.~Warner,
``Some Properties Of The Scalar Potential In Gauged Supergravity Theories,''
Nucl.\ Phys.\ B {\bf 231} (1984) 250.

\bibitem{Warner:vz}
N.~P.~Warner,
``Some New Extrema Of The Scalar Potential Of Gauged N=8 Supergravity,''
Phys.\ Lett.\ B {\bf 128} (1983) 169.

\bibitem{Weinberg:kr}
S.~Weinberg,
``The Quantum Theory Of Fields. Vol. 2: Modern Applications''

\bibitem{Wess:tw}
J.~Wess and B.~Zumino,
``Supergauge Transformations In Four-Dimensions,''
Nucl.\ Phys.\ B {\bf 70} (1974) 39.

\bibitem{Witten:1995ex}
E.~Witten,
``String theory dynamics in various dimensions,''
Nucl.\ Phys.\ B {\bf 443} (1995) 85
[arXiv:hep-th/9503124].

\end{thebibliography}
\end{document}